# Pattern Dynamics of a Coupled Map Lattice for Open Flow

Frederick H. Willeboordse[1] and Kunihiko Kaneko[2]

*Department of Pure and Applied Sciences*
*University of Tokyo, Komaba, Meguro-ku*
*Tokyo 153, Japan*

July 6, 1994

**Abstract**

The pattern dynamics of the one-way coupled logistic lattice which can serve as a phenomenological model for open flow is investigated and shown to be extremely rich. For medium and large coupling strengths, we find spatially periodic, quasiperiodic and chaotic patterns with temporal periodicity and analyze their stability with the help of a newly introduced spatial map. Criteria are established for predicting the down-flow bifurcation behavior of the coupled map lattice, and for selecting attractors through modulation of the boundary. For smaller coupling strengths we find a novel regime in which chaotic defects form a periodic lattice that is shown to be the result of a boundary crisis.

## 1 Introduction

In recent years, spatio-temporal chaos has been studied extensively, and the use of coupled map lattices has been demonstrated to be powerful for studying universal features in complex spatiotemporal dynamics [1]-[6]. Of particular interest is the existence of universality classes like pattern selection, frozen random patterns, spatiotemporal intermittency and traveling waves which have, for example, been discovered in the diffusively coupled logistic lattice (DCLL).

In this paper we study the pattern dynamics and spatiotemporal chaos in an open flow system, observed typically in fluid experiments such as pipe flows and also seen in many physical, chemical and biological systems. If one does not aim at modeling a specific physical system, but at distillating some underlying universal characteristics, one only needs the essential ingredients of open flow. In their most simplified form, these may be considered the possibility of local chaos, and the transport of information in the downflow direction. This idea yields, rather naturally, the one-way coupled logistic lattice (OCLL) [7]-[12] (see also [13]-[15] for some other open flow models), originally introduced in order to investigate the features of an open flow system from the viewpoint of dynamical systems. It is defined as

$$x_{n+1}(i) = (1-\epsilon)f(x_n(i)) + \epsilon f(x_n(i-1)), \tag{1}$$

[1]e-mail: frederik@complex.c.u-tokyo.ac.jp
[2]e-mail: kaneko@cyber.c.u-tokyo.ac.jp

where $n$ is the discrete time and $i$ discrete space. The parameter $\epsilon$ represents the strength of the coupling and $f(x)$ is the logistic map given by

$$x_{n+1} = f(x_n) = 1 - \alpha x_n^2, \tag{2}$$

where $\alpha$ is the nonlinearity. Unless mentioned otherwise, the left boundary is fixed, and (as immediately follows from eq. (1)) the right boundary is open. (For some other conceptually related studies see also [16]).

Previous studies [7] revealed that some of the main characteristics of the model (1) are: spatial bifurcation cascades in which the temporal periodicity of the lattice doubles in the downflow direction making space a kind of bifurcation parameter, and the importance of noise for the location of the bifurcation sites resulting in the absence of scaling relations. These results were in principle obtained for not too large coupling constants ($\epsilon \lesssim 0.5$), a regime in which most of the patterns are temporally non-periodic (the exception here being the zigzag pattern), although locally the lattice may very well be almost periodic with kinks traveling by. Recent studies of the diffusively coupled logistic lattice, however, have shown that larger values of $\epsilon$ may yield very interesting dynamics including traveling waves [17] and pattern selection at high nonlinearity [18]. In this report we will show that in the OCLL too, there is a host of fascinating phenomena for $\epsilon \gtrsim 0.5$, like purely spatial chaos, while for smaller values of the coupling constant, we find several new universality classes, including periodic lattices of chaotic defects.

The present paper is organized as follows. In section 2, the main pattern classes are outlined and phase diagrams are given, for both the temporally periodic and temporally non-periodic areas. In order to facilitate analysis, in section 3, we introduce a purely spatial map which can accurately reproduce the temporally periodic patterns, and discuss the notion of convective instability. In section 4, we apply the results of section 3 and show how the stability of the spatial map can be used to predict the bifurcation behavior of the coupled map lattice. For an interval around $\epsilon = 0.9$, we found a spatial bifurcation cascade to chaos which is analyzed with the help of the explicitly solvable spatial map corresponding to a temporal periodicity of one. We also employ a slight variation of the spatial map to associate temporally non-periodic lattices of periodic defects with a boundary crisis. Finally, in section 5 we show that (chaotic) attractors can be selected by appropriately modulating the boundary and find that the encountered phenomena again can be analyzed with the help of the spatial map as an application of the results of section 3. Section 6 is devoted to the discussion and the conclusions.

# 2 Outline of the Phenomenology

The pattern dynamics of the OCLL is extremely rich, but it can nevertheless be divided into two clearly distinct super-classes: spatio-temporal patterns and spatial patterns with perfect temporal periodicity.

A phase diagram displaying the regions in which the major classes dominate is given in fig. 1 (with a slight modification, adapted from [11]). The super-class of spatio-temporal patterns is indicated by STP, with remaining regions belonging to the super-class of spatial patterns with temporal periodicity being the three classes of spatial chaos (SC), spatial quasi-periodicity (SQP) and spatial periodicity (SP). A separate phase diagram indicating the various classes inside the STP region will be given below in section 2.2.

It should be stressed that the phase diagram is only intended to give a rough idea of the locations of the various classes and that it is not meant to be taken too literally. The boundaries between the marked areas are often not that sharp, and in some regions, several attractors may very well coexist. In such cases, only the dominant pattern is indicated.

## 2.1 Spatial Patterns with Temporal Periodicity

Examples of SC, SQP and SP are depicted with their return maps in figs. 2a),b), 2c),d) and 2e),f) respectively. The fixed boundary implies that the most upflow temporal periodicity is one in all cases. Further downflow, the system may bifurcate temporally (see also below), and the spatial pattern may change drastically. If fig. 2a), e.g., we first have some temporally period one homogeneous sites, then some temporally period two homogeneous sites followed by temporally period 4 and 8 spatially quasiperiodic sites before the temporally period 16 spatially chaotic pattern occurs (see for spatial chaos [19, 16]).

The occurrence of multiple attractors can basically be ascribed to either the initial conditions per se, or a combination of the initial conditions and a finite size effect. In order to clarify this distinction, first an essential feature of the OCLL needs to be pointed out: spatio-temporal period doubling [7]. This refers to the interesting phenomenon of temporal period doublings in the spatial direction without changing of the control parameters, i.e., within the same lattice, the temporal periodicity of a lattice site can be half the periodicity of a site further downflow, while the spatial pattern may completely have changed (hence the extreme sensitivity to noise in such cases). An example is shown in fig. 3. Exactly, at which lattice site such spatio-temporal bifurcations occur is often strongly dependent on the initial conditions. If, then, one investigates a finite system, sometimes the lattice may bifurcate before the final site and other times not. Consequently, the attractors may look rather different but really are not which is why this should be attributed to a finite size effect. On the other hand, if several patterns with the same temporal periodicity coexist, the initial conditions per se are in different basins of attraction.

Let us note that if one takes a closer look at eq. (1), it readily follows that for $\epsilon = 1$ the temporal periodicity must be one. For decreasing values of $\epsilon$, the temporal periodicity bifurcates and becomes rather high at the boundary of SC and STP (periodicities of up to 512 can quite easily be found).

## 2.2 Spatio-Temporal Patterns

As can be seen in the phase diagram of the previous subsection (fig. 1), below a certain value of $\epsilon$, patterns are no longer temporally periodic (with exception of the ZZ area). In this area, we would like to distinguish the three basic spatio-temporal phases roughly indicated in the phase diagram fig. 4 and shown in fig. 5. These are Traveling Pattern Selection (TPS), the three zigzag-like pattern types of Perfect Zigzag (PZ), Defect Lattices (DL), and Source Induced Intermittency (SII), and Spatio-Temporal Chaos (STC).

### 2.2.1 Traveling Pattern Selection (TPS)

In TPS, a clear downflow movement of a selected pattern can be seen in fig. 5a). This pattern contains strong remnant chaos, however, and domains sometimes split or merge. At first sight, TPS might appear to resemble the traveling waves of the DCLL somewhat, but the two phenomena are basically unrelated. Here, the motion of the domain walls is a consequence of the one sidedness of the model, the velocity is not quantized (i.e. there is basically one $\epsilon$ dependent velocity), and chaos is never completely suppressed, while in the DCLL the motion is caused by local phase-slips which yield an additive quantized speed (where the absence of a phase-slip implies a non-traveling attractor), and chaos may completely be suppressed [17].

### 2.2.2 Zigzag-like Patterns

For small, but not too small, values of the coupling constant $\epsilon$, zigzag-like patterns, characterized by a basic spatial and temporal periodicity of two, can be found up to the maximum nonlinearity $\alpha = 2$. The zigzag region can be subdivided into the four areas described below.

1) Perfect Zigzag Pattern (PZ).

The perfect zigzag pattern has exact and identical spatial and temporal periodicities. The most well known zigzag pattern is that with a periodicity of two. Without losing the zigzag structure, however, higher periodicities exist too. In fact, one of the possible routes to chaos when decreasing $\epsilon$ from within the zigzag area is a bifurcation cascade. Despite being given as an example of a different region in parameter space, fig. 2e) as such is a zigzag pattern[3].

Below the zigzag area, for certain intervals of the boundary condition, the zigzag regime temporally and spatially bifurcates to chaos for decreasing $\epsilon$ (see section 4.5 on the zigzag regime). At the end of the bifurcation cascade one might expect spatiotemporal chaos. It turns out, however, that for quite a big range of $\epsilon$ there is a chaotic zigzag regime in which an approximate spatial periodicity of two is maintained while all sites behave temporally chaotically. This is illustrated in fig. 5b), where per 2 time steps 60 successive states are overlaid. For other intervals of the boundary condition, we observed spatio-temporal quasiperiodicity. This is likely caused by too large a mismatch of the boundary value and the zigzag solution. That is to say, certain values of the boundary cause such strong chaotic motion at site $i = 1$ that the lattice is prevented from reaching the periodic attractor.

It is interesting to note that within or near the chaotic zigzag area, window-like spatially and temporally periodic parameter regions can be found. For example, $\alpha = 1.70$ and $\epsilon = 0.0675$ yields spatial and temporal periodicities of four while $\alpha = 1.70$ and $\epsilon = 0.08$ yields chaotic zigzag patterns.

2) Defect Lattices (DL).

[3] Although the appearance of a snapshot of a zigzag pattern in the SP area may be identical to that of a zigzag pattern in the zigzag area, we believe that it is nevertheless not inconsistent to classify them in separate groups since all the patterns of the zigzag area (including the temporally non-periodic ones) conceptually form one entity.

Above the zigzag pattern we have observed the rather unexpected occurrence of defect lattices [12] of which an example is shown in fig. 5c). In this regime, when starting from random initial conditions, a lattice is formed that consists of localized defects which have a (usually) predetermined number of lattice sites between them. In fig. 5c), e.g., the distance between defects is 11 sites. We find that when approaching the zigzag region from the DL region, the distance between defects diverges logarithmically as a function of $\epsilon - \epsilon_c$ with $\epsilon_c$ the largest value of $\epsilon$ at which all initial conditions are attracted to the zigzag pattern. The DL attractor always coexists with the perfect zigzag pattern, and consequently, there are initial conditions (although the basin is rather small) for which the system is either partially or entirely attracted to the zigzag pattern. A more detailed account of this regime is given in section 4.6.

3) Source Induced Intermittency (SII).

For values of the nonlinearity $\alpha \gtrsim 1.76$, instead of (but based on) the DL, we find a type of intermittency which we would like to call Source Induced Intermittency since intermittent bursts always originate at the site following a defect.

In this regime, some sections of the lattice select the zigzag pattern and others not as can be seen in fig. 5d). The size and location of the zigzag islands within the chaotic sea changes continuously. In section 4.7, the mechanism of this regimes is outlined in somewhat more detail.

4) Spatio-Temporal Intermittency (type II) (STI II)

For a rather small area at the transition from chaotic zigzag patterns to STC, we observed the occurrence of type II spatio-temporal intermittency which is characterized by the possibility of spontaneous bursts within laminar sections of the lattice. Such a transition from a periodic pattern to STC via STI is rather general in spatially extended systems [1, 5, 20, 4]. In this case the intermittent behavior is caused by the global existence of a tiny leak from one chaotic band to the other. Since the band structure is still prevalent, after a site leaks from one band to the other, there is still a strong tendency for a site to return to the basic period two motion. Spontaneously created defects do not continue to grow while moving in the down-flow direction. An example of STI is given in fig. 5e).

### 2.2.3 Spatio-temporal Chaos (STC)

An example of spatio-temporal chaos is given in fig. 5f). Spatial structures such as small ordered islands can hardly be observed here, although some spatial correlations can be inferred from the many 'rounded' sections. Spatial correlations decay exponentially within a few sites. The figure only shows the STC above the zigzag pattern, but the same is basically also true for the STC below the zigzag pattern (i.e. instead of the rounded sections, there are some zigzag like sections). Indeed, in the latter case, the probability of obtaining zigzag sites decreases drastically as the parameter $\alpha$ or $\epsilon$ departs from the STI region.

### 2.2.4 Relationship with the diffusively coupled logistic lattice

At this stage it might be worthwhile to outline the similarities and differences between the zigzag areas of the DCLL and OCLL. For this regime, the parallel is especially relevant since for the period 2 case, the solutions of the two CMLs are identical due to the spatial symmetry (of course, their stabilities are different in general).

Except for the just mentioned period 2 zigzag pattern, the main similarities occur at the lower transition from periodicity to chaos for decreasing $\epsilon$. In both cases, depending on the boundary conditions, there is a quasiperiodic and a period doubling route to chaos, and as a consequence of the band structure of the attractor we first encounter chaotic zigzag patterns and then (type-II) STI [4].

Above the zigzag pattern, we discovered the DL in the OCLL. Since the zigzag pattern as such is stable in the DCLL, one might expect some similar self-organization to also occur in the diffusively coupled lattice. This is, however, not the case. Instead, depending on the parameters, we have Brownian motion of defects, or defects with possibly some remnant chaos which do not have a minimum distance between them[4]. The underlying mechanism, however, is the same for both the DCLL and the OCLL. The resulting phenomena only differ due to the effects of the asymmetrical coupling.

In the OCLL with a fixed boundary condition, the source of the SII is an upflow, pinned and persistent defect from which bursts emanate (see section 4.7). The pinning of the source is not possible in the DCLL or in an OCLL with periodic boundary conditions, although in the latter case, non-persistent SII like dynamics may be observed as transients.

For the dynamics of the DCLL, frustration may play a very important role. In lattices with an even system size $N$, for example, all defects above the zigzag area will eventually disappear. In this sense, Brownian motion is a transient phenomenon. For lattices with an uneven system size, however, one defect will live forever. Due to the one-way coupling, such frustration cannot affect the OCLL. Nevertheless, in an OCLL with periodic boundary condition, there can very well be a mismatch between the system size and the wavelength.

## 2.3 Characterization of the phases

The various regimes can somewhat be distinguished by considering the spatial correlation function given by

$$C(r) = < \frac{\frac{1}{N}\sum_i x_n(i)x_n(i+r) - < x_n(i) >^2}{< x_n(i)^2 > - < x_n(i) >^2} >, \tag{3}$$

and shown in fig. 6, where $<>$ is the spatial average. In a), which depicts TPS, we see that correlations decay over a distance of several wavelengths (approximately $4-6$ sites for the parameters used) to almost zero. Due to the spatial periodicity, infinite range correlations exist in both, defect lattices and chaotic zigzag patterns as can be seen in b) and d) respectively. As may be expected, there is only very little correlation with chaotic sites in b), yielding a modulation of $C(r)$ equal to the distance between defects (11 sites in this case). Fig. 6c) depicts the correlations for the zigzag islands regime which indeed shows a few short range correlations, but then drops to zero rather quickly. Finally, e) gives the case for spatio-temporal chaos where there is only little short range correlation before reaching zero.

In order to provide some overview, the various results are summarized in table 1.

With regard to previous results, we would like to note that two types of patterns are not represented as such in the phase diagram: the transmission of defects and the flow of random domains [21]. In the present research, we mainly observed the former either during transients in a pattern selection region, or near the boundary of TPS and temporally periodic patterns for certain boundary conditions. That is to say, for identical parameters $\alpha$ and $\epsilon$ we found both the transmission of defects and temporally periodic patterns depending on the value of the boundary. We observed the latter as an upflow spatial transient in the TPS regime.

# 3 Analytical Tools: Spatial Maps and Lyapunov Analysis

## 3.1 Spatial Maps

In this subsection, we will introduce a spatial map which turns out to be an extremely useful tool for the analysis of (almost) all temporally periodic and some temporally non-periodic patterns (see also [16]).

If a lattice site $x_n(i)$ of the OCLL has a temporal periodicity $k$, the equation

$$x_n(i) = F^k(x_n(i)) \tag{4}$$

must hold, where $F^k(x_n(i))$ is the $k$-th iterate of

$$F(x_n(i)) = (1-\epsilon)f(x_n(i)) + \epsilon f(x_n(i-1)). \tag{5}$$

Hence we can formally define an implicit spatial map corresponding to eq. (4) as

$$G^k(x(i)) = -x(i) + F^k(x(i)). \tag{6}$$

With the help of eq. (6) a lattice can in principle be generated by supplying the elements $x(i-1), \dots, x(i-k)$ as initial conditions and then successively finding those roots which correspond to actual solutions of the OCLL while shifting the spatial index $i$ (please note the absence of the time index $n$).

At first instance, one might expect there to be two major problems associated with the above scheme. First, an effective computer algorithm for calculating $F^k(x(i))$ is necessary since one cannot just iterate eq. (4), and second, an equation like eq. (6) will usually have a large number of roots.

The first problem can easily be solved by realizing that $F^k(x(i))$ is essentially nothing but an OCLL with the initial conditions $x(i-1), \dots, x(i-k)$ to be iterated $k$ times. Since the final values of the sites $x(i-1), \dots, x(i-k)$ have no further use, one can furthermore save oneself some overhead by only calculating the relevant parts of the lattice.

The second problem is more tricky, but we will now prove the surprising fact that for sufficiently large $\epsilon$ there is only one root. The derivative of eq. (6) is given by

$$G^{k\prime}(x(i)) = -1 + (1-\epsilon)^k(-2\alpha)^k \prod_{m=0}^{m=k-1} K^m(x(i)), \tag{7}$$

where $K^m(x(i))$ is the $m$-th iterate of $K(x(i)) = (1-\epsilon)f(x(i)) + \epsilon f(x(i-1))$, and $K^0(x(i)) = x(i)$. If $G^{k\prime}(x(i)) < 0 \forall x(i)$, there is at most one root.

Since $|\prod_{m=0}^{m=k-1} K^m(x(i))| \leq 1$ this will certainly be the case if $(1-\epsilon)^k(-2\alpha)^k < 1$, i.e., if

$$\epsilon > \epsilon_c(\alpha) = 1 - \frac{1}{2\alpha}. \tag{8}$$

In appendix A, it is shown that this condition can be improved to $\epsilon > \epsilon_c(\alpha) = 1 - \frac{3}{4\alpha}$ in the case of even $k$, and to $\epsilon > \epsilon_c(\alpha) = 1 - \frac{1}{\alpha}$ in the case of spatially chaotic patterns, a line only slightly above the one separating STP from SC.

It is not surprising that the differences between functions $G^k(x^i)$ above and below (the true value of) $\epsilon_c$ become more pronounced for larger $\alpha$. Let us therefore illustrate this for maximum nonlinearity in fig. 7, where $G^8(x^i)$ is plotted versus $x^i$ for ten different sets of random initial conditions. The frequent occurrence of multiple zeros in a) can clearly be seen, while in b) the function is nearly a straight line with slope -1. In the limit $k \to \infty$ the function will of course exactly be a line if condition (8) holds since the second term on the right hand side of eq. (7) goes to zero.

Numerical evidence furthermore indicates that $\epsilon_c(\alpha)$ coincides with the boundary between temporally periodic states and temporally nonperiodic states. We therefore would like to conjecture that **the occurrence of the temporally periodic patterns is a consequence of the disappearance of destabilizing multiple roots** .

For many sets of parameters, including the ones used in fig. 2, it was verified that the patterns generated with the help of the spatial map eq. (6) indeed possess all the relevant features of the patterns generated with the coupled map model (1). The only exception being the absence of spatio-temporal bifurcations. This is, of course, a direct consequence of the definition of the spatial map which needs the temporal periodicity (through $k$) as a parameter, and therefore can not spontaneously change it[4]. In order to obtain all the sections of fig. 2c), for example, one needs the $k = 1$, $k = 2$ and $k = 4$ spatial maps which generate two homogeneous and a quasiperiodic pattern, respectively.

Although our spatial map is most effective in generating the patterns of the OCLL, this does of course not imply that the stability of the two is identical. We therefore have a short look at Lyapunov analysis in the next subsection, and apply the results in the section 4.2 in order to link the stabilities of the spatial map and the OCLL.

## 3.2 Lyapunov Analysis

The usual Lyapunov spectrum, which yields the stability of the system in the stationary frame, can be determined by taking the eigenvalues of the product of Jacobi matrices,

[4] A possible exception is formed by subharmonics.

i.e.,

$$\lambda^i = \lim_{n\to\infty} \frac{1}{n} \ln \left[ i^{\text{th}}\text{eigenvalue of } (J_{n-1}J_{n-2}...J_0) \right], \tag{9}$$

where the Jacobi matrix at time $n$ is given by

$$(J_n)_{i,j} = \frac{\partial x_{n+1}(i)}{\partial x_n(j)}. \tag{10}$$

In this case the upper triangular matrix elements are all zero, and eq. (9) yields

$$\lambda^i = \log(1-\epsilon) + \frac{1}{T} \sum_{n=1}^{n=T} \log f'(x_n^i), \tag{11}$$

where $T \to \infty$. From this it readily follows that for large coupling constants, all the stationary Lyapunov exponents are negative due to the $\log(1-\epsilon)$ term, and that therefore this cannot be the reason for the bifurcations in the OCLL which may or may not lead to spatial chaos.

For systems in which flow plays an important role, the distinction between so-called absolute and convective instability is an essential concept [13, 22]. An instability which grows in the stationary frame is called absolute, and an instability which only grows in some moving frame convective. With regrad to eq. (11), this is to say that the stationary Lyapunov exponent might very well be negative, but that due to a convective instability the system as such is not stable (or at least not stable everywhere).

In order to measure the growth of a perturbation in a moving frame, Deissler and one of the authors (K.K.)[22, 8] introduced the the co-moving Lyapunov exponent which can be can be calculated through a product of Jacobi matrices defined as

$$J_n = \frac{\partial x_{n+1}(i_1 + [v(n+1)])}{\partial x_n(i_1 + [vn])}, \tag{12}$$

where $i_1$ is the lattice site at which the computation starts after discarding a spatial transient, and $[vn]$ the integer part of $vn$. Contrary to the usual determination of the Lyapunov spectrum [23], the size of the Jacobi matrix is not equal to the system size since a) the OCLL may bifurcate (temporally and/or spatially) and have many types of spatio-temporal patterns depending on the region in space, and b) the Jacobi matrix has to move in the downflow direction with the speed $v$.

Naturally, an accurate estimate of the Lyapunov spectrum can only be obtained if the spatio-temporal pattern does not bifurcate while taking the product of Jacobi matrices. Since such bifurcations do not occur in the spatial map, and since it is furthermore computationally much more efficient, the following spectra are all determined by using lattices generated by a spatial map unless mentioned otherwise.

The co-moving Lyapunov exponent is plotted as function of the velocity $v$ for several values of the coupling constant $\epsilon$ in fig. 8. For *all* the chosen values of $\epsilon$, the pattern has a spatial periodicity of two, and a temporal periodicity of one. As clearly can be seen, one of the curves has a positive region while another one is negative throughout.

# 4 Pattern Dynamics

## 4.1 Spatial Bifurcations

In general (but see also the subsection on stability below), the temporal periodicity of a large upflow section of the lattice is one, if the coupling $\epsilon$ is large enough. Within this temporally period one region, a spatial period doubling sequence to chaos can be observed where $\epsilon$ is the bifurcation parameter. Figure 9 depicts the state of the lattice before and after the $2 \rightarrow 4$ spatial saddle-node bifurcation.

In principle, one could now make a bifurcation diagram by choosing a certain spatial region (e.g. sites $i = 100$ to $i = 200$), plot the amplitudes of all the lattice sites and then adjust the coupling constant. After changing the coupling constant, some transient time should of course be discarded in order for the lattice to adjust to the new parameter and it should be checked that the temporal periodicity is still one. It turns out, however, that for regions sufficiently far downflow, only spatially periodic parts appear in the bifurcation diagram, and that many sections attain a higher temporal period (an explanation for this will be given in section 4.2).

As a first application, we would now like to employ the spatial map to nevertheless obtain a better idea of the bifurcation cascade. The OCLL has a unique solution for the temporally period one state, where

$$x_n(i) = (1-\epsilon)f(x_n(i)) + \epsilon f(x_n(i-1)) \tag{13}$$

must hold. The solution of eq. (13) immediately follows as

$$x(i) = f(x(i-1)) = \frac{-1 + \sqrt{1 + 4\alpha(1-\epsilon)(1 - \alpha\epsilon x(i-1)^2)}}{2\alpha(1-\epsilon)}, \tag{14}$$

where the temporal index was dropped. The shape of this function is plotted in fig. 10, for several values of $\epsilon$. In the limit $\epsilon \rightarrow 1$ this map reduces to a spatial logistic map as can immediately be seen by taking the limit in eq. (13), while for decreasing $\epsilon$ the hump becomes smaller and smaller until it completely disappears for $\epsilon = 0$, reflecting the fact that the lattice becomes a mere collection of single logistic maps without spatial information.

With the help of eq. (14) we can now generate the spatial pattern of the OCLL by starting with an initial value $x(0)$ and calculating successive iterates. Consequently, eq. (14) can be considered as a spatial map corresponding to an OCLL with a temporal periodicity of one, and a complete bifurcation diagram can easily be determined. For two values of $\alpha$ it is shown in fig. 11. Since for $\epsilon = 1$ eq. (14) reduces to a spatial single logistic map, it follows immediately that the bifurcation diagram must end in the state of the bifurcation diagram of the regular single logistic map with the corresponding $\alpha$, and thus be incomplete for $\alpha < 2$. The single hump function of eq. (14) naturally leads one to expect the windows visible in the bifurcation diagram. Since eq. (14) is the solution of the temporally period one OCLL, these windows should also occur in the OCLL. As mentioned above, however, the bifurcation diagram of the OCLL has many holes, and hence it is not completely trivial that they are indeed present as can be seen in fig. 12.

It is notable that the fixed point of eq. (14) is identical to the one of the single logistic map, $x^* = \frac{-1+\sqrt{1+4\alpha}}{2\alpha}$, independent of $\epsilon$, and stable if

$$\epsilon < \epsilon_s = \frac{1+4\alpha}{2(1+4\alpha-\sqrt{1+4\alpha})}. \tag{15}$$

The minimum value $\epsilon_s^{\min} = 0.75$ coincides with maximum nonlinearity, and accordingly, we always should have a stable solution for medium coupling strengths. On the other hand, we also know that the fixed point corresponds to the homogeneous state of the OCLL which is unstable for all chaotic values of $\alpha$.

## 4.2 Linking the Stabilities in the OCLL and the Spatial Map

Thus far we have shown that our spatial maps can effectively be used for generating the associated spatial patterns of the OCLL. We have already noted, however, three issues that need to be addressed; (i) the OCLL may or may not temporally bifurcate while of course a spatial map is only associated with one temporal periodicity, (ii) the bifurcation diagram of temporally period one patterns of the OCLL has many holes while the one of the spatial map does not, and (iii) the temporally period one spatial map has a stable period one solution for large regions of parameter space while the homogeneous state of the OCLL is unstable for such $\alpha$.

These phenomena can be related to the question of stability. Here we study the convective stability of our system, with the help of co-moving Lyapunov exponents. In fig. 13, the maximum co-moving Lyapunov exponent is plotted versus $\epsilon$ for the case of the $k = 1$ spatial map. The spatial patterns correspond to the ones of the bifurcation diagram fig. 11, which are a homogeneous fixed point, a period doubling cascade of spatially periodic patterns and spatial chaos. The non-homogeneous spatially periodic patterns occur approximately in the range $0.81 \lesssim \epsilon \lesssim 0.98$, where the existence of large regions with positive Lyapunov exponents are clearly visible. It should be stressed that in these regions the positive Lyapunov exponents are not associated with chaotic behavior, but with periodic patterns.

On the x-axis of fig. 13, some intervals are marked with diamonds, triangles and circles. In these regions the OCLL has a temporally period one pattern with a spatial periodicity of 2,4 and 8 respectively. Here it was verified (in an OCLL with $N = 10000$) that these patterns are stable and do not bifurcate downflow. As is expected, a negative maximum comoving Lyapunov exponent in the spatial map indicates that the same pattern is stable in the OCLL, and that thus the pattern will not bifurcate somewhere downflow [5].

[5] The large dips around $\epsilon \simeq 0.88$ and $\epsilon \simeq 0.96$ correspond to superstable patterns and fall within the regions in which the pattern is stable in the OCLL. Since there is a complete bifurcation cascade to chaos, this implies that in the OCLL spatial patterns with a spatial periodicity arbitrarily close to spatial chaos can exist. Due to the existence of periodic windows in the bifurcation sequence, stable trajectories with low spatial periodicities can also be arbitrarily close to spatial chaos in parameter space. Accordingly, there is a fractal structure of stable $k = 1$ spatially periodic trajectories in the chaotic region, and without detailed analysis of the bifurcation diagram, it is impossible to predict whether a parameter will yield a stable or unstable pattern.

At this stage it is natural to investigate whether a similar correspondence between the stability of the spatial map and the OCLL can also be established for temporal periodicities larger than one. In fig. 14, the possible maximum comoving Lyapunov exponents of the $k = 2$ and $k = 4$ spatial maps are plotted versus the coupling strength $\epsilon$. Due to the existence of multiple attractors, for every value of $\epsilon$, 20 lattices were generated from random initial conditions. In comparison with fig. 13, it can clearly be seen that the $k = 1$ attractors are also attractors of the $k = 2$ spatial map, and that there is an extra dip around $\epsilon = 0.94$. This dip indeed corresponds to a temporally period 2 and spatially period 4 pattern in the OCLL which was found to be stable. It is also interesting to note, that the patterns of the $k = 1$ spatial map which have a positive maximum co-moving Lyapunov exponent (associated with unstable patterns in the OCLL) do not appear as one of the solutions of the $k = 2$ spatial map.

For all the above calculations of the Lyapunov exponents, we basically only used the spatial map to easily generate a spatial pattern, which was used for the computation of the Jacobi matrices. In this way, the stability obtained, is that of the OCLL. Naturally, the spatial map itself has a stability too, which, in general, will be different. Of course a spatial Lyapunov exponent ($\lambda_{spa}$) is larger than zero for spatially chaotic patterns, equal to zero for spatially quasiperiodic patterns, and smaller than zero for spatially periodic patterns. This was verified by choosing the spatial map as a dynamical system and by measuring the divergence rate of two nearby orbits there.

All in all, we find that our numerical results provide sufficient numerical evidence for the following conjectures:

**Conjecture 1**

*a) A positive Lyapunov exponent in the spatial map ($\lambda_{spa} > 0$) implies a positive maximum co-moving Lyapunov exponent ($\lambda_{max} > 0$) in an open flow system.*

*b) A spatial pattern with a positive maximum co-moving Lyapunov exponent ($\lambda_{max} > 0$) will bifurcate temporally at some lattice site downflow*[6].

The plausibility of conjecture 1a) can be argued by considering a small perturbation of the state of a lattice. If spatial chaos were stable, any small perturbation in the OCLL would have to decay in average over time at all lattice points. At the same time, by assuming the shadowing property [24], with the help of the spatial map, we can generate a spatial pattern which is arbitrarily close to the original spatial pattern, and also a periodic solution of the OCLL. Consequently, a perturbation chosen as the difference of the above two solutions cannot decay in time, contradicting the initial assumption of stability (we have to admit that this does not exclude the special case of marginal stability).

The reverse of this argument does not hold, and although one might at first expect it nevertheless to be true too (after all, even if the magnitude of a Lyapunov exponent changes, qualitative aspects like its sign seem to remain the same) this is not the case. Although not that common, we found a few counter-examples in which spatially quasiperiodic patterns had a positive maximum co-moving Lyapunov exponent in the OCLL.

Conjecture 1b) is closely related to the fact that any perturbation will, at some point downflow, grow to $O(1)$ for any $\lambda_{max} > 0$, however small it may be. The growth of such a perturbation can be estimated by $\delta_n \propto \exp(\lambda_{\max} n)\delta$, where $\delta$ is the amplitude of the perturbation. In order to find the maximal size of the perturbation $j$ sites downflow from a noise source, we need to replace $n$ by $j/v_{max}$, where $v_{max}$ is the velocity at which the co-moving Lyapunov exponent attains its maximum value. We then obtain as the relationship between $\delta$ and the number of sites $j$ for which the perturbation remains below a certain threshold $j \propto \ln(1/\delta)$. Conjecture 1b) then follows from the fact that all physical systems include some noise. For the theoretical case of an infinite precision calculation without any noise, such a pattern could extend until infinity if the solution is linearly stable in a fixed frame (i.e. as long as the maximal stationary Lyapunov exponent $\lambda_0 < 0$).

[6]Of course, a pattern with $\lambda_{max} < 0$ is stable.

So far, we only discussed the stability of attractors as such. For a given value of $k$, however, several attractors (including subharmonics) may coexist, and we would now like to briefly consider the $k = 2$ spatial map as an example of such a case. Around $\epsilon \simeq 0.964$, three attractors coexist in the $k = 2$ spatial map. They are shown with their spatial return maps in fig. 15. Not surprisingly, the $k = 1$ attractor with its negative maximum co-moving Lyapunov exponent is spatially periodic. The $k = 2$ attractor with the larger positive Lyapunov exponent is chaotic, and the remaining attractor is quasiperiodic. Unlike expected, however, the maximum co-moving Lyapunov exponent of the quasiperiodic attractor is not zero but slightly positive (we will come back to this below). The basins of attraction are plotted in fig. 16, where the black squares mark initial conditions that are attracted to the temporally and spatially periodic attractor. The basin of the periodic attractor is intermingled with the basin of the quasiperiodic attractor, and has a fractal structure. To illustrate this the insets show blow-ups of smaller scales. The boundary with the basin of the chaotic attractor, which was not separately marked, is connected. It covers the white areas in the lower left and upper right corners (since the initial conditions are symmetric in $x(0)$ and $x(1)$, only the upper right quadrant is shown). Consistent with conjecture 1, the OCLL was only found to select the periodic attractor.

## 4.3 Spatial Quasiperiodicity

Many of the spatially quasiperiodic patterns we found have a maximum comoving Lyapunov exponent equal to zero. Some spatially quasiperiodic patterns, however, have a slightly positive maximum co-moving Lyapunov exponent, which due to the small value often yields patterns that exist for very large regions of the lattice. In fig. 17, the value of the comoving Lyapunov exponent versus the velocity is plotted for both situations. The insets show the associated patterns with their return maps.

There is a big difference between the the two types of quasiperiodic patterns as far as their sensitivity to noise is concerned. For the case of the attractor with a positive maximum (co-moving) Lyapunov exponent, the numerical result is shown in fig. 18, where we plotted the distance in space from a local noise source before the temporally periodic lattice is destroyed versus the amplitude of the noise at the source. In the case of the attractor with a zero maximum Lyapunov exponent, however, there seems to be a threshold below which the lattice will not be destroyed [7].

[7]Although a zero maximum comoving Lyapunov exponent would in principle imply that induced noise could live forever, one possible explanation for our result might be that the perturbation vector is not

## 4.4 Spatial Chaos

All spatially chaotic patters have a positive maximum co-moving Lyapunov exponent ($\lambda_{max}$) according to conjecture (1a), and conjecture (1b) states that such a pattern must bifurcate. We therefore believe that spatially chaotic patterns do not last over large domains in the lattice. The stationary Lyapunov exponent $\lambda_0$, however, is generally smaller than zero (see section 3.2), and even in the presence of some noise, a spatially chaotic pattern can exist for some section of the lattice, before it grows to a macroscopic order according to $exp(\lambda_{max} n\delta)$. This not only makes it possible to observe spatial chaos in numerical simulations, but also suggests that spatial chaos could be observed in experiments.

In the presence of any tiny noise spatial chaos will be destroyed by local noise, and will be replaced by a spatiotemporally irregular pattern downflow. It is interesting to note that thus far we have not found any differences between the original spatio-temporal chaos and the irregular patterns which are the result of a destroyed spatial chaos. In the Fourier spectrum, for example, there are no traces whatsoever of the original temporal periodicity.

## 4.5 Zigzag Regime

Thus far we have shown that the spatial map $G(x)$ is generally applicable as long as it has a single root. If more than one solution exists, the problem of which root to use becomes non-trivial, but, in principle, there should be no reason why our method could not be used, and indeed, we will now use the zigzag regime as an example to illustrate the application of the spatial map to situations where multiple roots exist.

When decreasing $\epsilon$ from values which yield DL, the number of domains with strong remnant chaos decreases gradually until it becomes zero at the upper boundary of the zigzag area, where chaos is completely suppressed. For not too small $\alpha$, the pattern then has a spatial and temporal periodicity of two with the two temporal states exactly out of phase, and the spatial map eq. (6) is in principle applicable. Since the values of $\epsilon$ in this regime are small, however, there will, in general, be multiple roots, as can be seen in fig. 19, where $G^k(x)$ is plotted for $k = 2$ and $k = 4$ respectively. Consequently, the spatial map cannot be used in the same way as before to create a spatial pattern. For example, in fig. 19a), there are three roots for all initial conditions. Let us denote these roots as L, M and R, respectively. We then have an infinite number of solutions of the OCLL formed by all the possible combinations of the three symbols, and can generate these solutions by selecting the appropriate root of the spatial map. Despite the existence of these solutions, the OCLL only seems to chose one, namely the series $\overline{\mathrm{LR}}$, where the overline indicates repetition (of course the phase is arbitrary, so this is identical to $\overline{\mathrm{RL}}$). Since the middle root M is identical to the $k = 1$ solution, which is identical to the homogeneous solution that can be shown to be be unstable by a simple calculation, it may be expected that it does not appear in the OCLL. Even so, there still is an infinite number of combinations of L and R.

always exactly aligned in the zero direction since the information has to move down the lattice, and that thus small noise is damped.

We believe, that the reason why the OCLL only selects the LR combination is again related to stability. First we would like to note that series with many identical successive symbols are expected to be unstable (again) since the homogeneous solution is unstable. Therefore, we only considered some basic sequences and plotted their comoving Lyapunov exponents versus the velocity $v$ in fig. 20. As can clearly be seen, except for the zigzag pattern, all patterns not only have a positive maximum comoving Lyapunov exponent, but also a positive stationary Lyapunov exponent. Thus it seems natural that they are never realized in the OCLL, as opposed to the patterns for large $\epsilon$ which have a negative stationary and a positive maximum comoving Lyapunov exponent that can be realized in the OCLL sufficiently close to the boundary but later bifurcate. Consequently, we would like to propose the following conjecture

**Conjecture 2** *Symbol sequences which yield a pattern with a negative Lyapunov exponent can be realized in an open flow system. (If the maximum comoving Lyapunov exponent is also negative, the pattern will not bifurcate downflow, while it will bifurcate if the exponent is positive, according to conjecture 1.)*

For decreasing values of $\epsilon$, after the period two zigzag pattern, we again found a spatial bifurcation cascade to chaos. Contrary to the cascade for large $\epsilon$, however, here a spatial bifurcation coincides with a temporal bifurcation, and it should be noted that not only the spatial and temporal periodicities are the same, but that also the sequence and numerical values of the spatial and temporal phases are identical.

With regard to the spatial map this means that the situation becomes slightly more complicated as can be seen in fig. 19b), where the number of roots depends on the initial conditions. It turns out that again the pattern of the OCLL can be reproduced if one selects the appropriate roots. If we number these from left to right as L1,L2,L3,M,R1,R2,R3, the sequence corresponding to the OCLL is L1,R3,L3,R1. This is identical to the sequence of roots of the stable period 4 single logistic map (although for the present value of $\alpha$ the period four fixed points are unstable of course), and we have indications that also for higher periods, this correspondence holds.

In order to check whether our conjecture on the relation between temporal bifurcations downflow and positive maximum comoving Lyapunov exponents also holds in the present case, we used the LR rule for generating $k = 2$ zigzag patterns and determined their stability. The results are shown in fig. 21, where the opaque diamonds indicate the region in which the OCLL selects a zigzag pattern (when starting from random initial conditions) which does not bifurcate downflow. Interestingly enough, this region is rather small ($0.112 \lesssim \epsilon \lesssim 0.113$) and coincides with only slightly negative maximum comoving Lyapunov exponents. For ($0.155 \lesssim \epsilon \lesssim 0.165$), the spatial map frequently yields a homogeneous solution which is of course unstable in the OCLL[8].

[8]Between ($0.113 \lesssim \epsilon \lesssim 0.167$) we usually observed our defect lattices in which some sites show strong remnant chaos while others are periodic. At first this might appear to be somewhat in contradiction with our conjecture, but this is not the case since a stable zigzag pattern can very well exist in the almost the entire region with a negative maximum comoving Lyapunov exponent (i.e. for $0.112 \lesssim \epsilon \lesssim 0.164$) if suitable initial conditions are chosen. Indeed, the easiest way to achieve this is by sweeping $\epsilon$ from a value around $\epsilon \approx 0.112$. Starting from random initial conditions, the basin of the zigzag pattern is very small, however, and there clearly is some sensitivity to the value of the boundary (for fig. 21, we used $x(0) = 1$).

## 4.6 Defect Lattices

Above, we showed that the spatial map is also applicable in the case of the zigzag pattern. In fig. 21, there is no indication whatsoever though as to why random initial conditions would yield defect lattices for $0.113 \lesssim \epsilon \lesssim 0.164$. However, as a nice confirmation of its usefulness, the spatial map can be associated with this phenomenon by considering the number of roots. In fig. 19a), it can be seen that smallest right hand maximum is rather close to zero. It turns out that for $\epsilon \approx 0.115$ it can become smaller than zero implying that for some initial conditions there is only one root. Translated to the OCLL this means that some lattice sites will have values which force the next site to the same root. That is to say sometimes an L solution will be followed by another L solution. Successive L, however, were shown to be unstable and thus (remnant) chaotic motion seems to be a reasonable consequence. At other times, an L solution will be followed by an R solution yielding a rather stable combination as long as the zigzag pattern has a negative maximum comoving Lyapunov exponent.

As for the chaoticity of the lattice sites, if we look at the (stationary) Lyapunov exponents in fig. 22 (see also section 3.2), we see that every defect is formed by only one chaotic site. In this case we have not ordered the exponents in the usual way such that $\lambda_1 \geq \lambda_2 \geq ... \geq \lambda_N$ since due to the upper triangle of the Jacobi matrix being zero, the index $i$ contains actual spatial information and corresponds to the (local) chaoticity of a site.

It is furthermore remarkable that when approaching $\epsilon_c$ from the DL regime, the damping rate does not decrease like in a usual critical phenomenon but remains constant. This can clearly be seen in fig. 23 where the distance between the zigzag pattern and the defect lattice is plotted for several values of the coupling constant. We can now use this fact to approximate the second iterate of eq. 1 as

$$x_{n+2}(i) = (1-\epsilon)f((1-\epsilon)f(x_n(i) + \epsilon f(x_2^*)) + f(x_1^*), \tag{16}$$

where $x_1^*$ and $x_2^*$ are the zigzag solutions which can easily be calculated as

$$x_{1,2}^* = \frac{1 \pm \sqrt{4(1-2\epsilon)^2\alpha + 4\epsilon - 3}}{2(1-2\epsilon)\alpha}. \tag{17}$$

In fig. 24 eq. (16) is plotted just above and below the critical value $\epsilon_c$ in a) and b) respectively. What we can infer from these figures is that above $\epsilon_c$ there are two distinct basins of attraction, one for the chaotic attractor and one for the zigzag attractor, while below $\epsilon_c$ there is only the basin for the zigzag attractor. A boundary crisis occurs at $\epsilon_c$. Thus we can associate the occurrence of a periodic lattice of chaotic defects with a boundary crisis through the following steps:

When starting from random initial conditions, some lattice sites will be attracted to the zigzag pattern and others to the chaotic attractor. If the zigzag sites were exactly on the attractor, eq. (16) would also hold exactly, and consequently the chaotic sites would remain in their basin forever. A chaotic site, however, cannot be followed by a site precisely on the periodic attractor due to the downflow coupling. The chaotic modulation of a site means for eq. (16) that leaks are created from one basin to the other. Hence sites following a chaotic site will be attracted to the periodic solution. Due to the negative Lyapunov exponent of this solution, the chaotic modulation will be damped more and more in the down flow direction, until it is too small to create leaks to the attractive basin. If then a site happened to be in the chaotic basin it will remain there. Since the damping rate is determined by the parameters, it is constant throughout the lattice, and thus the number of lattice sites necessary to damp chaotic modulation below the threshold that allows for leaks is predetermined. Thus there is minimum distance between defects. When the distance between two defects is shorter than this minimum, the downflow defect has to move further downflow. Consequently, virtually every site will at some stage be in the chaotic basin making it very probable that the distance between successive sites is equal to the minimum distance. Thus, in general, we eventually obtain a lattice in which the chaotic defects are evenly spaced.

We note that the presented scenario actually determines only the minimum distance between defects. This can easily be verified by taking a zigzag pattern as the initial condition, and adding some defects manually. If the distance between defects is smaller than the minimum distance, the down flow defect will move further down flow, while otherwise nothing will happen.

We would like to note that for $\alpha \lesssim 1.67$ the transition from the defect lattice to the zigzag pattern becomes less distinctive, i.e. large distances between defects are rather hard to observe, since it involves the zigzag pattern with a spatial and temporal periodicity of four. Depending on the boundary value, we also often see quasiperiodic behavior instead of self-organizing defects.

Below the zigzag regime, it is again possible to roughly analyze the dynamics of the OCLL with the map of eq. (16). When decreasing $\epsilon$ from $\epsilon_c$, we first have a bifurcation cascade to chaos, in which both the spatial and temporal periodicity double. Even when the dynamics has become chaotic though, the period two band structure is still maintained. Although it is not possible to directly employ eq. (16), we can still use it as an approximation for obtaining some qualitative information. As can be seen in figs. 25a) and b), if $\epsilon$ is not too small, it is impossible for a site to escape from the basin of attraction near the right hump, and a period 2 chaotic band structure is the result. Only when $\epsilon$ is lowered much further, a second boundary crisis occurs and the chaotic basin suddenly increases. With regard to the OCLL this means that at a certain moment the chaotic motion of an upflow site is strong enough to create leaks leading to spontaneous bursts and thus leads to (type-II) STI.

Of course we have to again note that the approximation of eq. 16 may be rather rough, in the case of chaotic zigzag patterns. Nevertheless, qualitatively this is not of essential importance.

Just as in the case for smaller $\alpha$ above the zigzag pattern, we also have encountered quasiperiodicity below the zigzag pattern. Depending on the boundary condition, we either observed a period doubling cascade to chaos in which both the spatial and temporal periodicity double simultaneously, or a route to chaos through quasiperiodicity.

## 4.7 Source Induced Intermittency (SII)

For values of the nonlinearity $\alpha > \alpha_c \approx 1.76$ we found a somewhat new type of spatio-temporal intermittency just above the regular zigzag area. In this regime, the mechanism which yields the DL is still at work but bursts may spontaneously be created and spread downflow while destroying the following zigzag sites. Due to this, in a space-time amplitude plot, it appears as if zigzag 'islands' dynamically appear, disappear and change.

Like in the DL region, the perfect zigzag attractor also stably exists in the entire SII area. Consequently, if a section of the lattice is on or near the zigzag attractor, bursts will not spontaneously occur. Bursts can only be created at the site following a defect, as can be argued by considering the same two-dimensional function eq. 16 as for the DL, and its return maps in fig. 24. In the DL area, once a site has entered the upper basin of the periodic fixed point, it will stay there as long as the previous site remains in the chaotic basin. In the SII area, however, the motion of the site in the chaotic basin is strong enough to sometimes create leaks from the periodic basin to the chaotic basin, yielding a burst which may propagate downflow.

Due to the one-sidedness of the coupling, the creation of a burst does not affect the defect itself, and hence defects will no be destroyed unless a burst created further upflow collides with it. Because we have a fixed boundary, this implies that all the way upflow, when starting from random initial conditions [9], there will always be a persistent defect somewhere which acts as a source of bursts (hence the name of this intermittency). An overview of the possible dynamical transitions of a site is given in table 2.

In fig. 26, the stationary Lyapunov spectrum for SII is depicted. Just like in the DL case, the horizontal axis represents space, and all the exponents thus give the local chaoticity. As clearly can be seen, only the exponent corresponding to the source and the one following it are positive [7]. This is due to the fact that all other sites are sometimes regular, (while contributing some negative value to the exponent), and sometimes chaotic, (contributing some positive value). The average of the two turning out to be negative.

Since bursts are not created spontaneously but are always created at the first site after a defect site, this type of intermittency is reminiscent of type I spatio-temporal intermittency (STI I) which does not allow for spontaneous bursts, and in which laminar regions can only change at their borders [1, 20]. Thus, in the STI I for finite lattices, the intermittency terminates after a huge number of steps [25], while the SII lasts for ever.

We will now briefly discuss the dynamics of the system when approaching the zigzag and DL regions from within the SII area and show that except for the approach of one special point, the average spatial size of the laminar regions does not diverge towards the boundaries of the SII region.

In the DL regime, the distance between defects diverges, when decreasing $\epsilon$. Since bursts can only be created near defects, at first one might expect the size of the laminar regions in the SII area to also diverge. This is not the case however, for the following reason. Due to the small change in $\epsilon$ the probability of a burst to occur does not change significantly when approaching the zigzag regime from above. Consequently, the average, time between bursts is more or less fixed and thus the average time the lattice has to

[9] With the exception of the measure zero probability (for $N \to \infty$) that all the randomly chosen initial conditions are very close to the zigzag pattern

relax to the DL state. Hence the average size of a zigzag domain does not diverge.

For $\epsilon < \epsilon_c$, with $\epsilon_c$ the value of the coupling constant that yields zigzag patterns, all defects (including the source, although some remnant chaos may remain) will move downflow and disappear since any site can now reach the basin of the periodic attractor. As $\epsilon$ approaches $\epsilon_c$ from above, the speed to downflow goes to zero, and the transient time is increased.

When decreasing $\alpha$ towards the DL area, the probability to create bursts from defects goes to zero, and thus the average time between bursts diverges. Again, however, the spatial correlations do not diverge, since for constant $\epsilon$ and decreasing $\alpha$, the distance between DL defects remains more or less the same regardless of the frequency of the bursts. The only possibility for obtaining a divergence in the average size of the laminar sections of a lattice is when approaching the DL from the SII area staying just above the perfect zigzag region.

# 5 Modulation of the Boundary

In dynamical systems with multiple attractors of which one or several may be chaotic, being able to assign or influence the selection of a specific attractor (and thus obtaining the means of controlling chaos) is of utmost importance. This was our original motivation for investigating the effects of modulating the boundary. We will now show that not only attractors can be selected, but also that the spatial map can be used to predict in which cases this will be successful.

Thus far we only used the spatial map to generate lattices that correspond to patterns in the OCLL with a temporal periodicity of $k = 2^l, l \varepsilon \mathrm{N}$. Although this is a natural choice from the point of view of the coupled map lattice since other (non-trivial) periodicities have never been reported, there seems to be no a priori reason why the spatial map could not be used to generate patterns with $k \varepsilon N$. In fig. 27, patterns generated with $k = 3 - 8$ are shown for two sets of parameters $\alpha$ and $\epsilon$, with the corresponding co-moving Lyapunov exponents. Spatially periodic, quasiperiodic, and chaotic patterns are obtained depending on the period $k$.

It is interesting to note that the $k = 8$ case in a) is periodic as can clearly be seen in its return map given in b) despite the maximum comoving Lyapunov exponent being zero. This phenomenon can be associated with the downflow motion of the periodic pattern as shown in fig. 27b), obtained from the simulation of the OCLL. The speed is one lattice site per two time steps, and, since the shape of a domain remains the same, this implies that the entire lattice is invariant per 8 time steps. In fig. 27c), the attractors for $k = 3$ and $k = 6$ are identical and quasiperiodic, as is the $k = 8$ pattern which has a slightly positive maximum comoving Lyapunov exponent. The return maps of these two cases are given in fig. 27d).

As in the examples given, we found that for large regions in parameter space, some $k$ exist in which the maximum comoving Lyapunov exponent is either zero or negative. Reversely, we also found that for many given values of $k$, parameter regions exist corresponding to all three basic patterns (spatially and temporally periodic, spatially quasiperiodic with temporal periodicity, spatially chaotic with temporal periodicity). An example for

$k = 3$ is given in fig. 28.

## 5.1 Selection of Attractors

The question which arises now is whether these patterns are an artifact of the spatial map, or whether corresponding patterns also exist in the OCLL. In order to find an answer, we used the fact that perturbations grow in patterns that have positive maximum comoving Lyapunov exponents. Near the boundary, virtually all lattices start with a short homogeneous section. In the chaotic region of the single logistic map this yields a positive maximum comoving Lyapunov exponent and we should therefore be able to set a fundamental frequency for the lattice by modulating the boundary $x(0)$ periodically with a frequency $k$ and a small amplitude.

This idea turns out to work extremely well. Fig. 29 shows the final state of the lattice (sites 200–264) for a fixed boundary condition in a), and for a modulated boundary in b). As can clearly be seen, the unmodulated lattice is spatially chaotic while the modulated lattice is periodic. The fact that it is both, temporally and spatially periodic implies that the pattern is stable against local and global noise.

On the other hand, conjecture (1) is still valid here, and thus a spatially chaotic period 3 pattern bifurcates at some stage downflow. Naturally stepping through the sequence 3-6-12-24... .

Modulating the boundary is very effective in selecting an attractor if the pattern associated with the fixed boundary is unstable, and the pattern associated with the modulated boundary stable. In such cases, even a tiny modulation amplitude (e.g. in the order of 1e-10) will drastically affect the dynamics of the system. If two spatially and temporally periodic attractors coexist, however, the modulation amplitude will likely play an essential role in determining which one will be selected. It should be noted that, of course, a fixed boundary corresponds to a modulation frequency of 1.

By modulating the boundary, a large variety of patterns can be selected. Let us for example consider $\epsilon = 0.5$, a value of the coupling constant for which spatially and temporally periodic pattern are particularly common. The results of our simulations are given in table 3, where the periodicities of the patterns at site $i = 1280$ are shown as a function of the modulation frequency and the nonlinearity. All the patterns are temporally periodic, while nearly two thirds are also spatially periodic. In all cases, the spatial periodicity can be derived from the modulation frequency indicating the existence of strong correlations. This is quite different from the diffusively coupled logistic lattice where the spatial periodicity is usually unrelated to the temporal periodicity. The patterns which have a temporal and spatial periodicity of five, seem to be particularly stable since they can be observed in a much larger range around $\epsilon = 0.5$ than the other ones.

## 5.2 Inverse Bifurcations

In general, modulating the boundary yields a bifurcation cascade which has a basic frequency equal to the modulation frequency and which may stop at some point or continue until the pattern becomes temporally chaotic. If, however, a rather stable attractor with a temporal periodicity different from the modulation frequency exists, we observe the interesting phenomenon of the system (inversely) bifurcating to it as soon as the temporal periodicity equals the smallest common denominator. An example is shown in fig. 30, where (for increasing i) the lattice has the following temporal periodicities: $3 \to 6 \to 12 \to 24 \to 8$. The inverse $24 \to 8$ trifurcation around site 1125 can clearly be seen in the right half of the figure. The (final) attractor with a temporal periodicity of 8 is quasiperiodic, and its maximum comoving Lyapunov exponent is 0. This kind of scenario seems to be quite general, and $40 \to 8$ inverse multifurcations were also observed.

Here we would like to note that these results again form a nice confirmation of our conjecture in section 4.2 according to which patterns with a positive maximum co-moving Lyapunov exponent bifurcate.

# 6 Discussion and Conclusions

In this paper we have reported several new phases which we discovered by studying the one-way coupled logistic lattice. The first question arising then is whether these phases are unique to the present system or whether they are representatives of larger universality classes that can also be observed in other systems and experiments.

We believe that TPS, the zigzag pattern, defect lattices, and intermittency are excellent candidates for universal characteristics of open flow type systems that might be experimentally confirmed in fluid flows, electric convection in liquid crystals, josephson junction arrays (these would be a particularly suitable systems to research, since every junction can be associated with one site in our model rendering the thermodynamic limit completely irrelevant), or optical array systems [16].

In several regions of parameter space we found spatial chaos with temporal periodicity. In the case of infinite precision computation without noise, these patterns can in principle be infinitely long. In practical situations however, there will always be some noise which eventually leads to spatio-temporal bifurcations, and in the thermodynamic limit spatial chaos might not exist as such. Nevertheless, it could be observed for rather large sections of the lattice, and we speculate that it should also be possible to encounter spatial chaos in experimental systems whose phenomenology (or at least certain aspects of it) is described on a macroscopic scale sufficiently well by the OCLL.

In order to analyze the patterns of the OCLL, we introduced a novel class of spatial maps which turned out to be an extremely valuable tool. Among its great merits is that, due to the exact correspondence with the one-way coupled logistic lattice, it can be used for investigating the properties of the OCLL with minimal effort. For example, spatial maps can be used to accurately and efficiently determine the comoving Lyapunov exponents. This is often rather difficult, if not impossible, in the OCLL since patterns may bifurcate and not allow for a sufficient number of Jacobi matrices to be multiplied, or in the case of higher temporal periodicities require impractically large lattices since the desired patterns may only occur far downflow.

Another merit of the spatial map is that it allows us to formally define spatial chaos as a spatial sequence which has a positive Lyapunov exponent with respect to the spatial map. This is important since the relationship between the stabilities of the spatio-temporal and the purely spatial systems is not trivially given. Related to this it should also be interesting for future work to investigate the dimensions of the spatial attractors by examining the generated spatial sequences.

In order to link the dynamics of the OCLL with the stability of our spatial map we have proposed a conjecture according to which a pattern with a positive maximum comoving Lyapunov exponent is unstable in the one way coupled logistic lattice. All our numerical results support this conjecture which makes it possible to predict the downflow behavior of a very high-dimensional spatially extended system with the help of a low-dimensional map that is much easier to handle. We also gave a plausible argument for our conjecture, but further mathematical proof is still necessary. Nevertheless, we believe that the line of reasoning is general enough to assume that the conjecture may also hold in other open flow like systems. In this context it might be worthwhile to mention that spatial chaos was also found to be unstable in the optical array system by Ohtsuka and Ikeda [16].

The spatial map's advantage of corresponding to only one temporal periodicity also includes a restriction. Without actually computing the maximum comoving Lyapunov exponent for every pattern in the (spatio-temporal) bifurcation sequence of the OCLL it is not possible to draw conclusions on their stability. If the maximum exponent is negative, this poses no problem, since the OCLL will remain stable. If the maximum exponent is positive however, all one can do is compute it for the next higher periodicity, and a priori it is impossible to predict whether its pattern will be periodic, quasiperiodic or chaotic. At this moment, with regard to the thermodynamic limit, it remains unclear whether the spatially chaotic patterns will continue to temporally bifurcate, and thus become both spatially and temporally chaotic at some stage, or whether they finally all end up on a periodic or quasiperiodic attractor. We would like to note however, that due to the one way coupling we are assured that regardless of the thermodynamic limit, all of our results remain valid in the upflow section of a lattice. Since the coupled map lattice model in principle acts on a macroscopic or semi-macroscopic scale, this means that any universal properties should be applicable to actual physical systems of a finite size.

Although final conclusions on the existence of spatial chaos with temporal periodicity could not be drawn for the infinite size limit, for large regions in parameter space, rather long sections may indeed be purely spatially chaotic. In all cases investigated, even a small amount of noise eventually led to the destruction of the attractor. Nevertheless, before a small amount of noise has grown to order 1, there is a least some section of the lattice where its influence is still small. We therefore believe that it is not impossible to find our spatial chaos in physical systems.

Even though the temporal periodicity of the attractors of the OCLL can be very high, implying spatial maps of a correspondingly high power, there is only one solution for sufficiently large values of $\epsilon$. We found that this uniqueness of the solution itself plays an important role for the dynamics of the OCLL, since our simulations indicate that (with exception of the perfect zigzag pattern) non-uniqueness of the spatial map coincides with temporally non-periodic patterns in the OCLL.

In the zigzag region, where the OCLL has a perfect temporal periodicity and the spatial map multiple solutions, any spatial sequence generated by chosing arbitrary roots is in principle a solution of the OCLL. The only sequences of roots which were found to be stable, however, were regular ones that followed the sequence of a logistic map with the same temporal periodicity (in the zigzag case, e.g., we have a temporal periodicity of two and need to chose the left and right roots alternatingly). For future research, it might be an interesting question to see whether multiple stable solutions can coexist with more complex symbol sequences than the basic ones investigated here.

The construction of our spatial map is quite straightforward in the case of the OCLL. It is, however, not limited to this case but can also be applied to other systems including the diffusively coupled logistic lattice. So far, however, we have not observed spatial chaos or quasiperiodicity with temporal periodicity in the DCLL. This may be related to the fact that, in general, the comoving Lyapunov exponent $\lambda(v)$ has its maximal value $\lambda_{max}$ around $v \approx 0$. In the case of spatial chaos, however, we have that $\lambda_{max} > 0$ and thus that the stationary Lyapunov exponent $\lambda_0 > 0$. Accordingly, spatial chaos is unstable even in the stationary frame, rendering its observation impossible. It is therefore likely that the effectiveness of the spatial map and the the observation of spatial chaos are limited to rather asymmetric couplings.

In the region of spatially and temporally non-periodic patterns, we found the zigzag pattern to be located below two novel states. Above the zigzag pattern, for nonlinearities smaller than a certain critical value, we observed the periodic lattices of chaotic defects which were associated with the occurrence of a crisis. For larger values of the nonlinearity, a novel type of spatio-temporal intermittency was found, which is maintained by the bursts of a persistent defect, and hence called source induced intermittency.

We have demonstrated the possibility of selecting attractors through the modulation of the boundary, which may have interesting implications for the control of chaos. For some parameters, for example, there might not be any stable attractor along the regular bifurcation sequence, while there is one for an odd temporal periodicity. In such a case, chaos could be controlled solely by acting on the boundary. This is quite different from some other proposals for controlling chaos that require large numbers of feedback terms throughout the system. In actual physical applications the latter seems to be close to the impossible, while a modulation of the boundary can easily be achieved.

In this context, it might be interesting to point out that the OCLL can be related with the time-delayed map (introduced by one of the authors (F.W.) in ref. [26]). Control might then be achieved through adopting a delayed feedback by employing the delay-time as a variable which could indirectly induce the desired modulation frequency without overly artificial procedures.

Several of our results indicate that the system preferably selects a pattern with the lowest possible maximum comoving Lyapunov exponent. For example, the OCLL quickly bifurcates to a temporal periodicity of two when the $k = 1$ spatial map has a positive maximum comoving Lyapunov exponent, but remains stable even if, for identical parameters, the $k = 2$ spatial map has two other coexisting attractors with larger exponents. Finally, the ease with which patterns with an odd temporal periodicity can be selected through the modulation of the boundary, and the occurrence of inverse bifurcations also point into this direction. All in all we might be so bold as to speculate on the existence of a minimum expansion principle [10]. We also found many patterns for which the relation

[10] At first this might seem to be in conflict with the results presented in the section on the zigzag patterns where solutions with chaotic sites and non-chaotic sites coexist. However, even if a minimum expansion principle holds, a lattice site can be blocked from reaching the preferred state by a repelling basin separatrix, i.e. a situation reminiscent of a metastable state.

$\lambda_{max} > \lambda_{spa}$ seemed to hold. It would be rather interesting to see whether or in which cases this is true since that would allow us to formulate conjecture (1) quite a bit more strongly.

**Acknowledgements**

We would like to thank N.B. Ouchi and T. Yamamoto for fruitful discussions. One of the authors (KK) would like to thank S. Mizumi for his collaboration during the early stages of this work. This work was supported by a Grant-in-Aid of the Japan Society for the Promotion of Science (JSPS) under grant no. 93043, and a Grant-in-Aid for Scientific Research from the Ministry of Educations, Science and Culture, Japan under grant no. 05836006.

# A Improvement of the condition for single roots

If the periodicity of the spatial map is a multiple of two, the condition for a single root can be improved by calculating the worst case maximum of $\prod_{m=0}^{m=1} K^m(x(i))$:

$$\prod_{m=0}^{m=1} K^m(x(i)) < (1-(1-\epsilon)\alpha x^2(i))x = \frac{2}{3\sqrt{3(1-\epsilon)\alpha}}. \tag{18}$$

Inserting this into eq. (7) for $k = 2$ we obtain

$$\epsilon > \epsilon_c(\alpha) = 1 - \frac{3}{4\alpha}, \tag{19}$$

which enlarges the monotonic region in the phase diagram by about 50%.

In the case of spatial chaos further improvement can be achieved by considering that for sufficiently high iterates, the logistic map can be thought of as the generator of a random variable with a mean value of $<\bar{x}>= 1-\alpha/2$. For large enough $k$, we can then use the approximation

$$\prod_{m=0}^{m=k-1} K^m(x(i)) \approx \prod_{m=0}^{m=k-1} \left((1-\epsilon)f(\bar{x}_m) + \epsilon f(\bar{x}_{-m-1})\right), \tag{20}$$

where the subscript $m$ in $\bar{x}_m$ indicates that the variable needs to be randomly chosen every time it is used. Since we only need to be concerned with positive right hand sides in eq. (7), we have

$$\begin{aligned}&\prod_{m=0}^{m=k-1} \left((1-\epsilon)f(\bar{x}_m) + \epsilon f(\bar{x}_{-m-1})\right) \\ &\qquad < \prod_{m=0}^{m=k-1} \left((1-\epsilon)f(\bar{x}_m) + \epsilon\right) = \prod_{m=0}^{m=k-1} \left(1-(1-\epsilon)\alpha\bar{x}_m^2\right).\end{aligned} \tag{21}$$

If we furthermore assume the random variable to be uncorrelated, as is reasonable for high iterates, we obtain

$$< \prod_{m=0}^{m=k-1} \left(1-(1-\epsilon)\alpha\bar{x}_m^2\right) >= \left(<1-(1-\epsilon)\alpha\bar{x}_m^2>\right)^k = (1-(1-\epsilon)\alpha/2)^k, \tag{22}$$

where we used that the left hand term just a logistic map with $a' = (1-\epsilon)\alpha$. Inserting eq. (22) into eq. (7), the condition for monotony becomes

$$\epsilon > \epsilon_c(\alpha) = 1 - \frac{1}{\alpha}, \tag{23}$$

yielding a line which is only slightly above the line separating STP fron SC.

**Table Captions**

Tbl. 1. The various characteristics of the phases in the spatio-temporally non-periodic region. The row labeled with 'attractors' is to indicate whether in a regime several attractor types coexist or not. The row labels with $\lambda > 0$ indicates approximately how many positive stationary Lyapunov exponents each regime has. It should be noted that all regimes except for the periodic and quasiperiodic zigzag area have positive maximum co-moving Lyapunov exponents. Multiple entries in one column indicate that within the same regime various possibilities exist. It is furthermore notable that the number of positive stationary Lyapunov exponents is zero for all zigzag patterns, including the chaotic one. The STI II, which possibly is not a phase, but a transition regime, is included for reference.

Tbl. 2. The various possible dynamical transitions in the SII regime. The left column indicates the initial states, and the right column the states after the transition, and the conditions undeer which the transition can occur.

Tbl. 3. Periodicities of the patterns at site $i = 1280$ as a function of the modulation frequency of the boundary (vertical direction) and the nonlinearity (horizontal direction). The coupling constant is $\epsilon = 0.5$, and the amplitude of the modulation 1e-3. The temporal periodicity and spatial periodicity are indicated with **tp** and **sp**, respectively. A dash indicates that no periodicity could be detected.

## Figure Captions

Fig. 1. Phase diagram of the open flow model. The system size is $N = 384$. Only the predominant patterns are indicated and labeled as SP (spatially and temporally periodic), SQP (spatially quasiperiodic but temporally periodic), SC (spatially chaotic but temporally periodic), and STP (spatially and temporally non-periodic), respectively.

Fig. 2. The main pattern classes of the OCLL in the temporally periodic region. The system size is $N = 1000$ and $\alpha = 1.50$. a) Spatial chaos (SC) with temporal periodicity. The coupling constant is $\epsilon = 0.55$ and the temporal periodicity is 16 for $158 \leq i \leq 846$. b) Spatial return map corresponding to a). c) Spatial quasiperiodicity (SQP) with temporal periodicity. The coupling constant is $\epsilon = 0.6$ and the temporal periodicity is 4 for $88 \leq i \leq 1000$. d) Spatial return map corresponding to c). e) Spatial periodicity (SP) with temporal periodicity. The coupling constant is $\epsilon = 0.975$, the temporal periodicity is 1 and the spatial periodicity 8 for $1 \leq i \leq 1000$. f) Spatial return map corresponding to e). The first 200 lattice sites were discarded as spatial transients in the spatial return maps.

Fig. 3. Spatio-temporal bifurcations. The nonlinearity is $\alpha = 1.45$ and $\epsilon = 0.5$. The numbers indicate the temporal periodicity.

Fig. 4. Phase digram for the spatio-temporal patterns. As in fig. 1, the regions are only intended to give a rough indication of the location of the various universality classes. For the meanings of the abbreviations, please see fig. 5.

Fig. 5. The main universality classes in the STP region. a) Traveling Pattern Selection (TPS): the nonlinerity is $\alpha = 1.65$ and $\epsilon = 0.23$, every 4th time step is plotted. b) Chaotic zigzag pattern below the perfect zigzag pattern: the nonlinearity is $\alpha = 1.70$ and $\epsilon = 0.08$, every 2nd time step is plotted. c) Defect Lattice (DL): the nonlinearity is $\alpha = 1.72$ and $\epsilon = 0.1275$, every 2nd time step is plotted. The distance between defects is 11 lattice sites. d) Source Induced Intermittency (SII): the nonlinearity is $\alpha = 1.8$ and $\epsilon = 0.17$, every 8th time step is plotted. e) Spatio-Temporal Intermittency of type II. The nonlinearity is $\alpha = 1.7$, and the coupling constant is $\epsilon = 0.061$. Every fourth time step is plotted. f) Spatio-Temporal Chaos (STC): the nonlinearity is $\alpha = 1.8$ and $\epsilon = 0.25$, every 4th time step is plotted.

Fig. 6. Spatial correlations of the various spatio-temporal patterns. a) Traveling pattern selection, $\alpha = 1.65$, $\epsilon = 0.23$. b) Defect lattice, $\alpha = 1.72$, $\epsilon = 0.1275$. c) Chaotic zigzag pattern, $\alpha = 1.70$, $\epsilon = 0.08$. d) Zigzag islands (SII), $\alpha = 1.80$, $\epsilon = 0.17$. e) Spatio-temporal chaos, $\alpha = 1.80$, $\epsilon = 0.25$.

Fig. 7. Plots of the function $G^8(x)$ for values of the coupling constant below and above $\epsilon_c(\alpha)$. The nonlinearity is $\alpha = 2.0$, while the coupling constant is $\epsilon = 0.3$ in a), and $\epsilon = 0.7$ in b).

Fig. 8. The co-moving Lyapunov exponent $\lambda(v)$ for the $k = 1$ spatial map as a function of the velocity $v$. The nonlinearity is $\alpha = 1.5$, and the product of 5000 Jacobi matrices was taken. In all the cases, the patterns have a spatial periodicity of 2.

Fig. 9. Spatial bifurcation in the OCLL. The state of the lattice before and after the spatial bifurcation is depicted. The temporal periodicity is 1, while the spatial periodicity is 2 in a), and 4 in b). The nonlinearity is $\alpha = 1.5$ and the system size is $N = 1000$, although only the first 64 sites are displayed. The coupling strength is $\epsilon = 0.9$ in a) and $\epsilon = 0.97$ in b).

Fig. 10. The function values of the map eq. (14) for several coupling strengths. The nonlinearity is $\alpha = 1.5$. As can clearly be seen, except for $\epsilon = 0$, this is a single hump function.

Fig. 11. Spatial bifurcation cascades to chaos for eq. (14). In a) the nonlinearity is $\alpha = 1.5$, and in b) the nonlinearity is $\alpha = 2.0$. The amplitudes of sites 1000–1200 are plotted in the $y$-direction.

Fig. 12. Spatial window in the temporally period one OCLL. The system size is $N = 1000$, but only the first 64 are shown. Due to the high stability of this pattern, there is virtually no spatial transient. The nonlinearity is $\alpha = 1.50$, and $\epsilon = 0.995$.

Fig. 13. The maximum co-moving Lyapunov exponent computed with the help of the $k = 1$ spatial map versus the coupling constant $\epsilon$. The nonlinearity is $\alpha = 1.5$, and for each value of $\epsilon$, the product of 5000 Jacobi matrices was taken. Regions in which the OCLL has a temporal periodicity of one at lattice site 1000 are indicated with their markers on the $x$-axis. The spatial spatial wavelengths are 2,4 and 8 sites respectively.

Fig. 14. The possible maximum co-moving Lyapunov exponents of the spatial map versus the coupling constant $\epsilon$. For every value of $\epsilon$, the maximum comoving Lyapunov exponent of 20 patterns generated from random initial conditions was plotted. The nonlinearity was $\alpha = 1.5$, and the product of 5000 Jacobi matrices was taken. In a), $k = 2$, and in b), $k = 4$.

Fig. 15. Return maps of the three coexisting attractors in the spatial map $k = 2$ with $\alpha = 1.5$ and $\epsilon = 0.964$. The spatially periodic attractor has a temporal periodicity of 1, while the two others have a temporal periodicity of two.

Fig. 16. The basins of attraction of the three attractors of fig. 15. The black dots indicate initial conditions that yield the spatially and temporally periodic attractor. Initial conditions in the upper right and lower left corner (where the insets are) yield the chaotic attractor, while the remaining initial conditions yield the quasiperiodic attractor.

Fig. 17. Comoving Lyapunov exponent versus the velocity $v$ for two quasiperiodic patterns. The lower line represents $\alpha = 1.60$, $\epsilon = 0.6$ and $k = 8$, while the upper

line represents $\alpha = 1.50$, $\epsilon = 0.6$ and $k = 4$. The calculations were performed on lattices generated by the spatial map, and the product of 5000 Jacobi matrices was taken.

Fig. 18. The influence of local noise on the quasiperiodic pattern of fig. 17 with a positive maximum co-moving Lyapunov exponent, $\alpha = 1.50$ and $\epsilon = 0.6$.

Fig. 19. Plots of the function $G^k(x(i))$ for small values of the coupling constant $\epsilon$ for 30 random initial conditions. a) The nonlinearity is $\alpha = 1.7$, $\epsilon = 0.11$, and $k = 2$. b) The nonlinearity is $\alpha = 1.7$, $\epsilon = 0.10$, and $k = 4$.

Fig. 20. Comoving Lyapunov exponent for patterns generated with the $k = 2$ spatial map using roots as indicated by the symbols. The nonlinearity is $\alpha = 1.7$, and the coupling constant is $\epsilon = 0.114$.

Fig. 21. Maximum comoving Lyapunov exponent versus $\epsilon$ for patterns generated with the $k = 2$ spatial map by the LR rule. The nonlinearity is $\alpha = 1.70$, and for each datum, the product of 5000 Jacobi matrices was taken. The boundary was fixed to 1. The opaque diamonds indicate the region where the OCLL selects a zigzag pattern which does not bifurcate downflow.

Fig. 22. Stationary Lyapunov exponents for DL, $\alpha = 1.72$ and $\epsilon = 0.1275$. The horizontal axis represents space, and the Lyapunov exponents represent the local chaoticity of a site.

Fig. 23. Distance between the defect lattice and the zigzag pattern. The numbers in the graph indicate the values of $\epsilon$ for the corresponding lines. The nonlinearity was $\alpha = 1.72$.

Fig. 24. Plots of eq. (16) just before and just after the critical point $\epsilon_c$. The nonlinearity is $\alpha = 1.72$ and the coupling strength is $\epsilon = 0.1269$ in a), and $\epsilon = 0.1268$ in b).

Fig. 25. Plots of eq. (16) before and after the boundary crisis. The nonlinearity is $\alpha = 1.72$ and the coupling strength is $\epsilon = 0.04$ in a), and $\epsilon = 0.03$ in b).

Fig. 26. Stationary Lyapunov exponents for SII, $\alpha = 1.80$ and $\epsilon = 0.17$. The horizontal axis represents space, and the Lyapunov exponents represent the local chaoticity of a site.

Fig. 27. Comoving Lyapunov exponents for several values of $k$. The numbers indicate the values of $k$, and the insets depict lattice sites 5436-5500. The product of 5000 Jacobi matrices was taken. a) $\alpha = 1.7$, $\epsilon = 0.5$. b) The $k = 8$ pattern of a) and its spatial return map. c) $\alpha = 1.9$, $\epsilon = 0.6$. d) Return maps of the patterns with parameters indicated in the figure.

Fig. 28. Comoving Lyapunov exponent for patterns generated with the $k = 3$ spatial map. For SQP we have $\alpha = 1.80$, $\epsilon = 0.6$, for SP $\alpha = 1.80$, $\epsilon = 0.7$ and for SC $\alpha = 1.80$, $\epsilon = 0.8$.

Fig. 29. Effect of modulating the boundary. The nonlinearity is $\alpha = 1.8$ and the coupling strength is $\epsilon = 0.7$. In a), the boundary is fixed to 0.0, and the pattern is spatially chaotic while having a temporal periodicity of four. In b), the boundary is modulated by a period 3 sawtooth-like wave with an amplitude of 1e-2. The pattern is spatially periodic (period 9) and has a temporal periodicity of 3. In both cases, lattice sites 192–256 are shown.

Fig. 30. Inverse bifurcation in the OCLL. The nonlinearity is $\alpha = 1.7$ and the coupling constant $\epsilon = 0.6$. The boundary is modulated with a period 3 sawtooth wave whose amplitude is 0.01. Initially, the lattice steps through the bifurcations 3-6-12-24, but then inversely bifurcates to 8. At the left side of the figure, 6 successive states of the lattice are overlaid, while at the left side 3 states sampled per eight time steps are overlaid.

| Phase | TPS | DL | SII | Zigzag | (STI II) | STC |
|---|---|---|---|---|---|---|
| Characterisics | remnant chaos | periodic defects | ordered islands | spatio-temp. period 2 (band) | (bursts) | no order |
| C(r) (range) | short | mod. $\infty$ | short | $\infty$ | (power-law) | very short |
| attractors | single | multiple | multiple | single | (single) | single |
| $\# of \lambda > 0$ | $> 0, = 0$ | $= \#$ defects | 2 | 0 | (O(N)) | $\approx N$ |

Frederick H. Willeboordse and Kunihiko Kaneko Table 1.

| | | | |
|---|---|---|---|
| zigzag | $\rightarrow$ | defect | only when colliding with a burst approaching from upflow |
| zigzag | $\rightarrow$ | burst | not possible |
| defect | $\rightarrow$ | zigzag | only when colliding with a burst approaching from upflow |
| defect | $\rightarrow$ | burst | finite probability |
| burst | $\rightarrow$ | zigzag | always possible |
| burst | $\rightarrow$ | defect | possible if upflow sites zigzag |

Frederick H. Willeboordse and Kunihiko Kaneko Table 2.

| $\alpha \Rightarrow$ | 1.65 | | 1.70 | | 1.75 | | 1.80 | | 1.85 | | 1.90 | |
|---|---|---|---|---|---|---|---|---|---|---|---|---|
| freq. $\Downarrow$ | tp | sp | tp | sp | tp | sp | tp | sp | tp | sp | tp | sp |
| 1: | 8 | 4 | 32 | - | 32 | - | 32 | - | 8 | 4 | 8 | 4 |
| 2: | 16 | - | 32 | - | 8 | 4 | 8 | 4 | 32 | - | 8 | 4 |
| 3: | 12 | 6 | 24 | - | 24 | - | 48 | - | 48 | - | 24 | 12 |
| 4: | 8 | 4 | 8 | 4 | 8 | 4 | 8 | 4 | 8 | 4 | 8 | 4 |
| 5: | 5 | 5 | 5 | 5 | 10 | 10 | 5 | 10 | 10 | 10 | 40 | - |
| 6: | 12 | 6 | 24 | - | 48 | - | 48 | - | 48 | - | 8 | 4 |
| 7: | 28 | - | 28 | 14 | 28 | - | 28 | - | 28 | 14 | 56 | - |
| 8: | 32 | 16 | 8 | 4 | 8 | 4 | 8 | 4 | 8 | 4 | 8 | 4 |
| 9: | 18 | 9 | 36 | - | 18 | 27 | 18 | 18 | 18 | 18 | 36 | - |
| 10: | 5 | 5 | 5 | 5 | 5 | 40 | 20 | - | 40 | 20 | 40 | - |

Frederick H. Willeboordse and Kunihiko Kaneko Table 3.

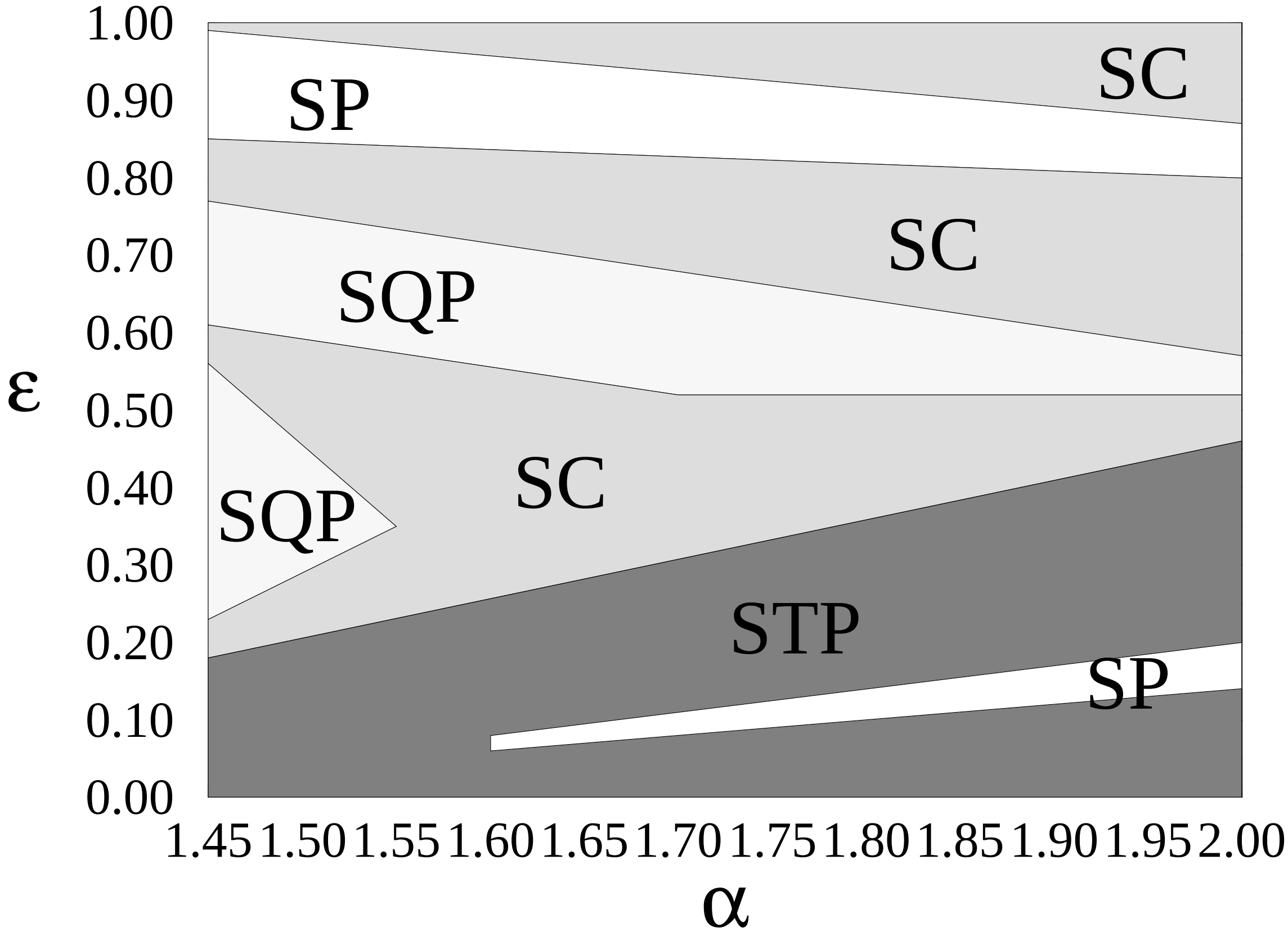

Frederick H. Willeboordse and Kunihiko Kaneko Fig. 1.

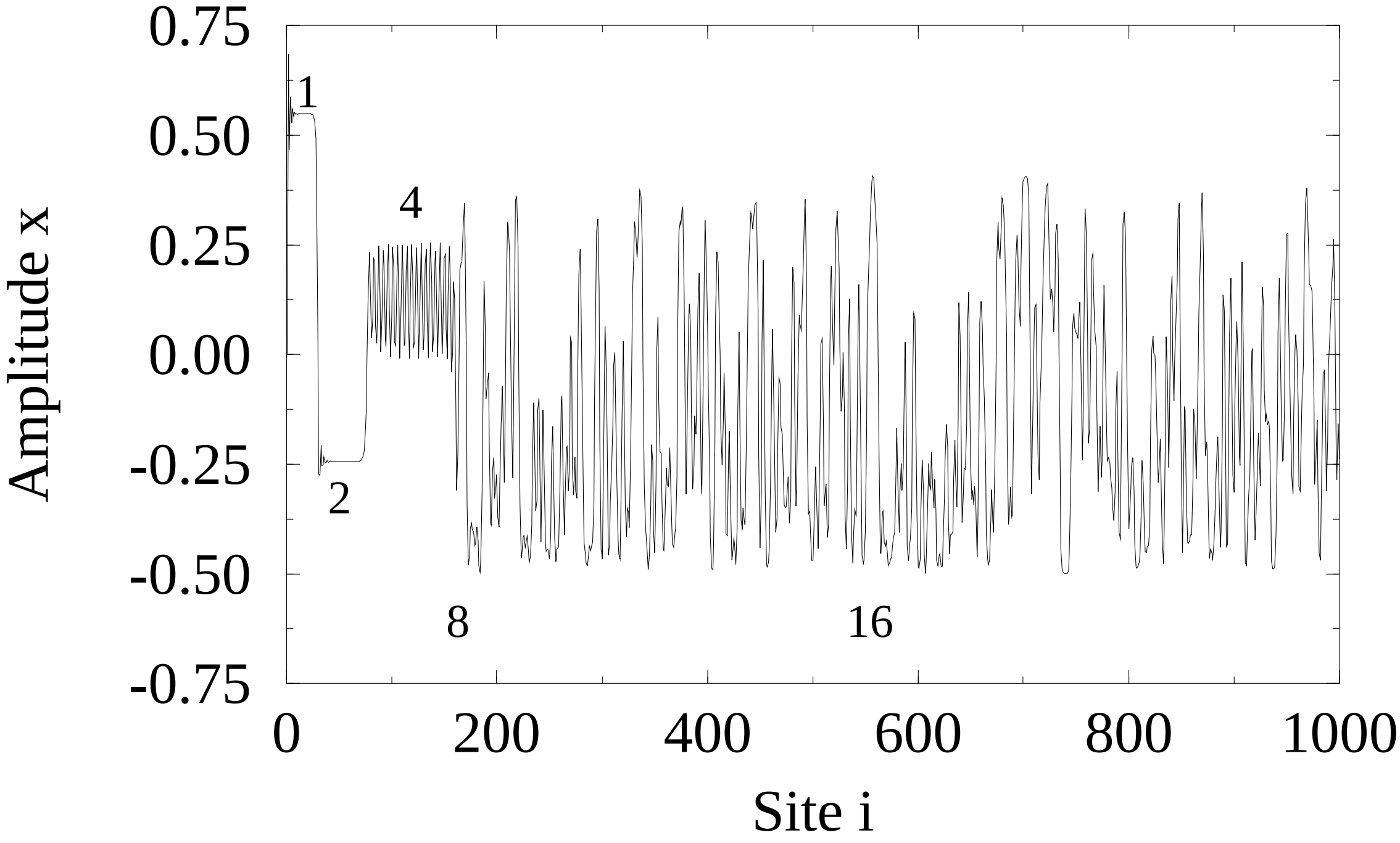

Frederick H. Willeboordse and Kunihiko Kaneko Fig. 2a).

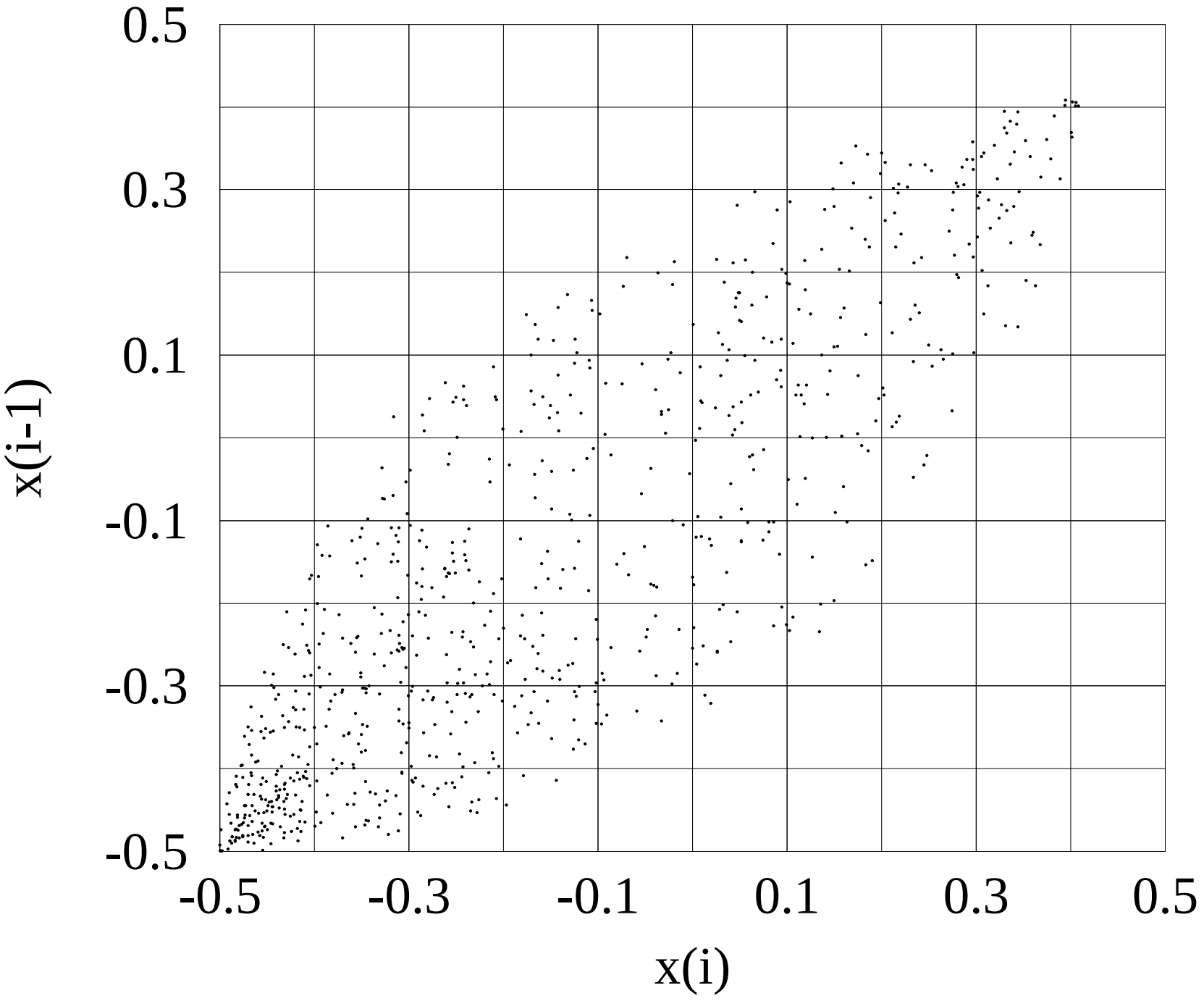

Frederick H. Willeboordse and Kunihiko Kaneko Fig. 2b).

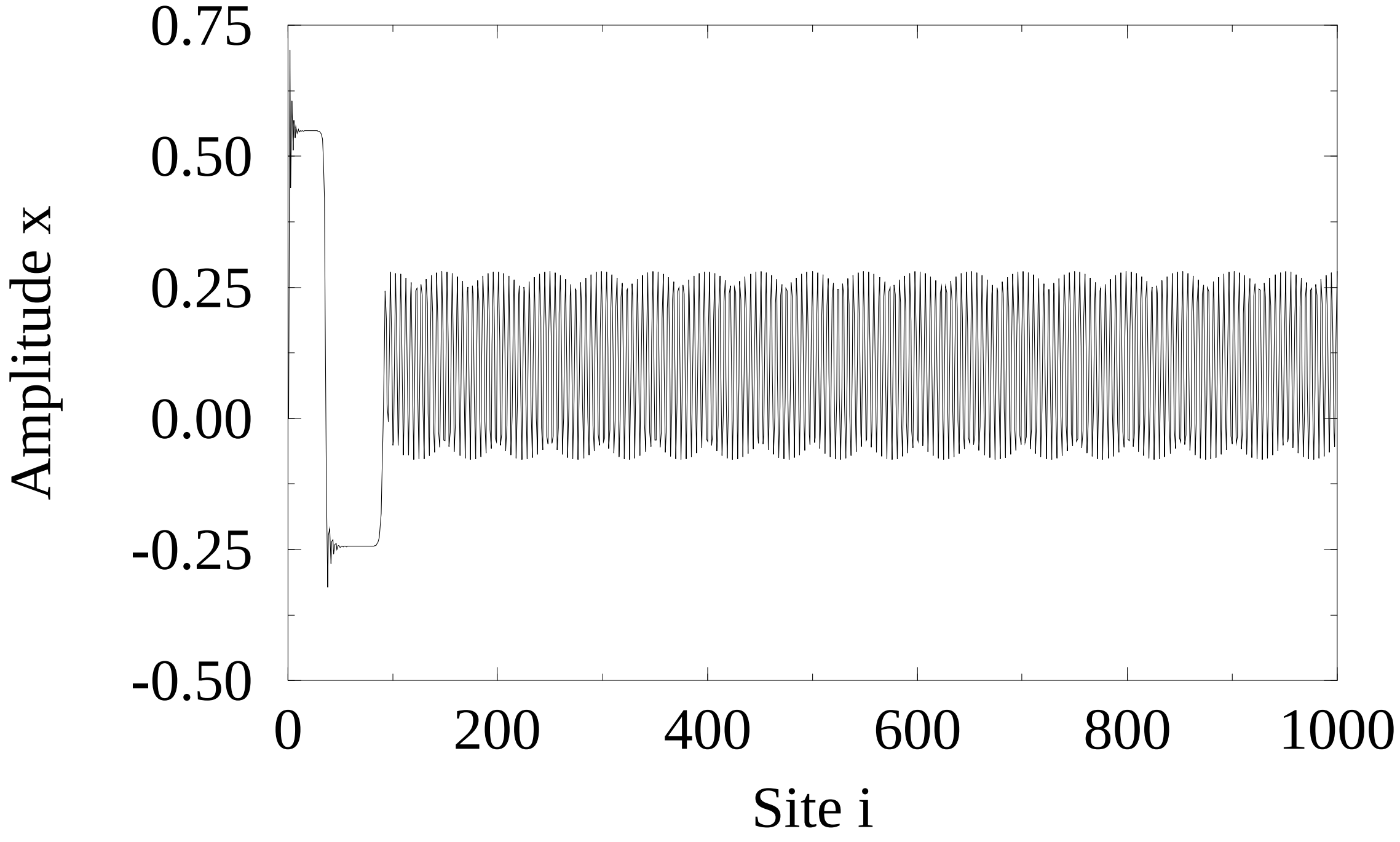

Frederick H. Willeboordse and Kunihiko Kaneko Fig. 2c).

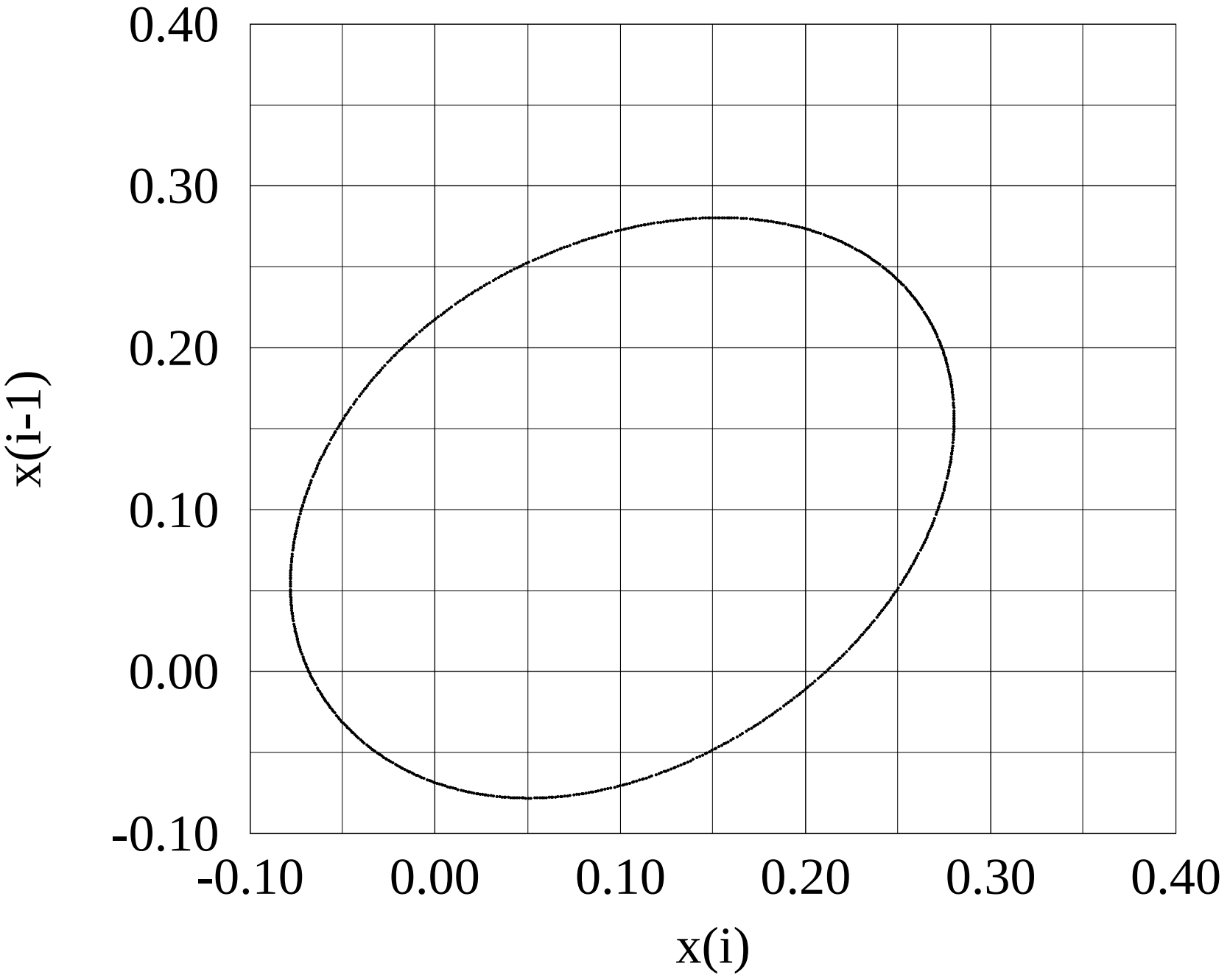

Frederick H. Willeboordse and Kunihiko Kaneko Fig. 2d).

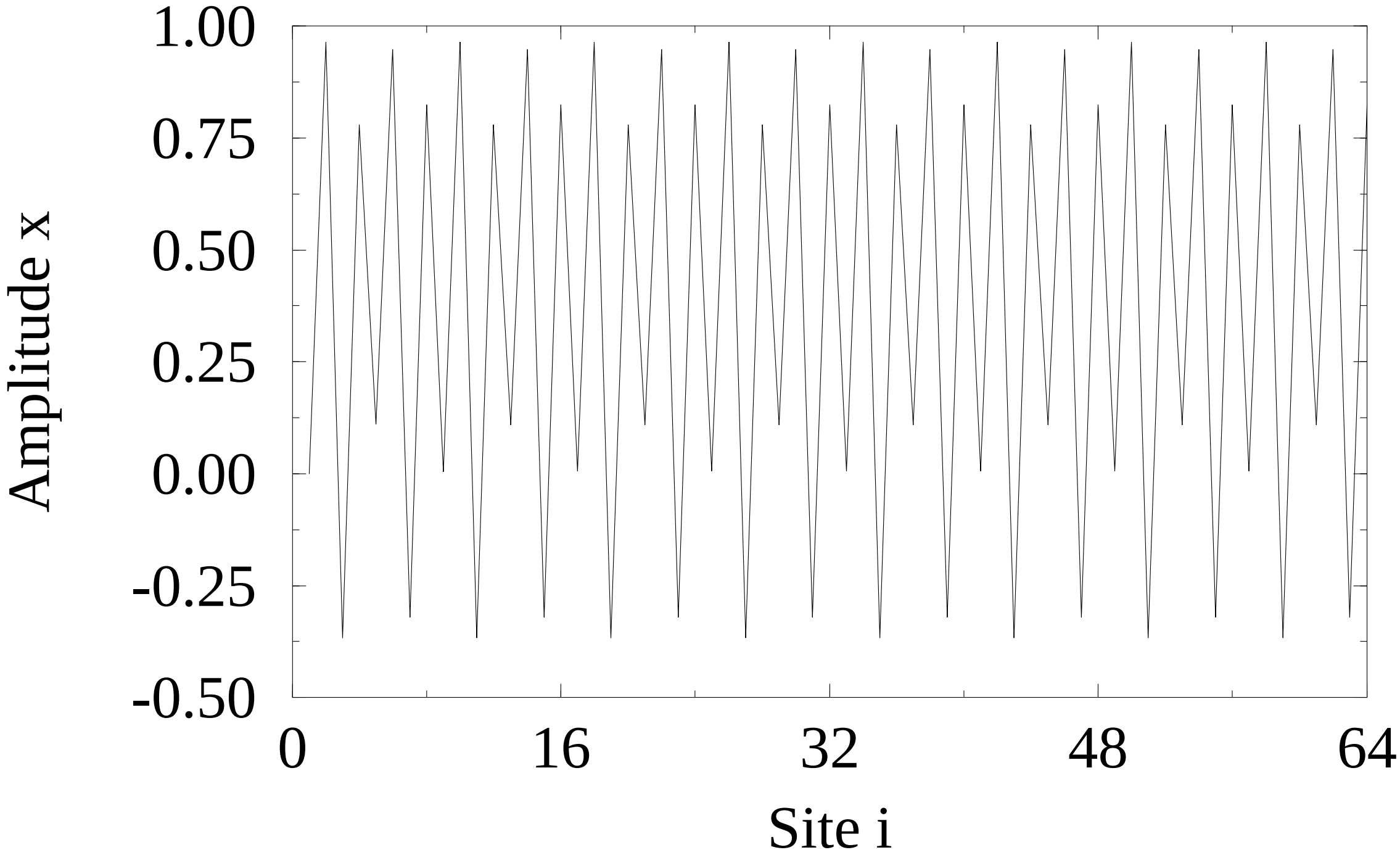

Frederick H. Willeboordse and Kunihiko Kaneko Fig. 2e).

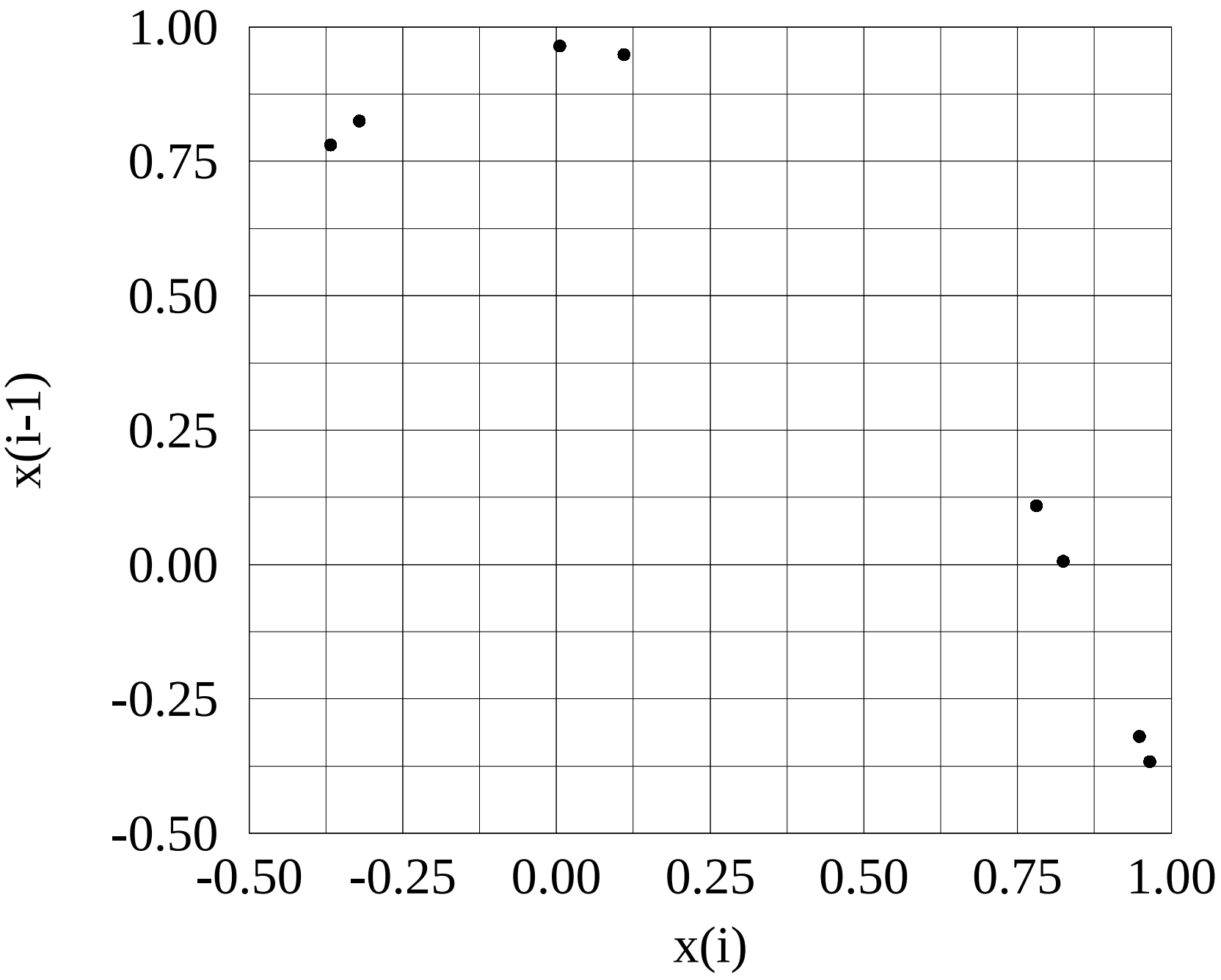

Frederick H. Willeboordse and Kunihiko Kaneko Fig. 2f).

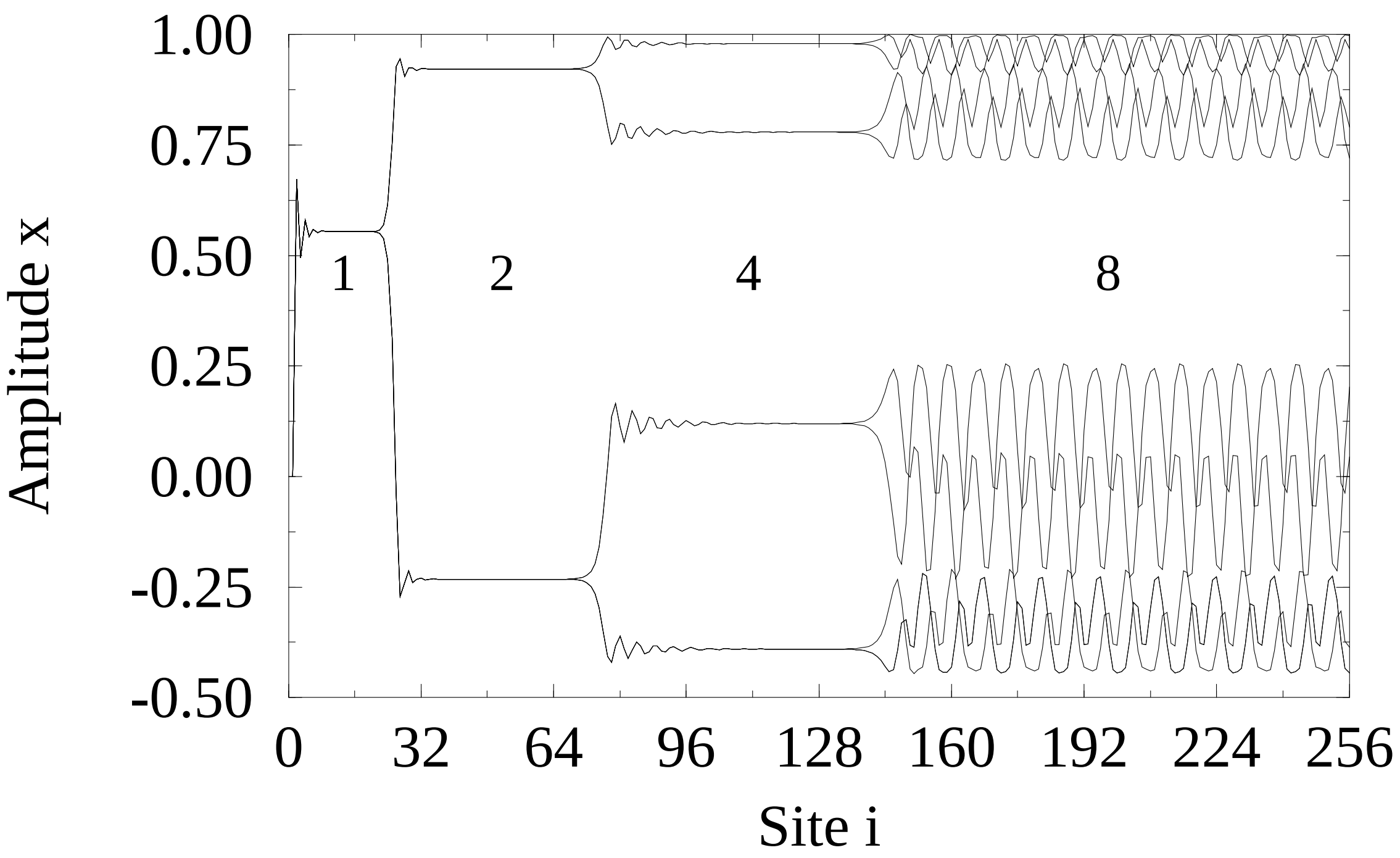

Frederick H. Willeboordse and Kunihiko Kaneko Fig. 3.

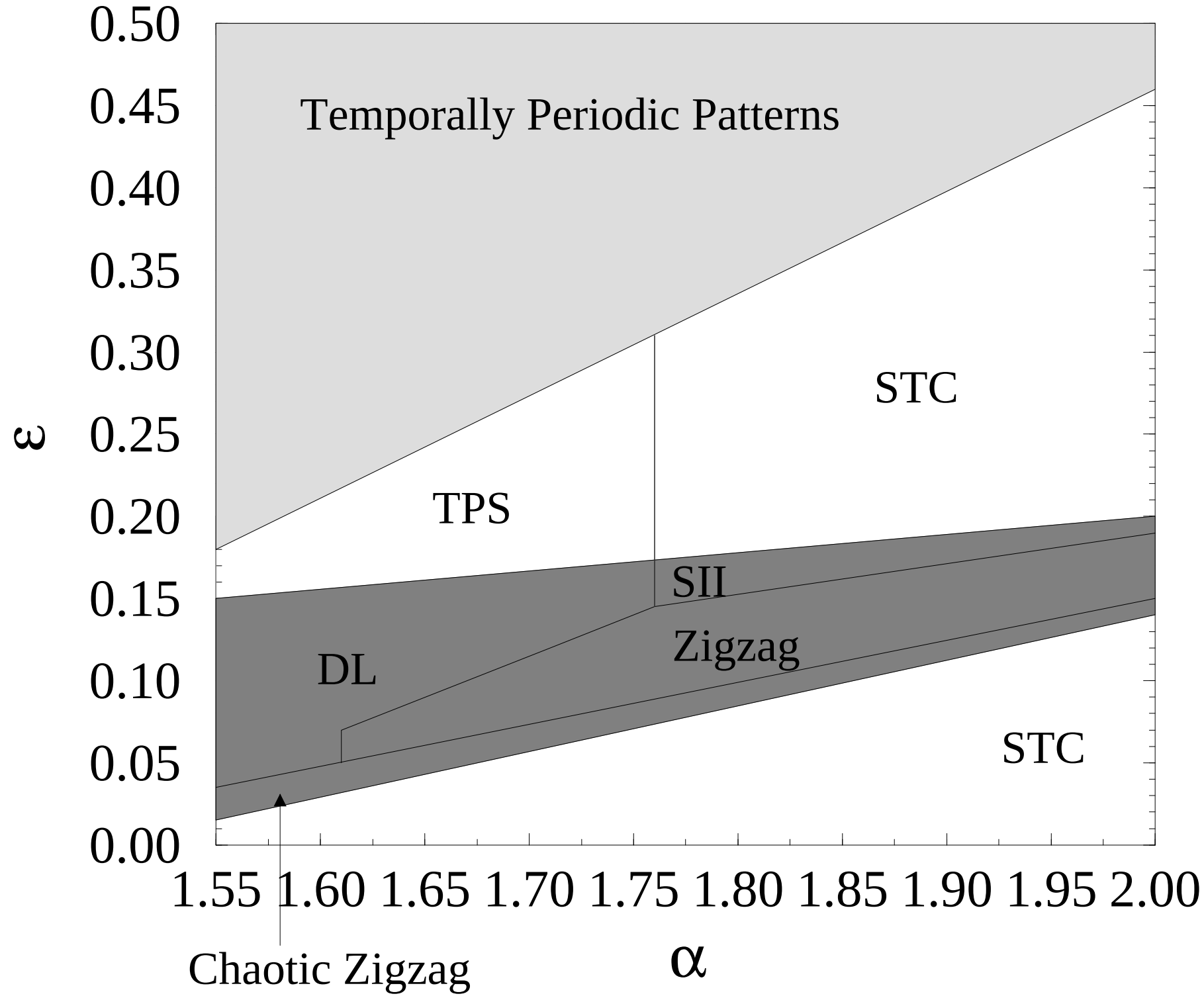

Frederick H. Willeboordse and Kunihiko Kaneko Fig. 4.

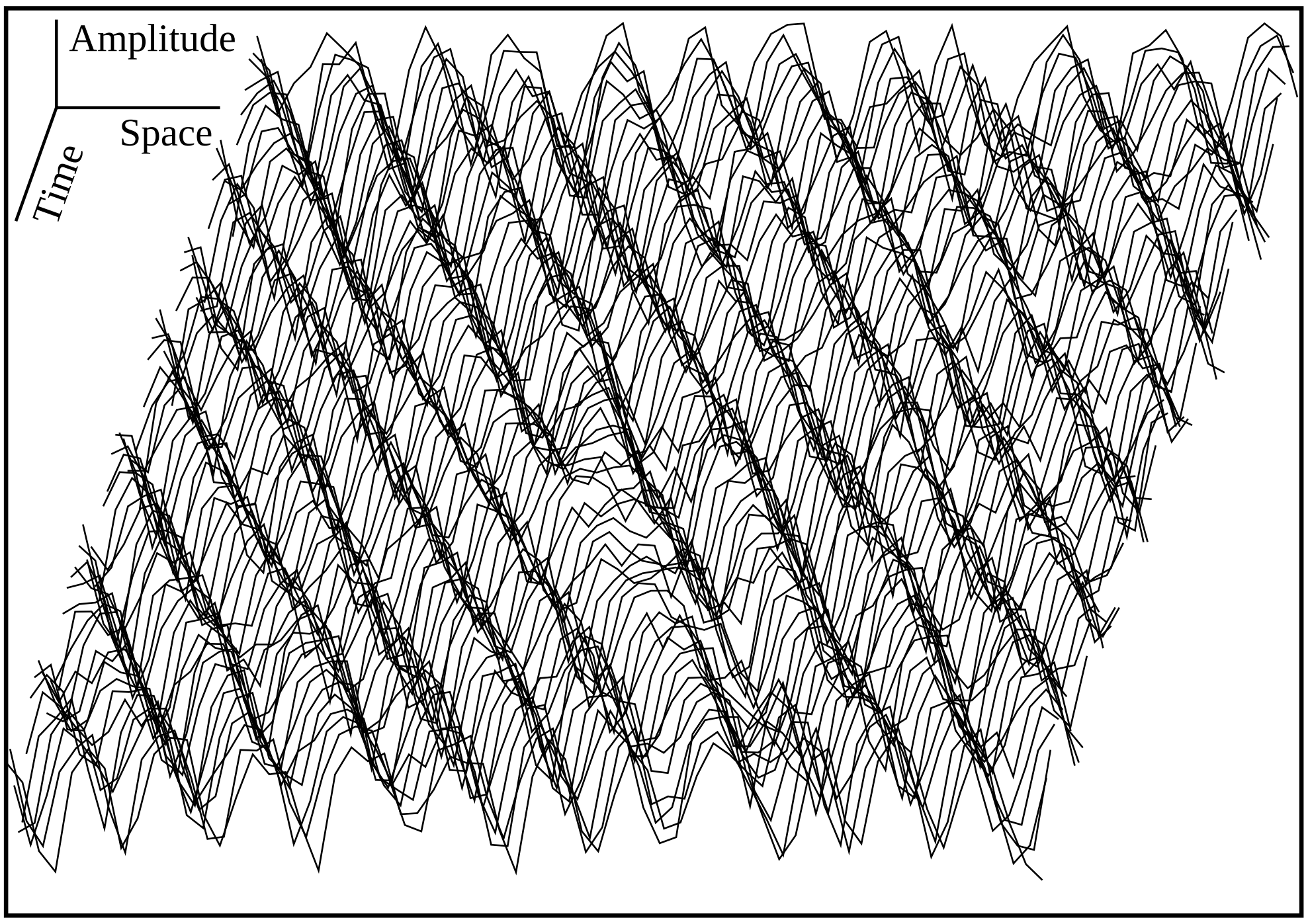

Frederick H. Willeboordse and Kunihiko Kaneko Fig. 5a).

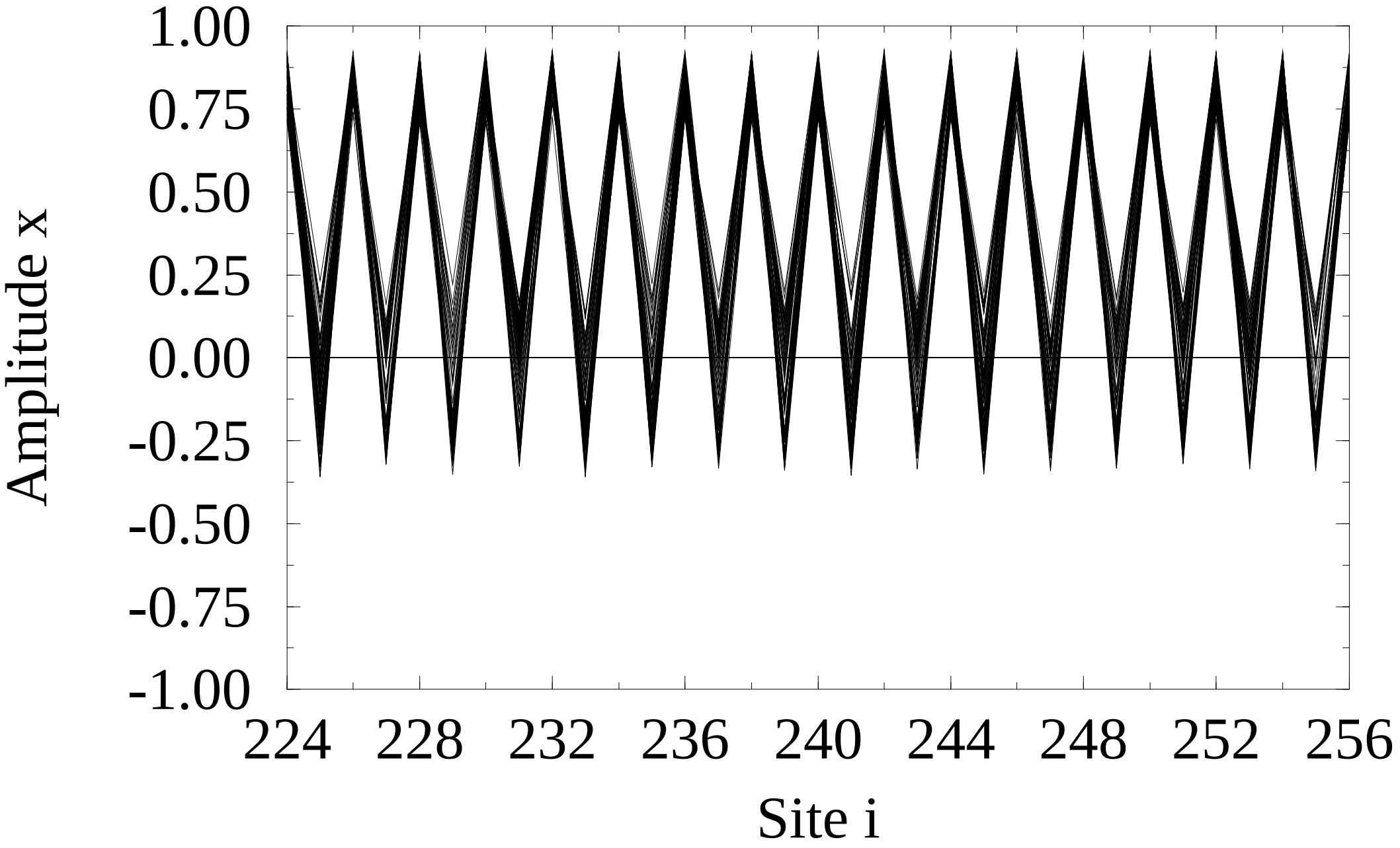

Frederick H. Willeboordse and Kunihiko Kaneko Fig. 5b).

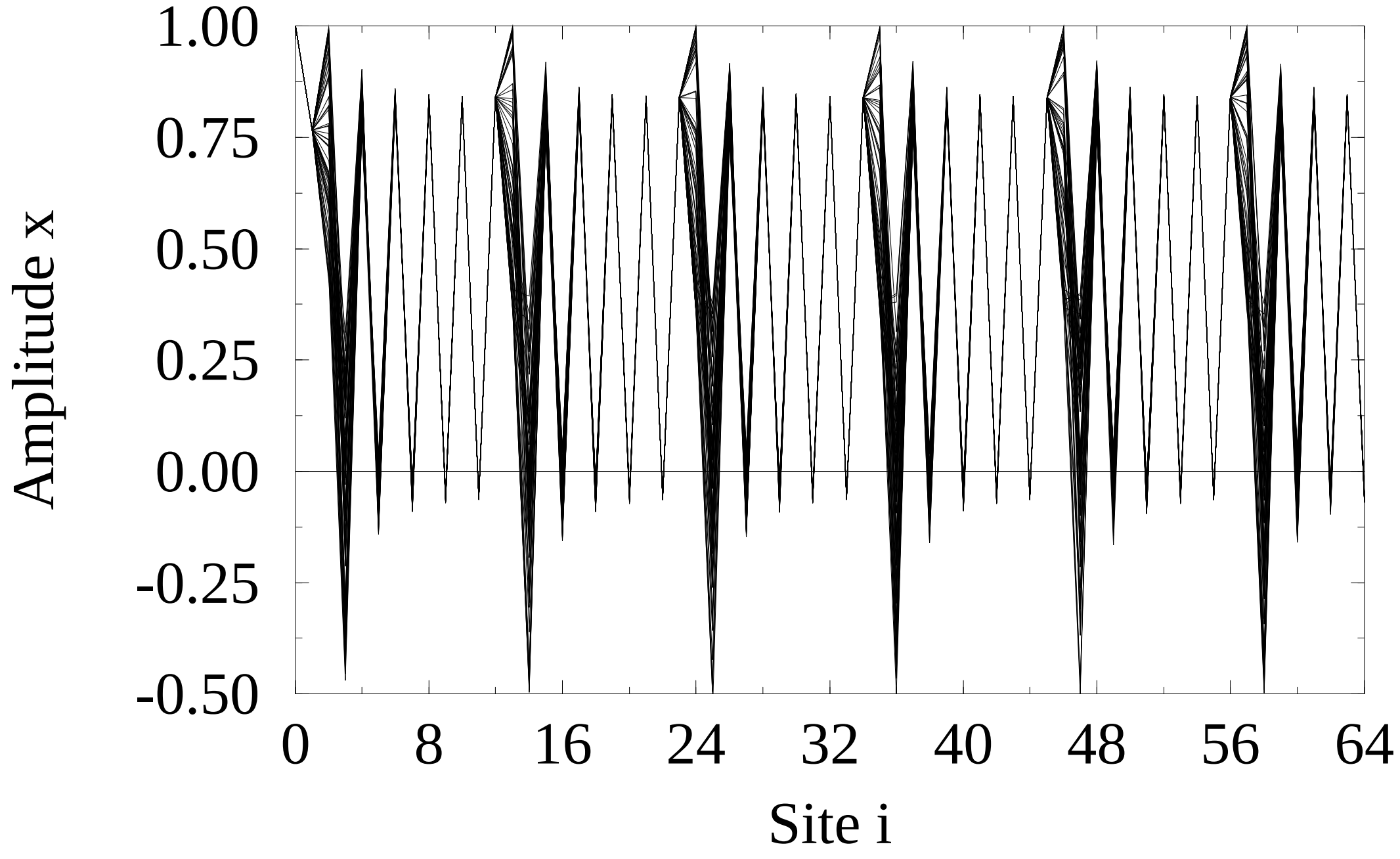

Frederick H. Willeboordse and Kunihiko Kaneko Fig. 5c).

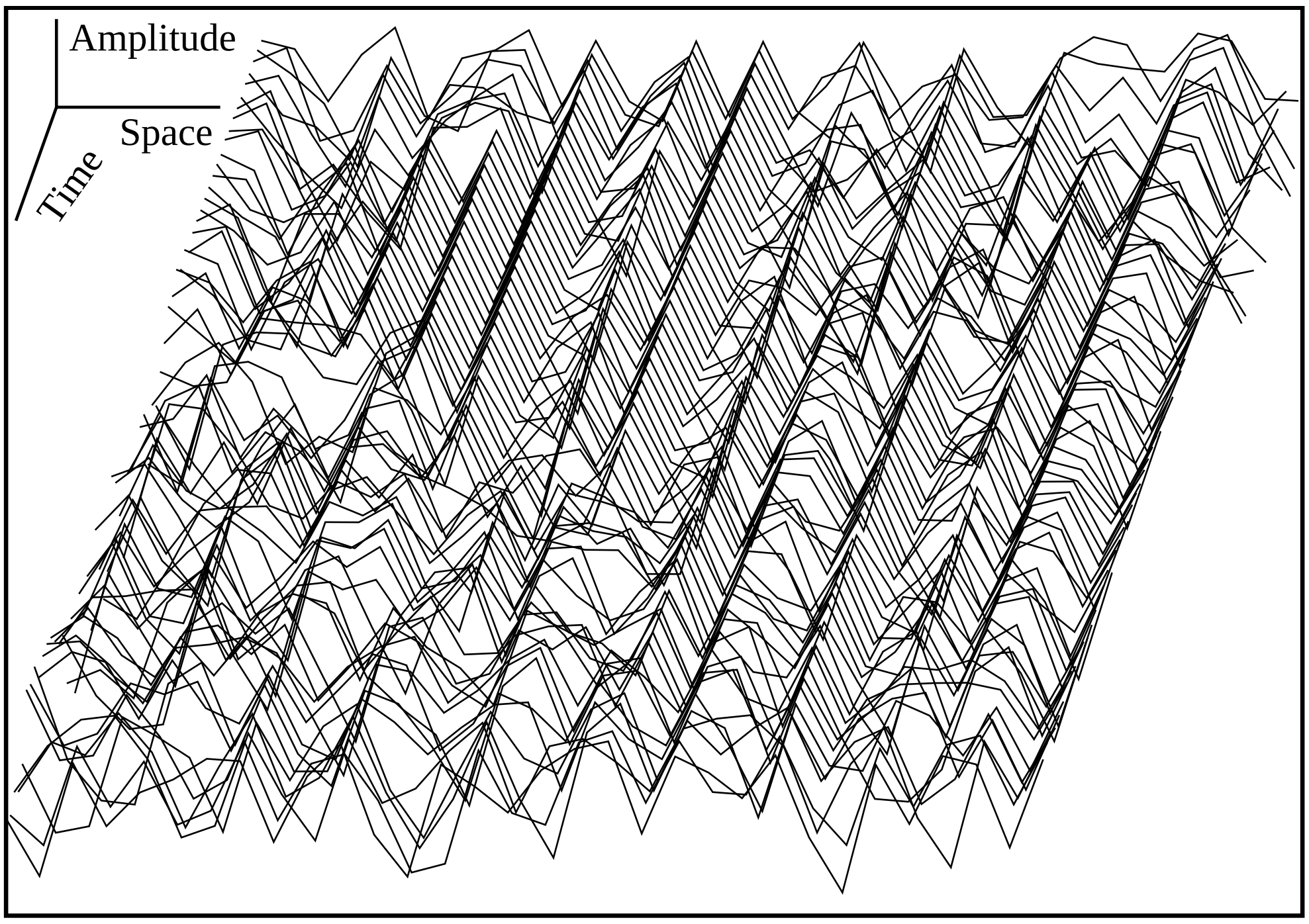

Frederick H. Willeboordse and Kunihiko Kaneko Fig. 5d).

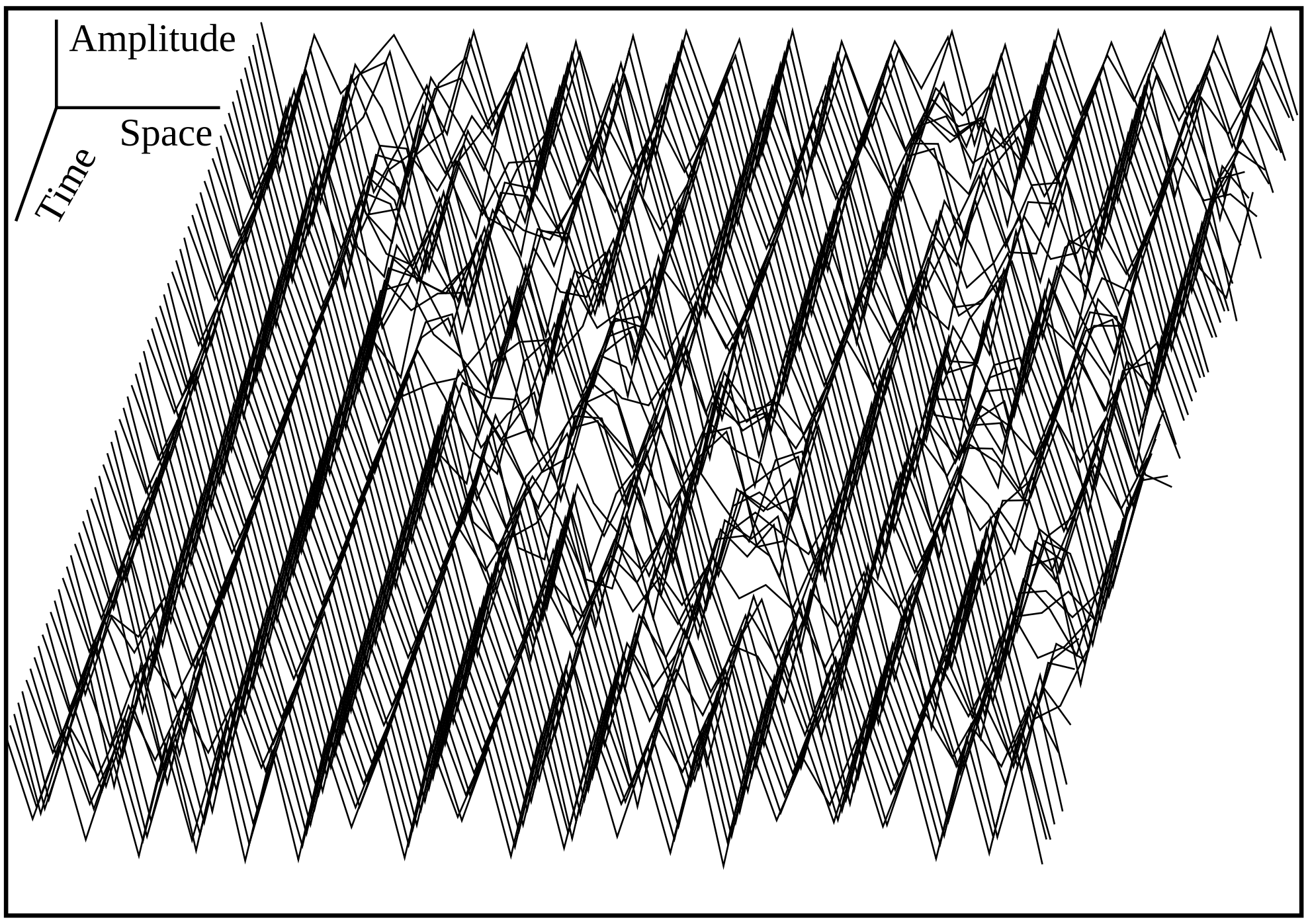

Frederick H. Willeboordse and Kunihiko Kaneko Fig. 5e).

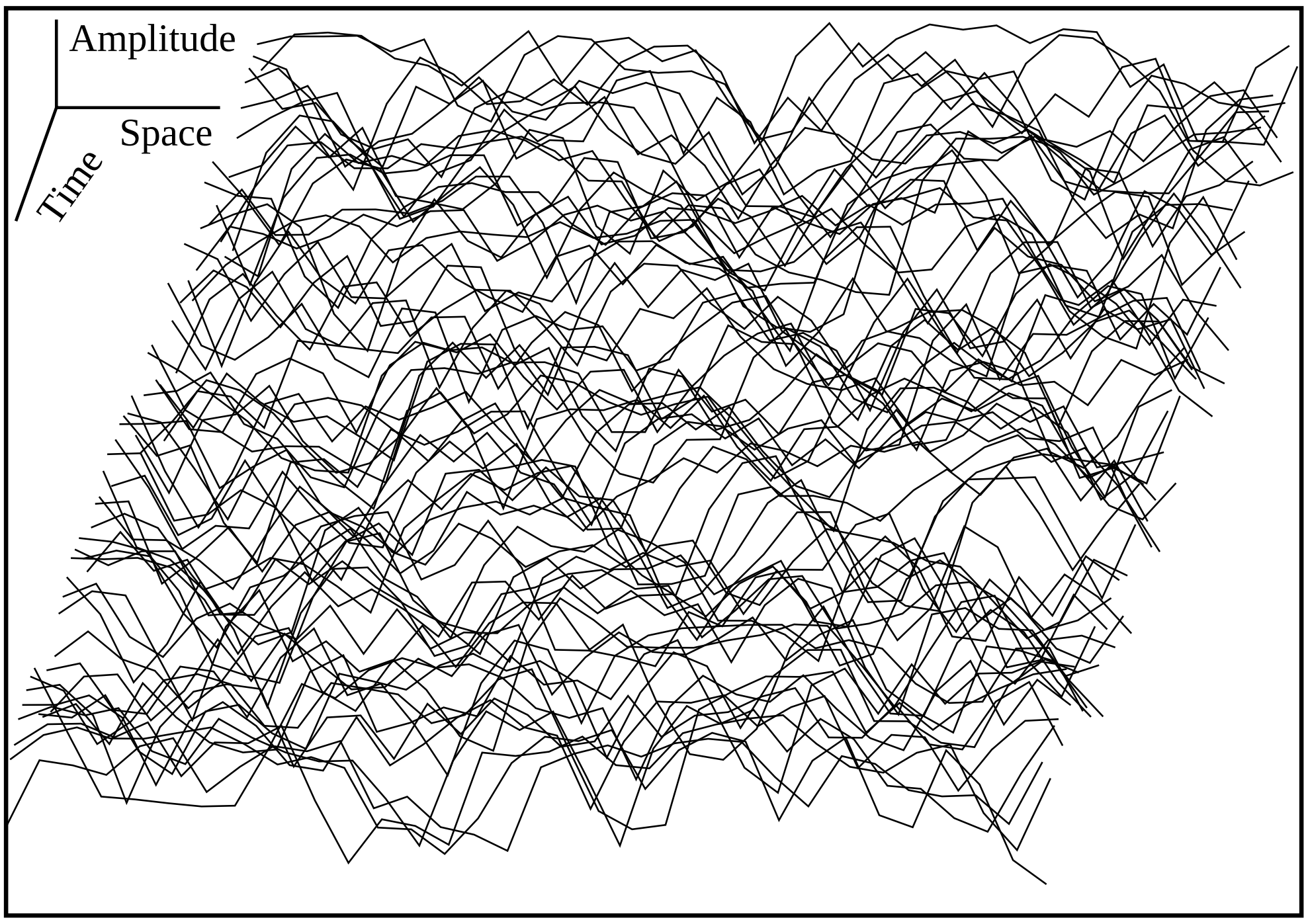

Frederick H. Willeboordse and Kunihiko Kaneko Fig. 5f).

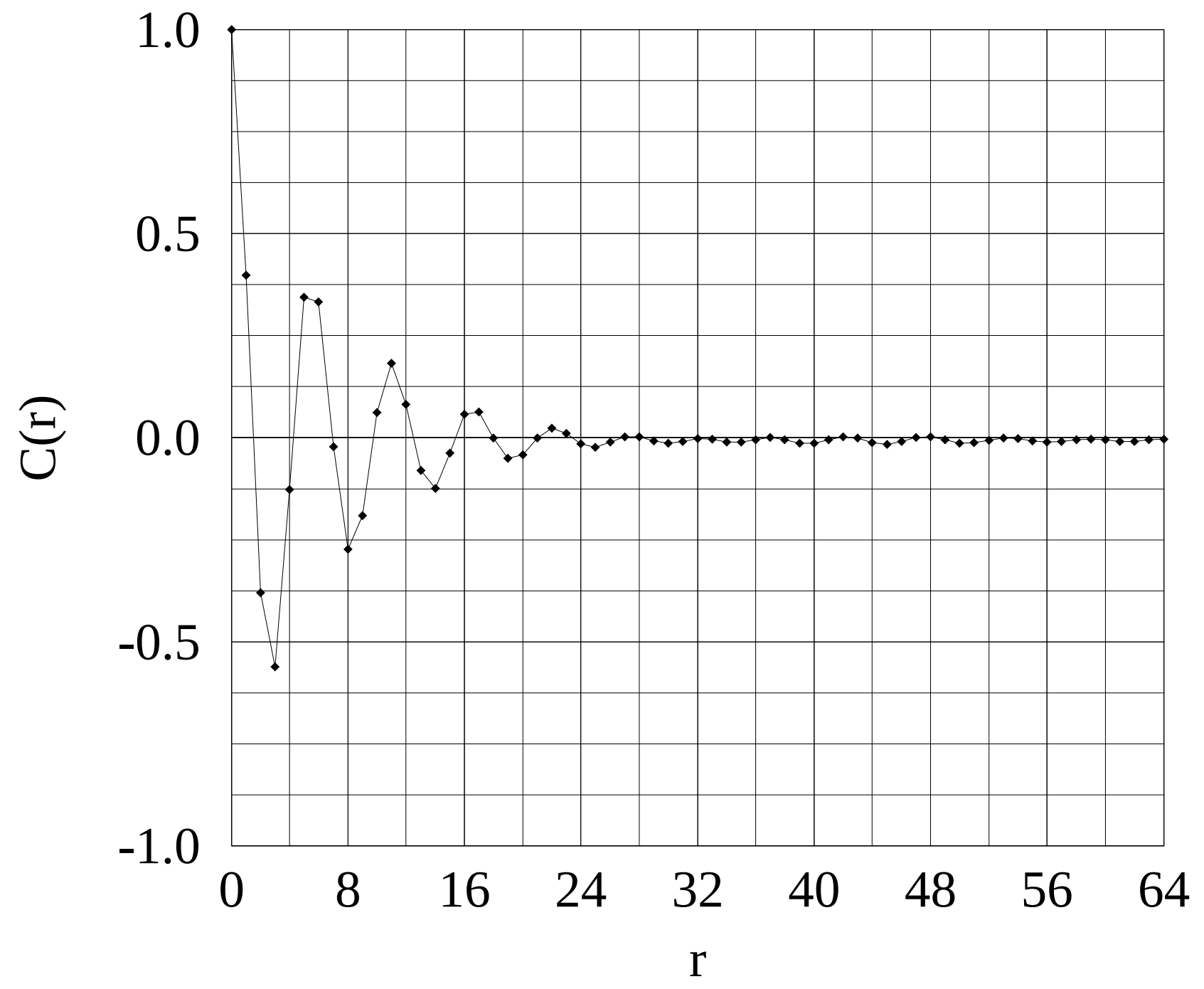

Frederick H. Willeboordse and Kunihiko Kaneko Fig. 6a).

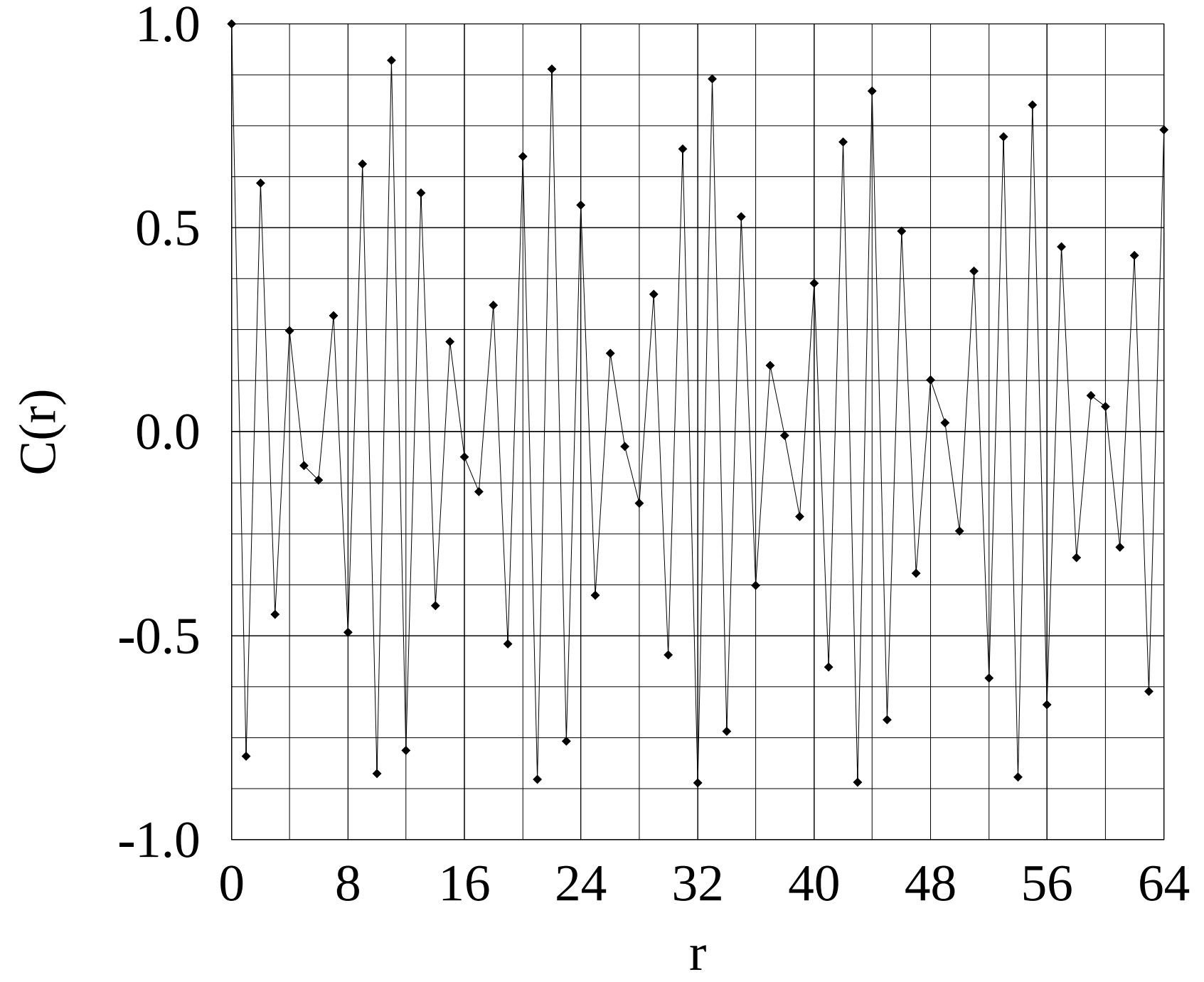

Frederick H. Willeboordse and Kunihiko Kaneko Fig. 6b).

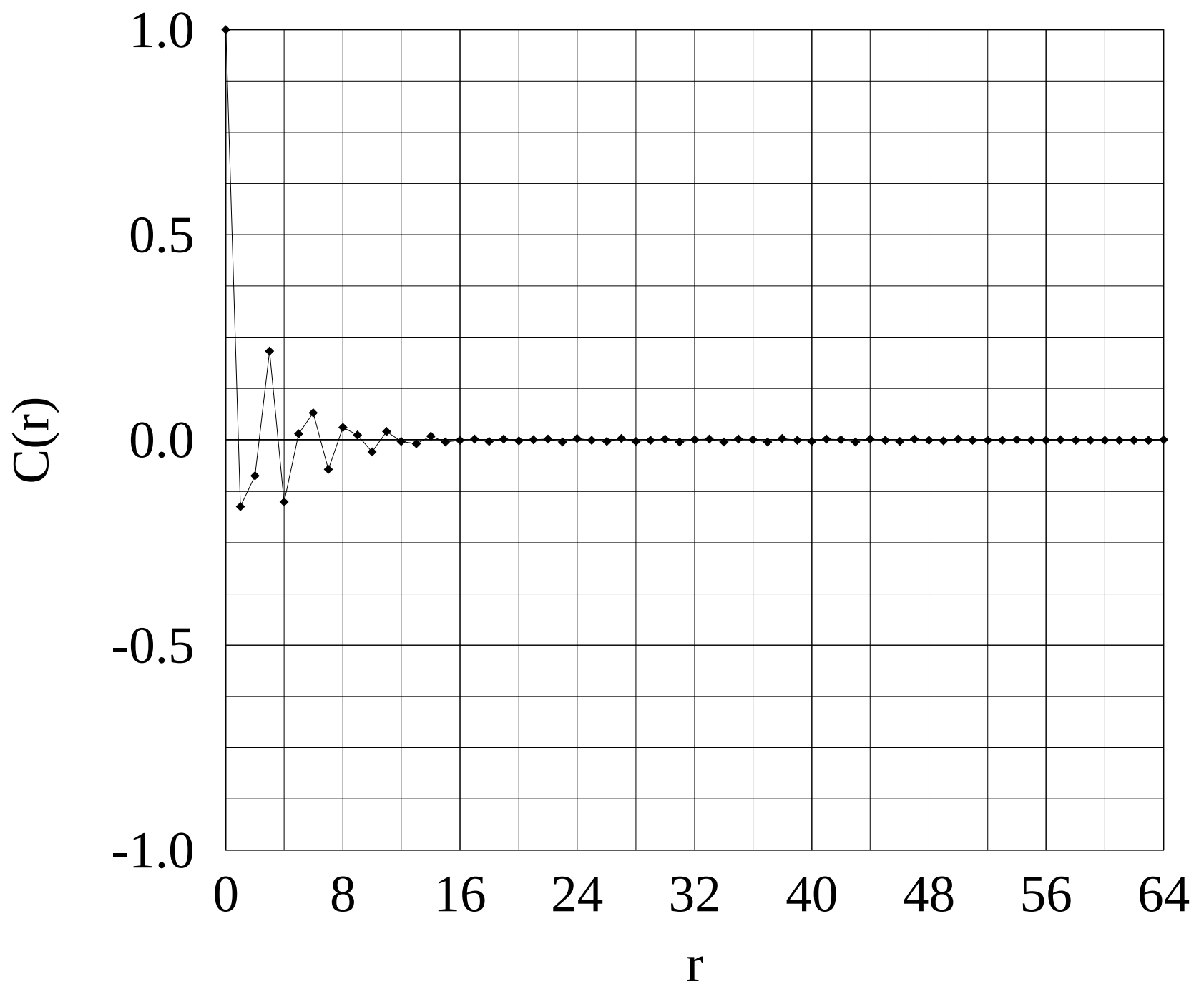

Frederick H. Willeboordse and Kunihiko Kaneko Fig. 6c).

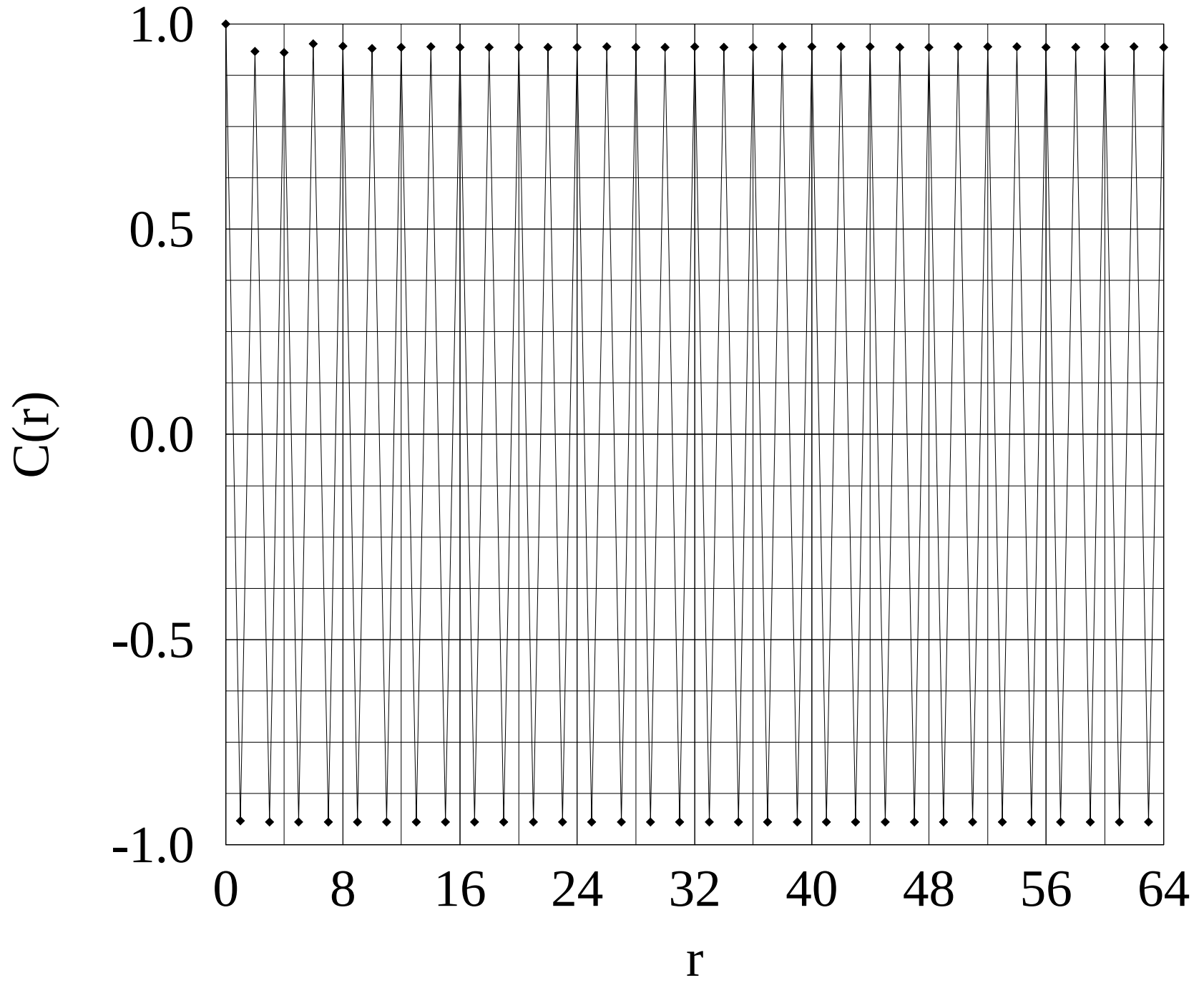

Frederick H. Willeboordse and Kunihiko Kaneko Fig. 6d).

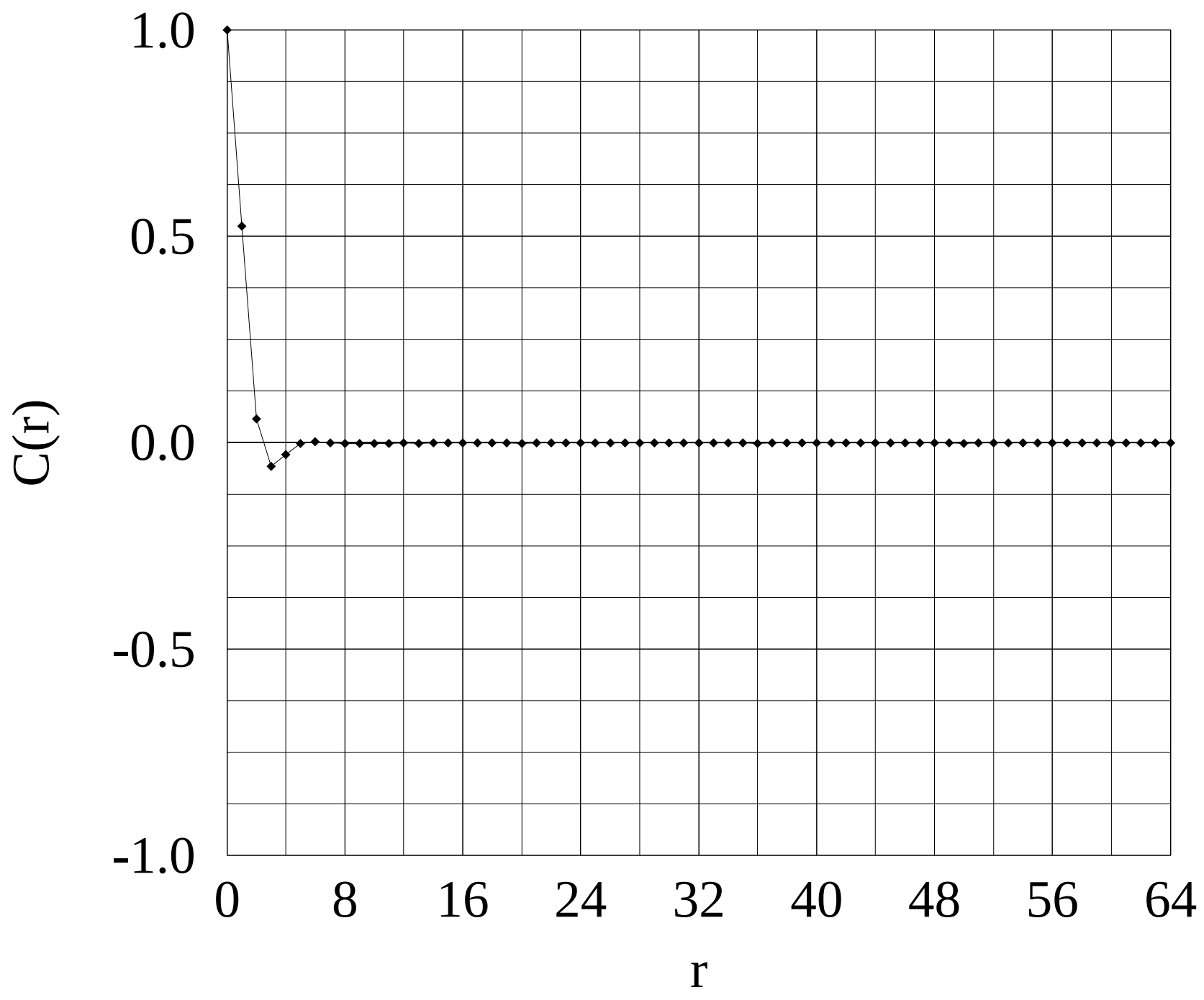

Frederick H. Willeboordse and Kunihiko Kaneko Fig. 6f).

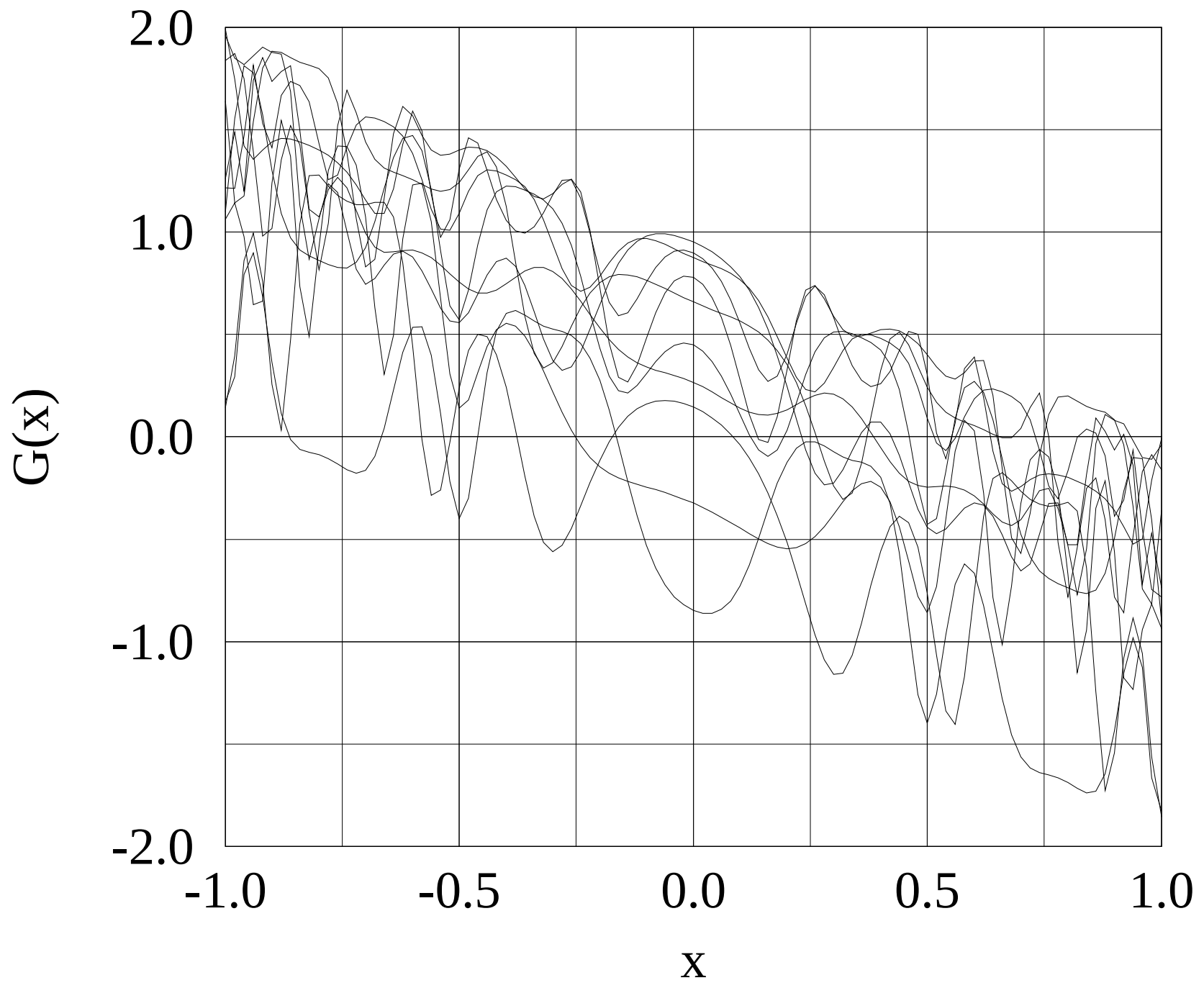

Frederick H. Willeboordse and Kunihiko Kaneko Fig. 7a).

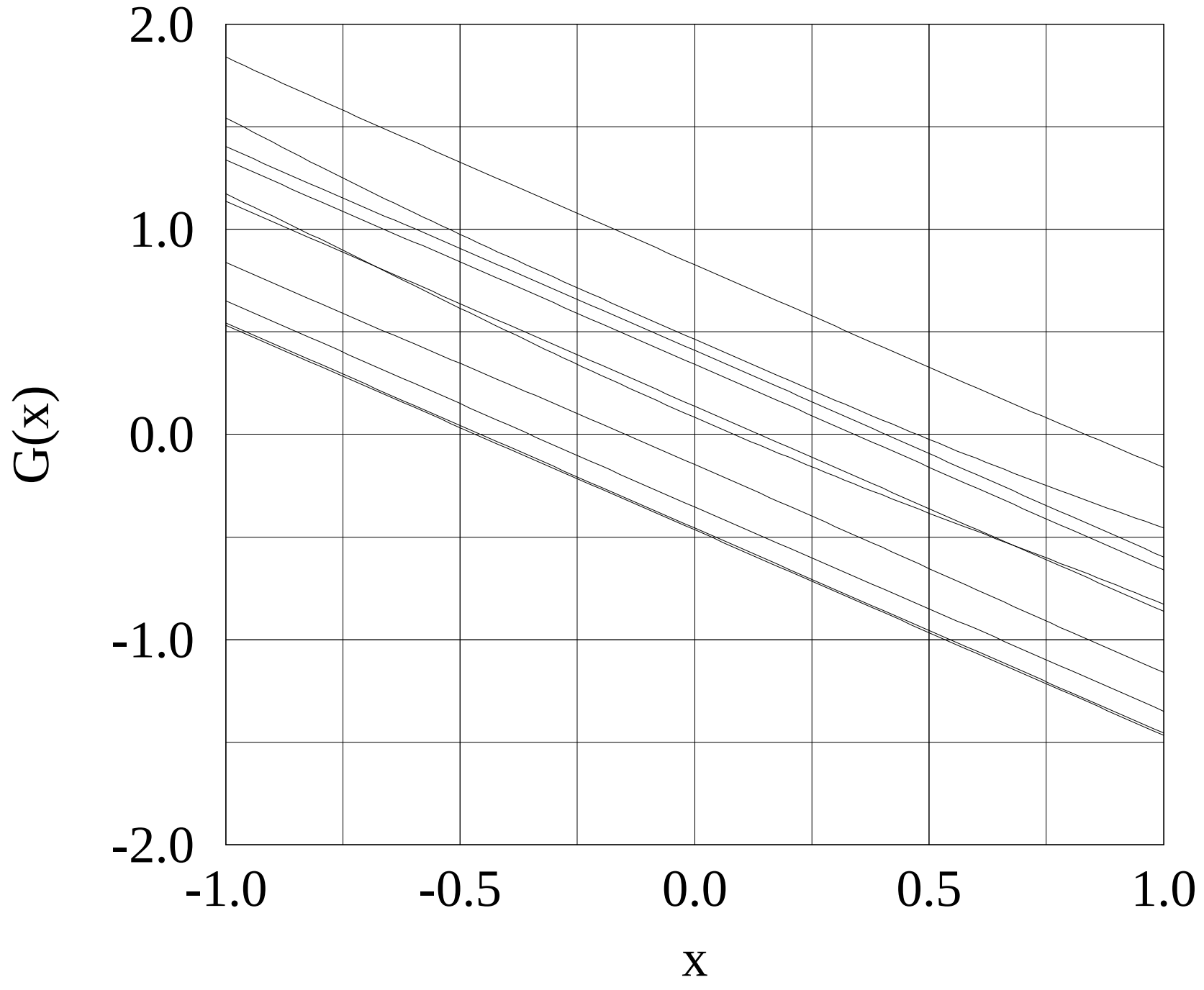

Frederick H. Willeboordse and Kunihiko Kaneko Fig. 7b).

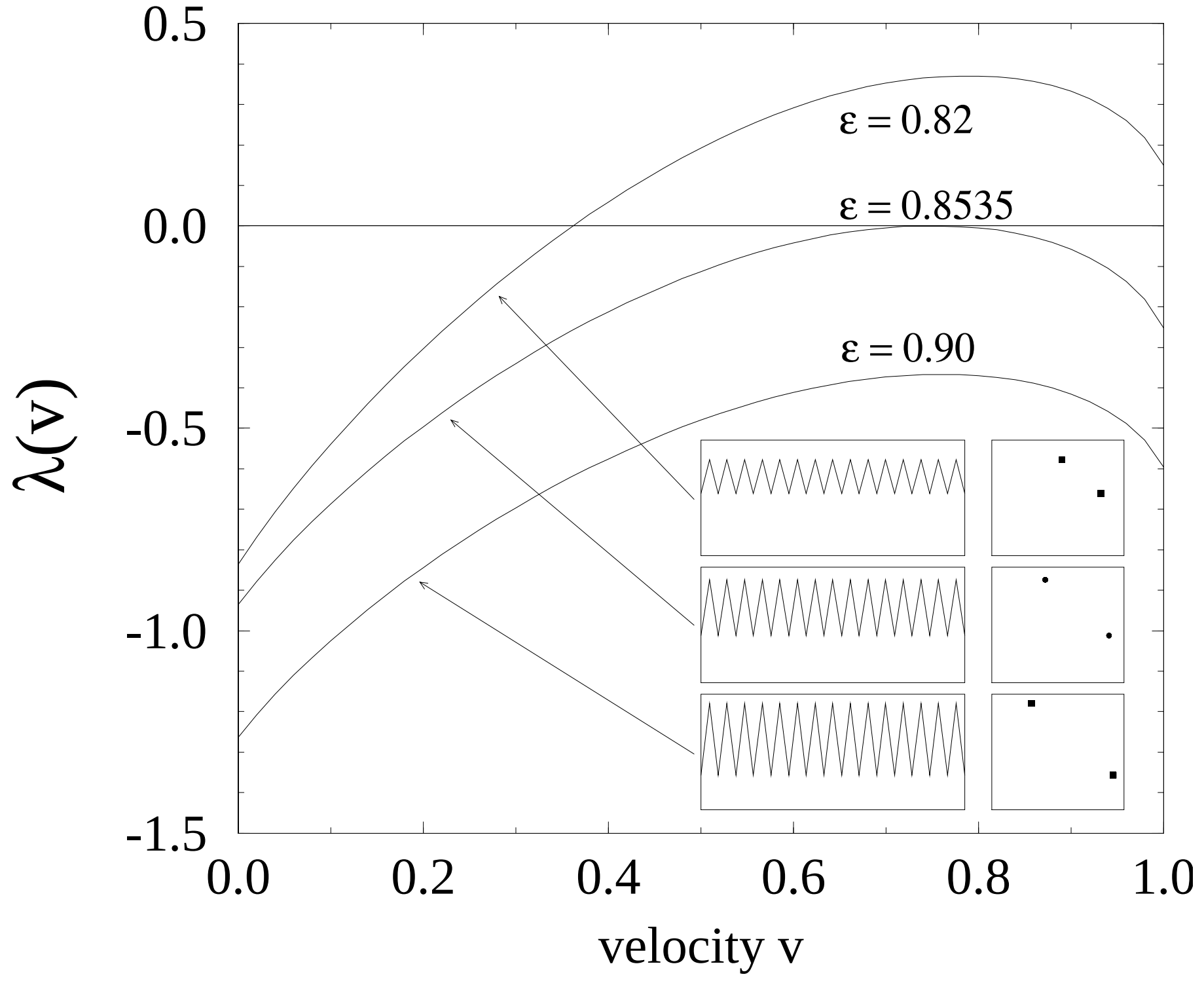

Frederick H. Willeboordse and Kunihiko Kaneko Fig. 8.

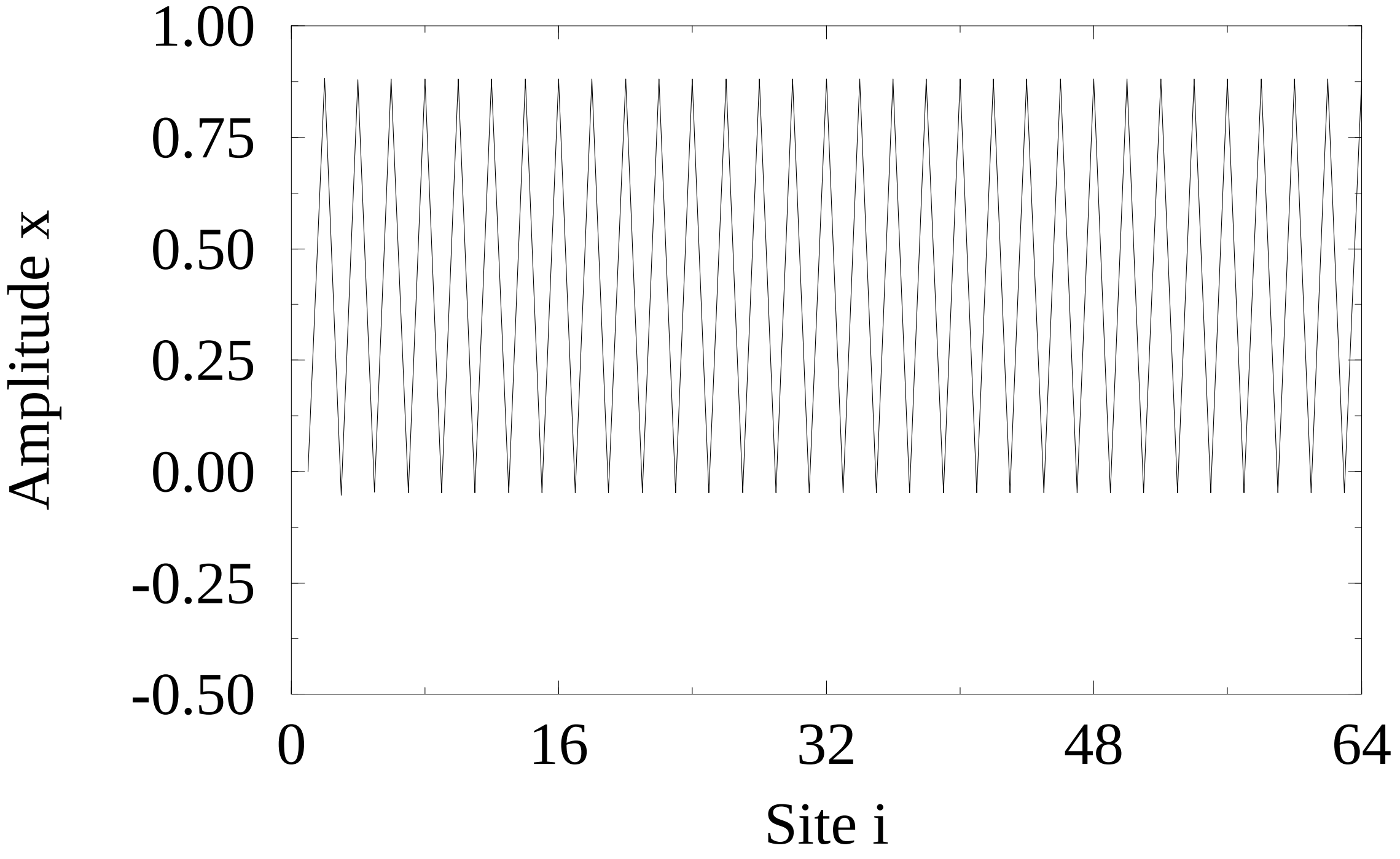

Frederick H. Willeboordse and Kunihiko Kaneko Fig. 9a).

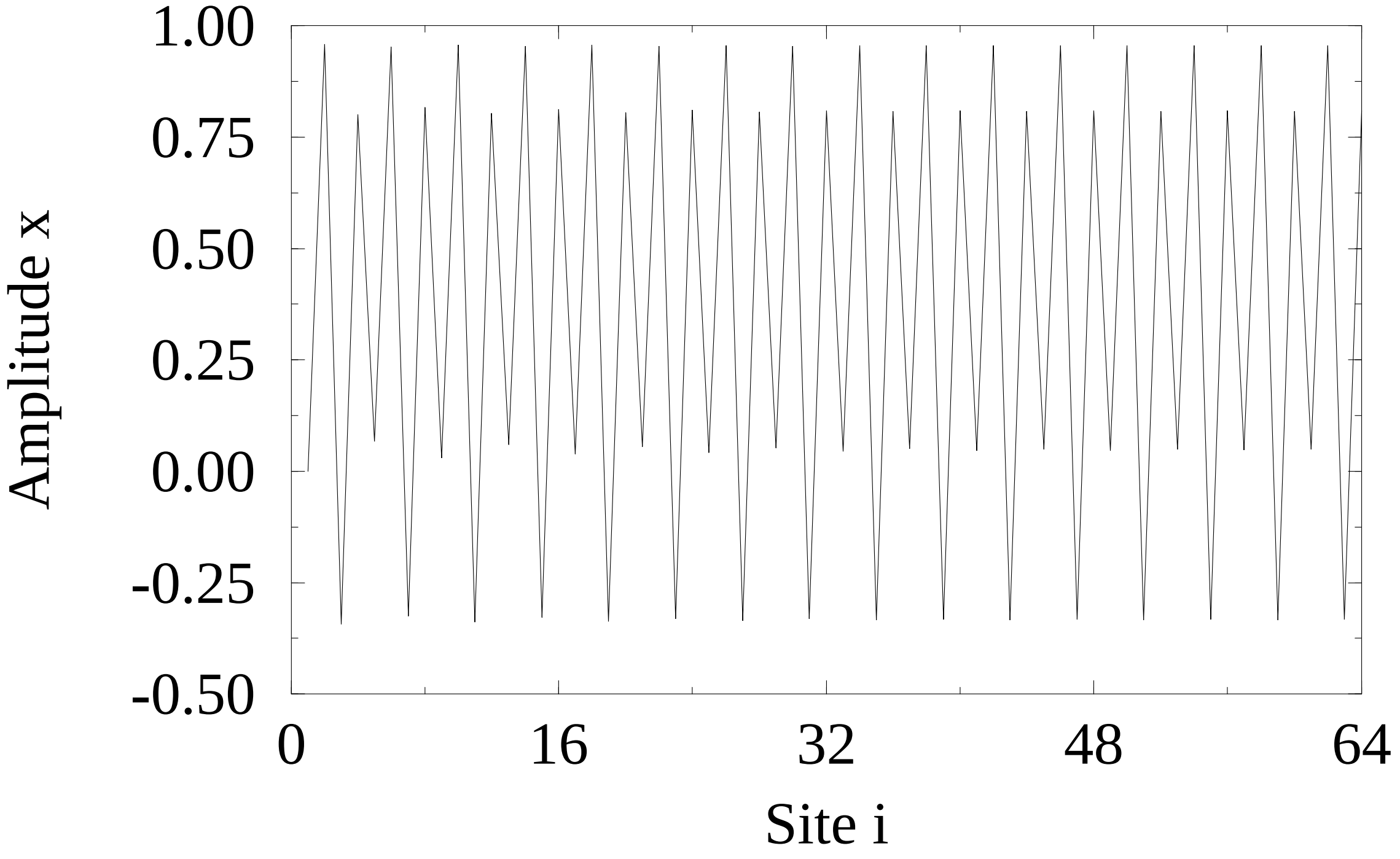

Frederick H. Willeboordse and Kunihiko Kaneko Fig. 9b).

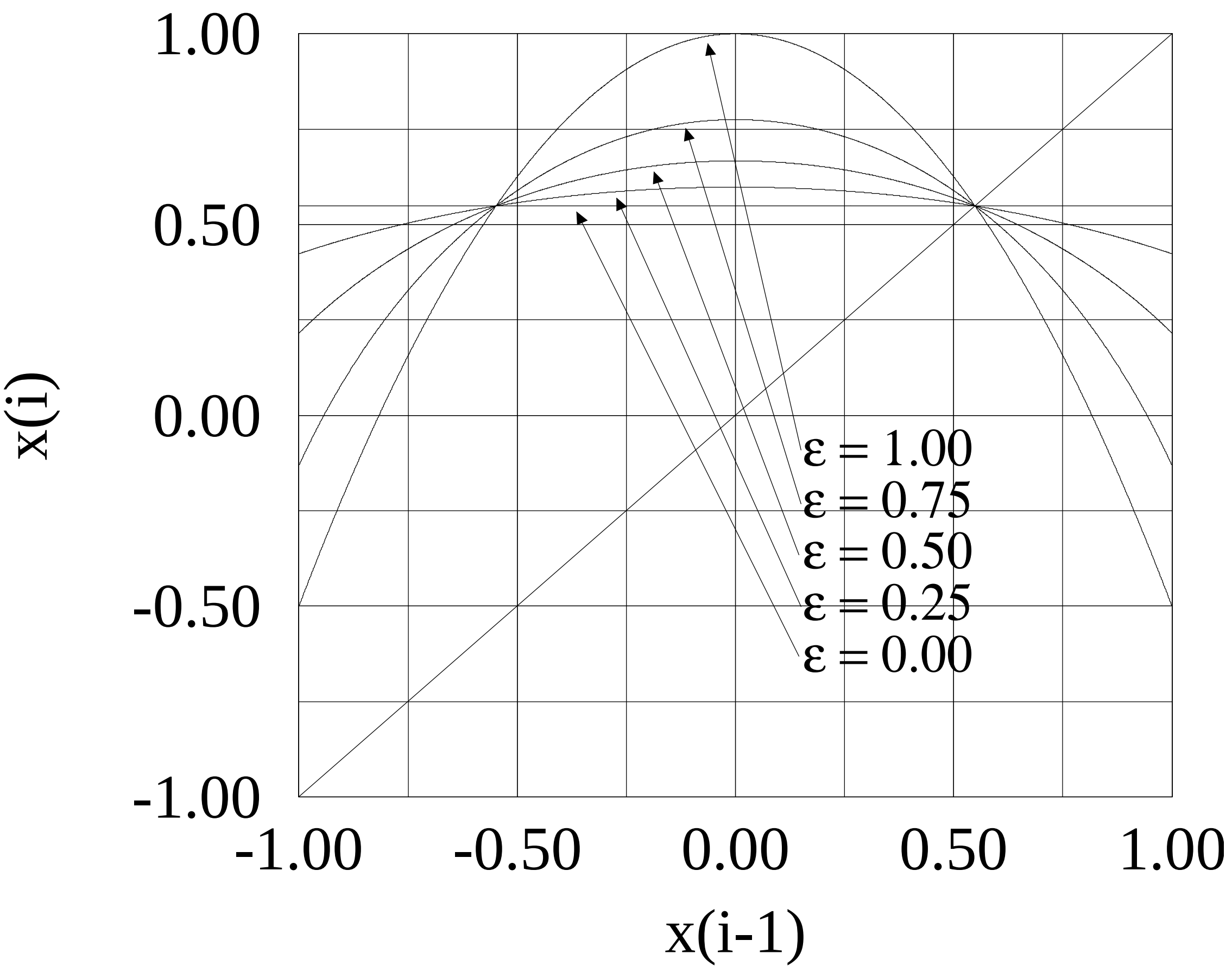

Frederick H. Willeboordse and Kunihiko Kaneko Fig. 10.

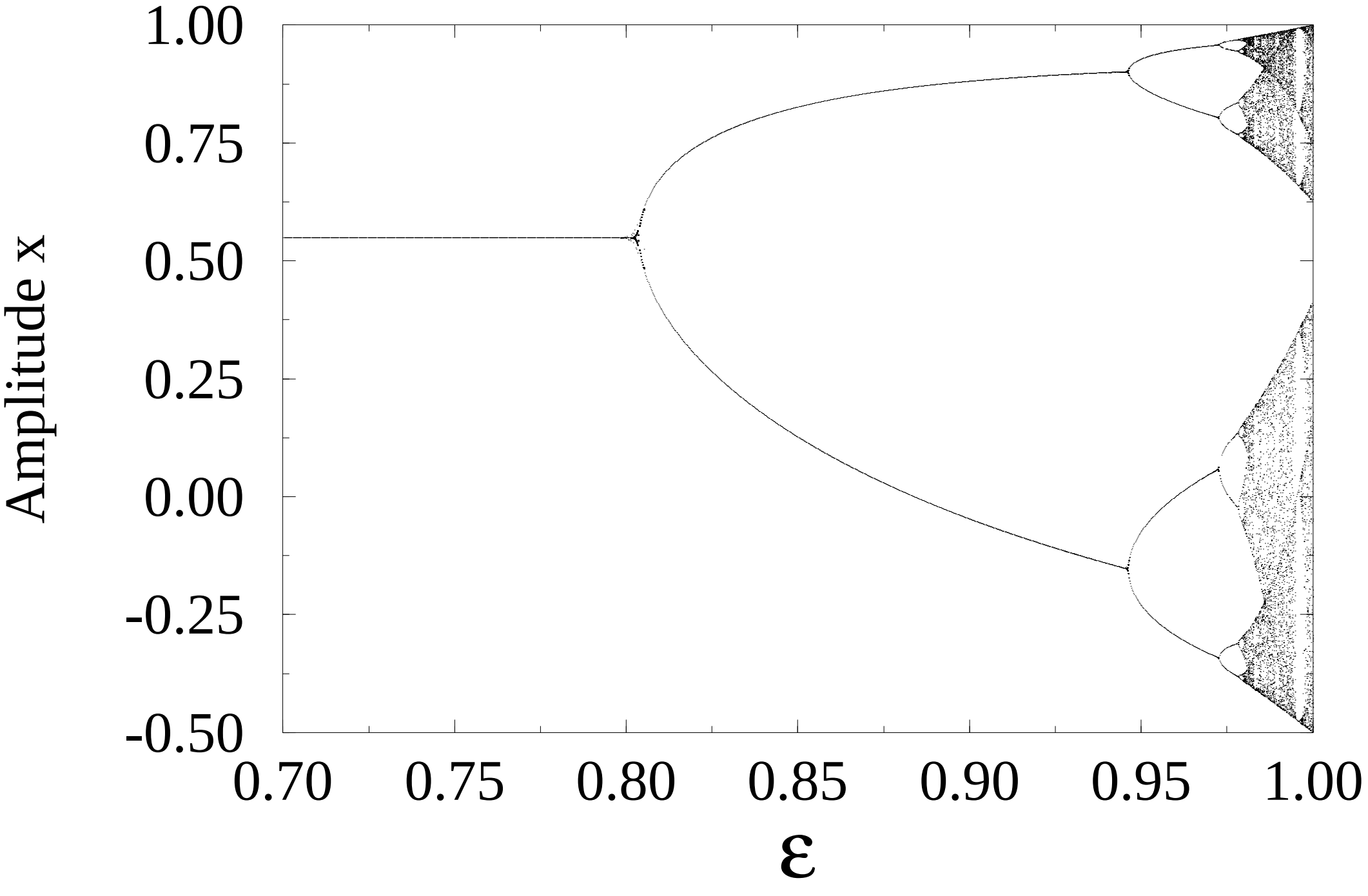

Frederick H. Willeboordse and Kunihiko Kaneko Fig. 11a).

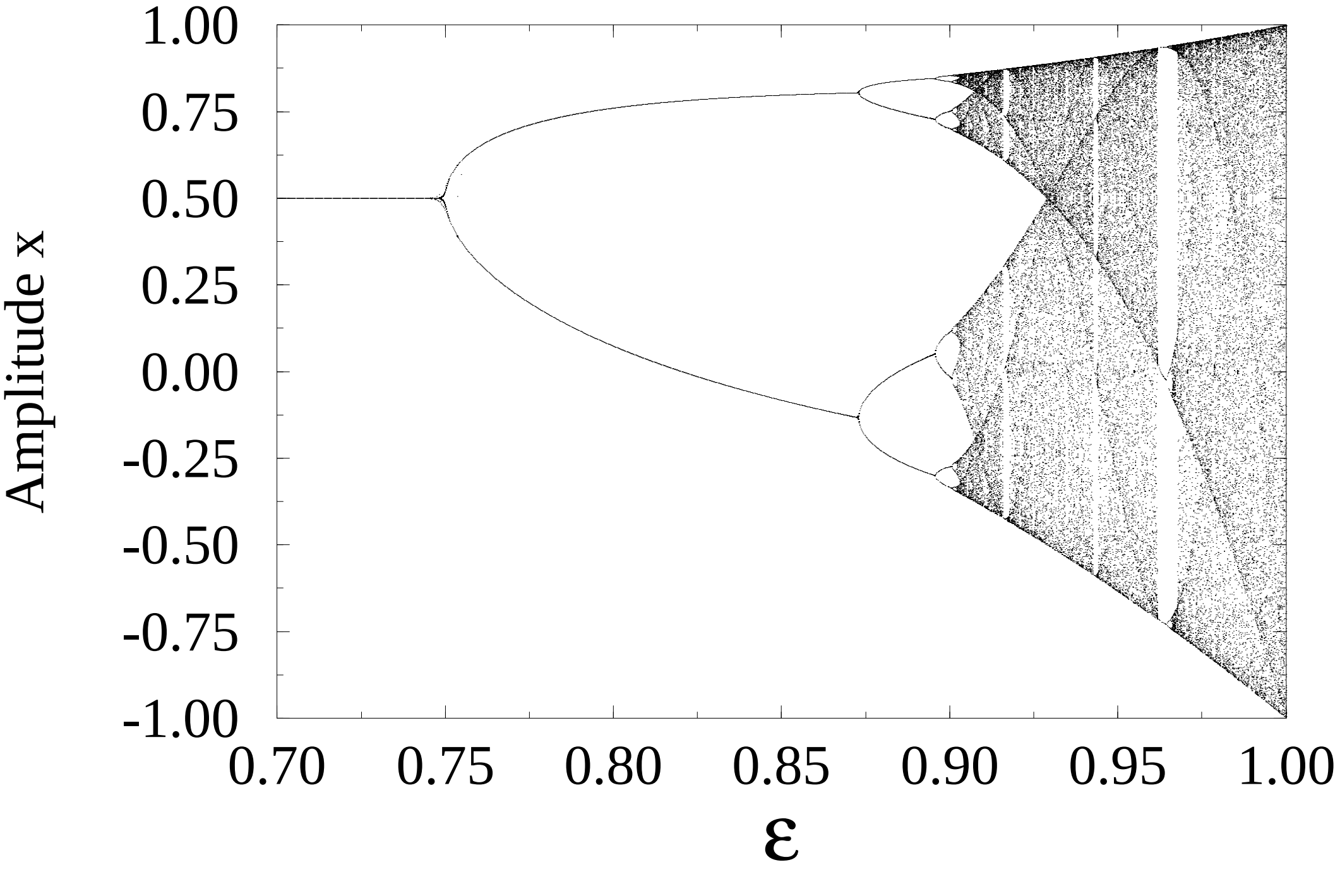

Frederick H. Willeboordse and Kunihiko Kaneko Fig. 11b).

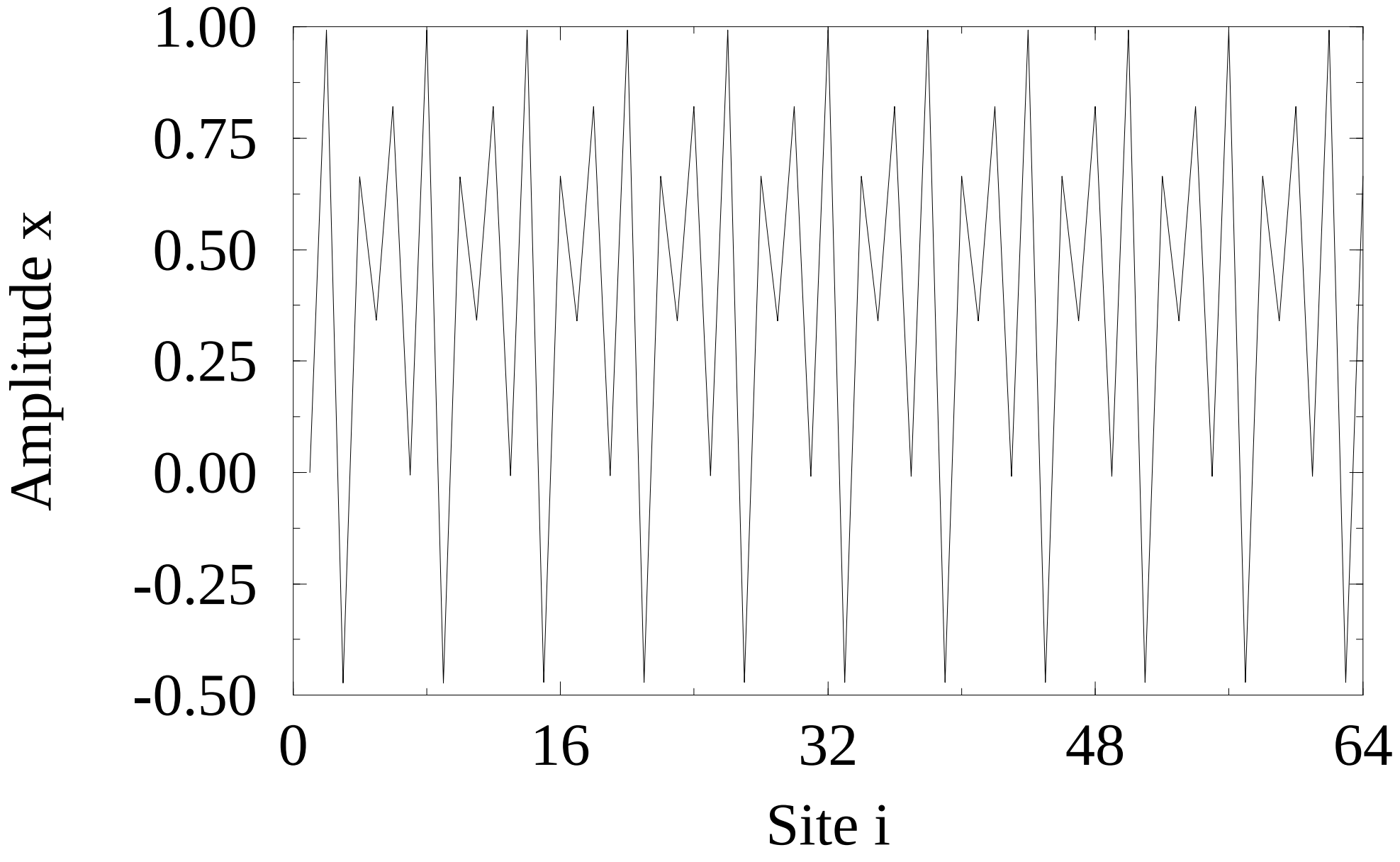

Frederick H. Willeboordse and Kunihiko Kaneko Fig. 12.

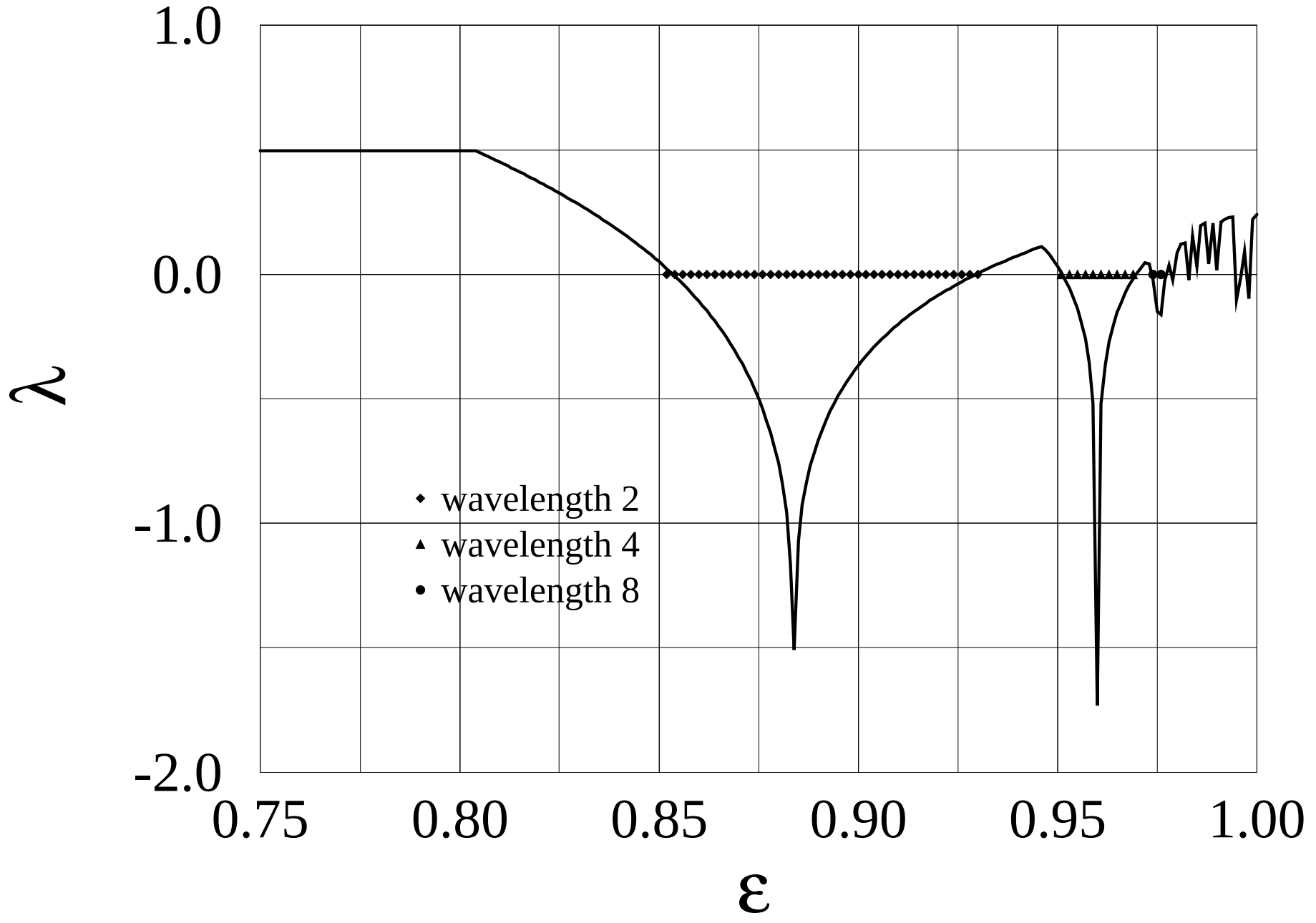

Frederick H. Willeboordse and Kunihiko Kaneko Fig. 13.

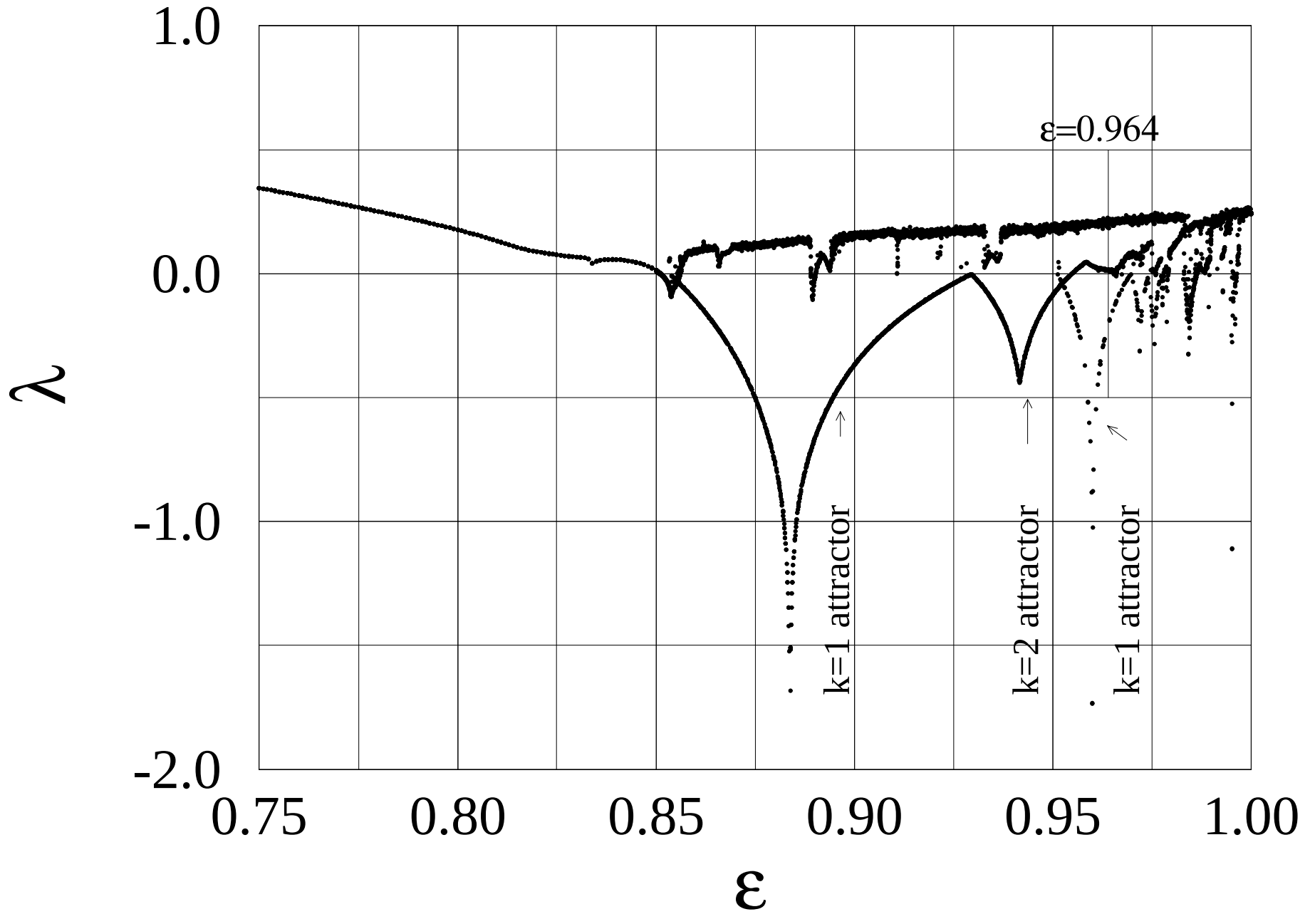

Frederick H. Willeboordse and Kunihiko Kaneko Fig. 14a).

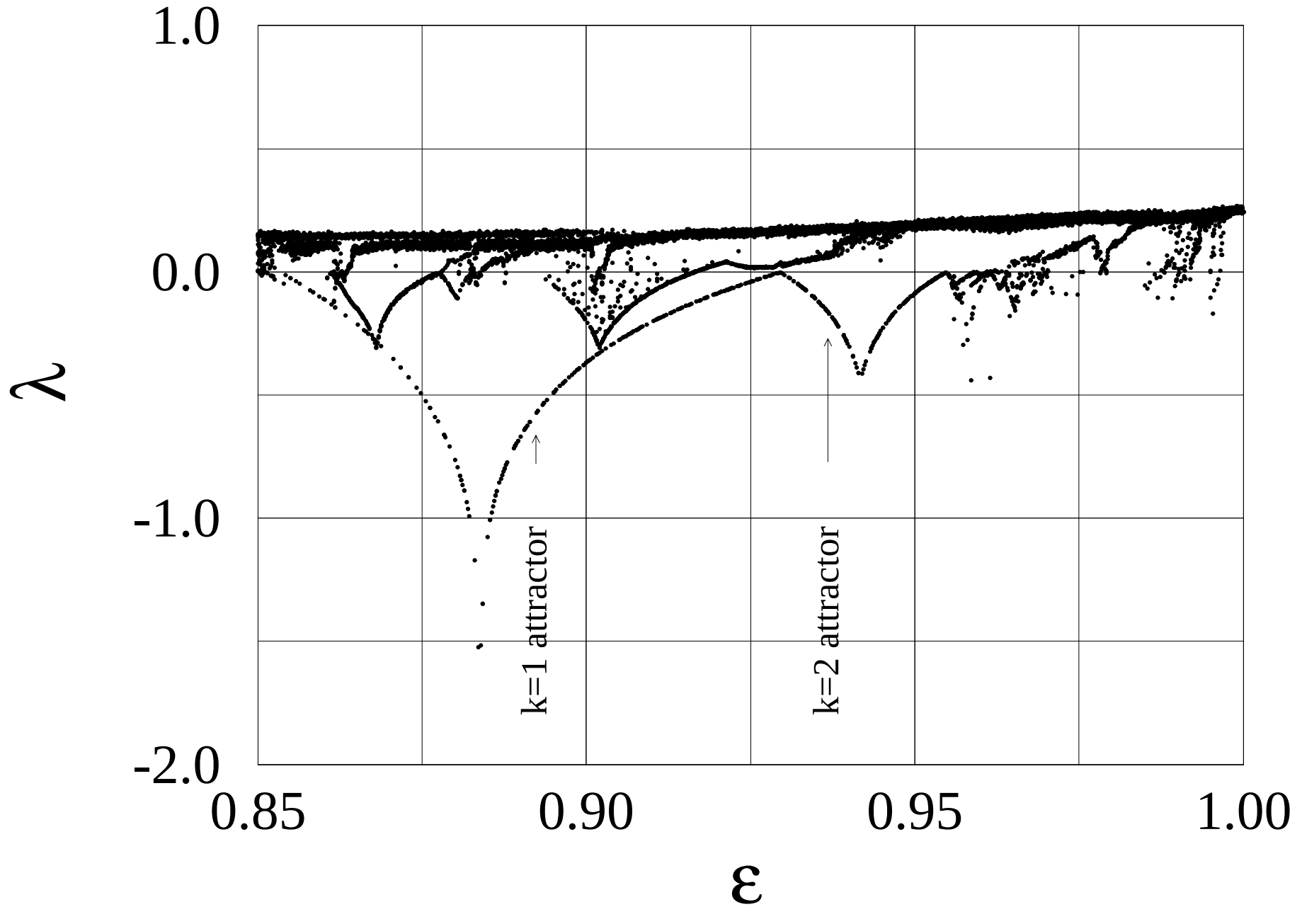

Frederick H. Willeboordse and Kunihiko Kaneko Fig. 14b).

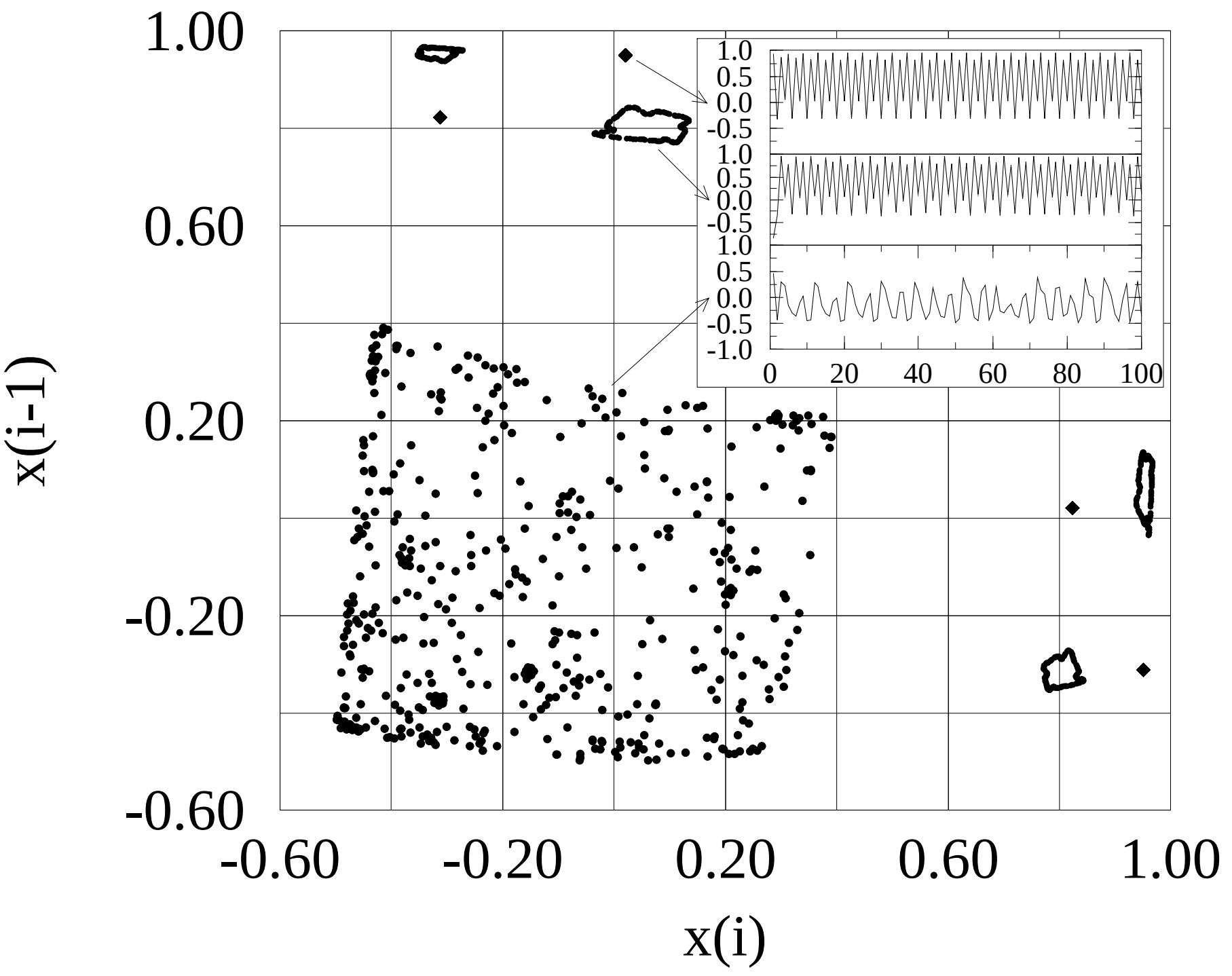

Frederick H. Willeboordse and Kunihiko Kaneko Fig. 15.

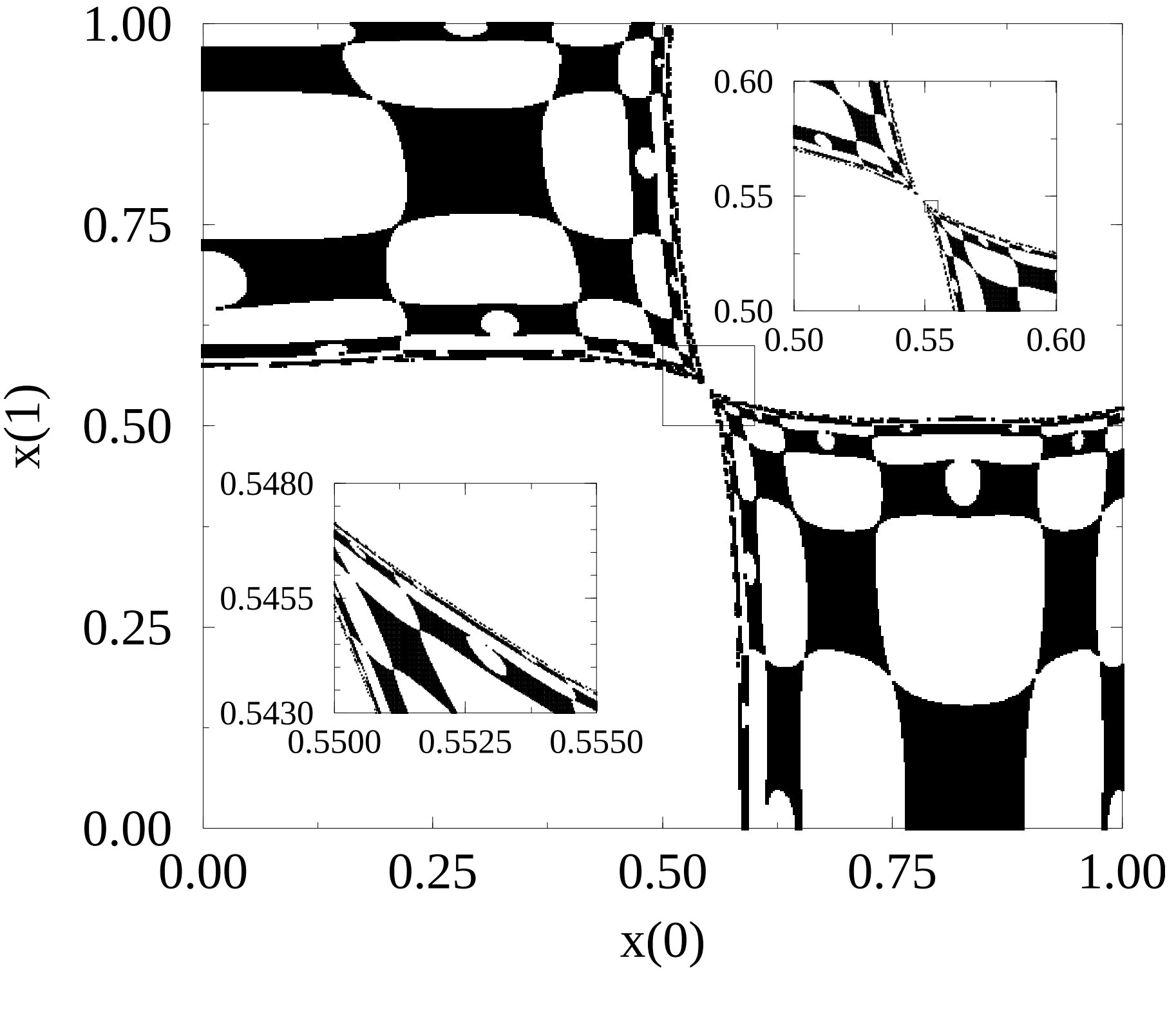

Frederick H. Willeboordse and Kunihiko Kaneko Fig. 16.

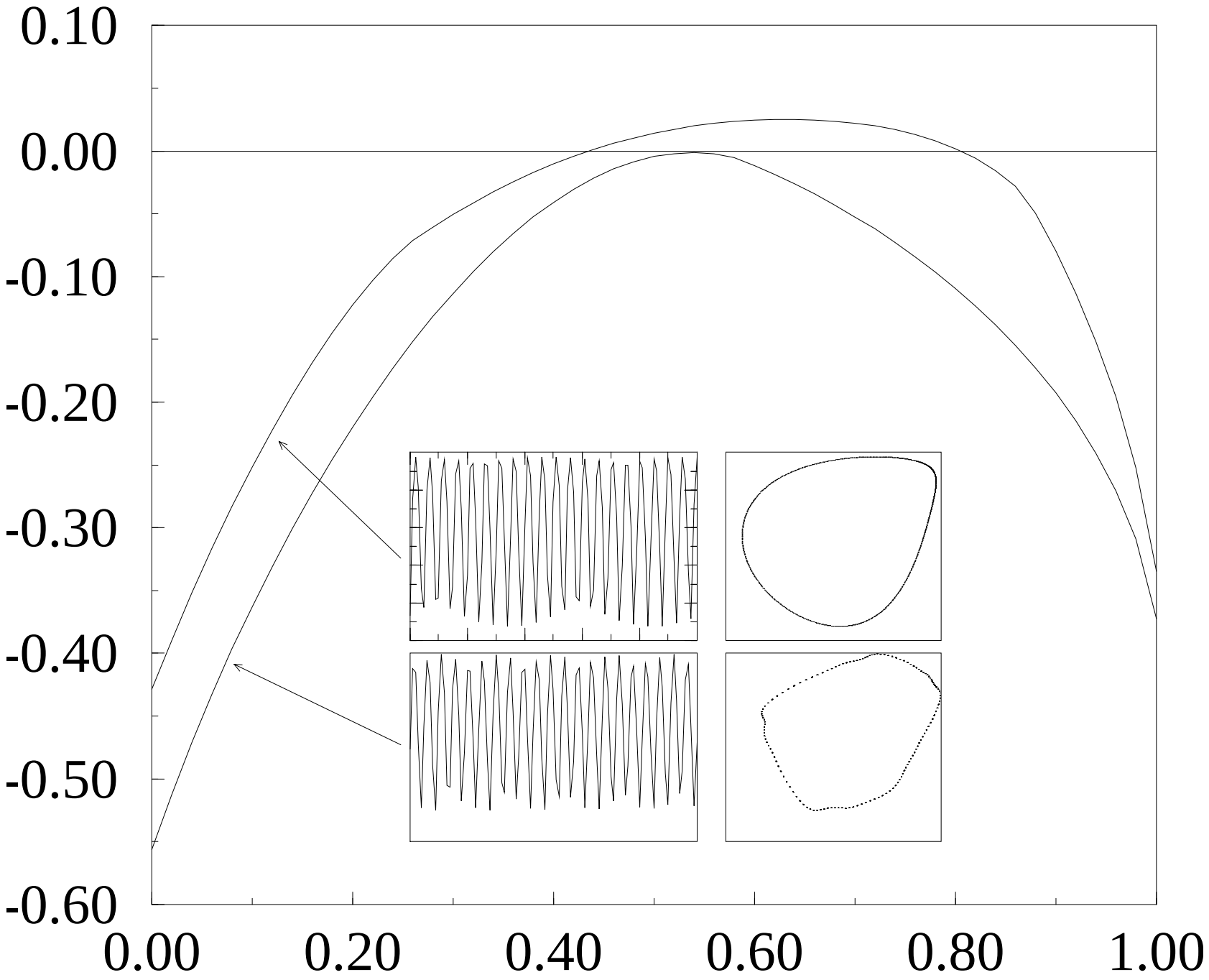

Frederick H. Willeboordse and Kunihiko Kaneko Fig. 17.

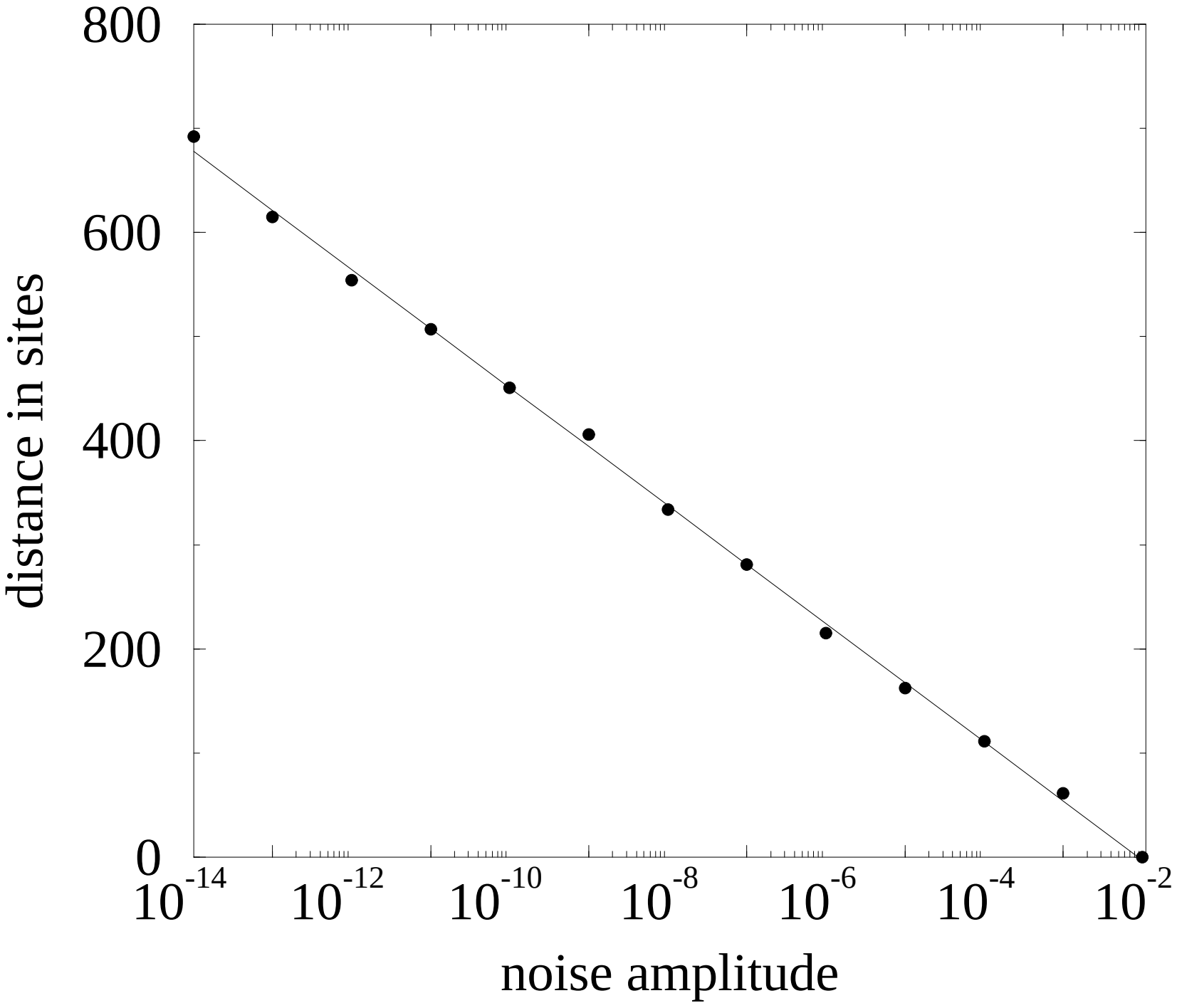

Frederick H. Willeboordse and Kunihiko Kaneko Fig. 18.

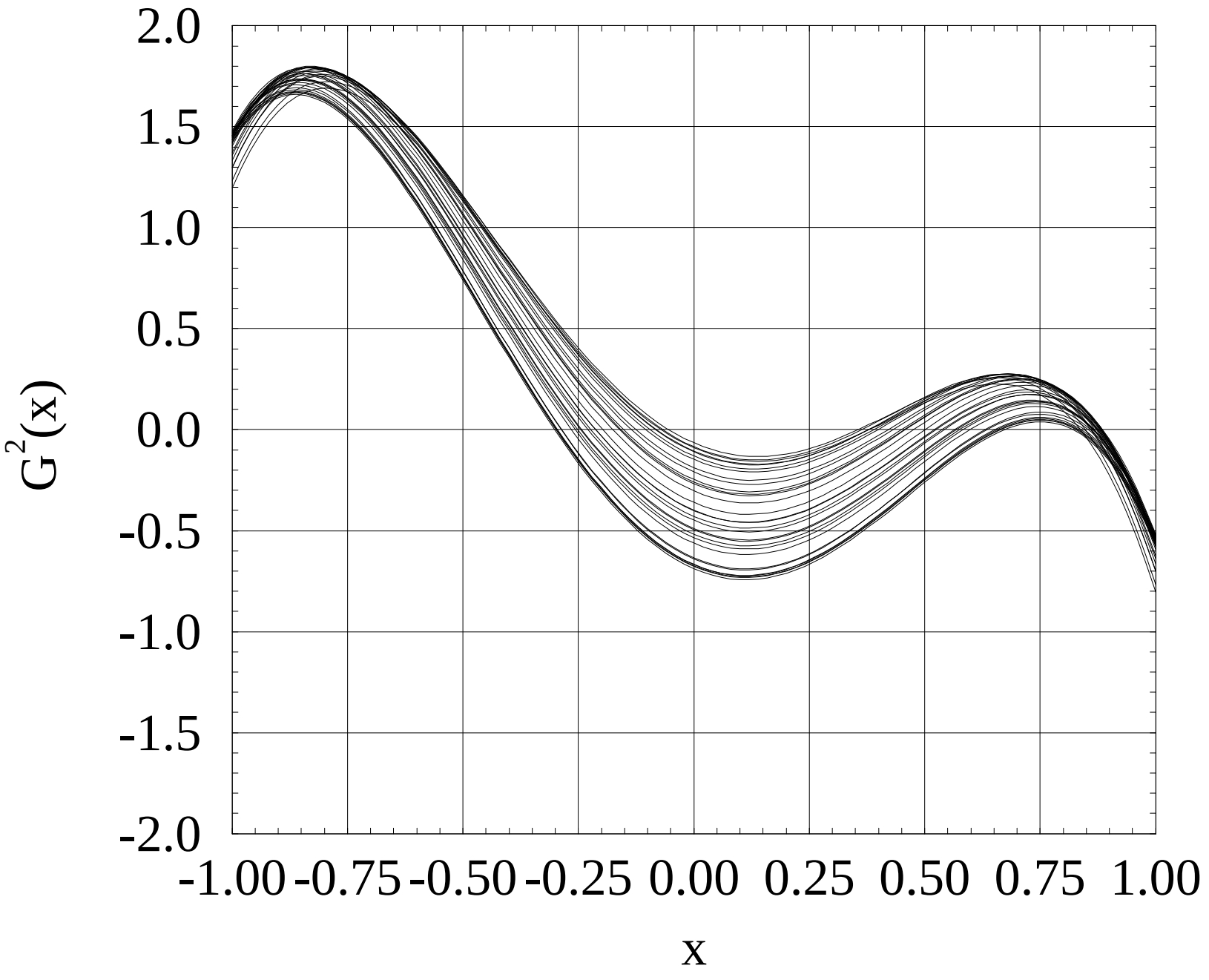

Frederick H. Willeboordse and Kunihiko Kaneko Fig. 19a).

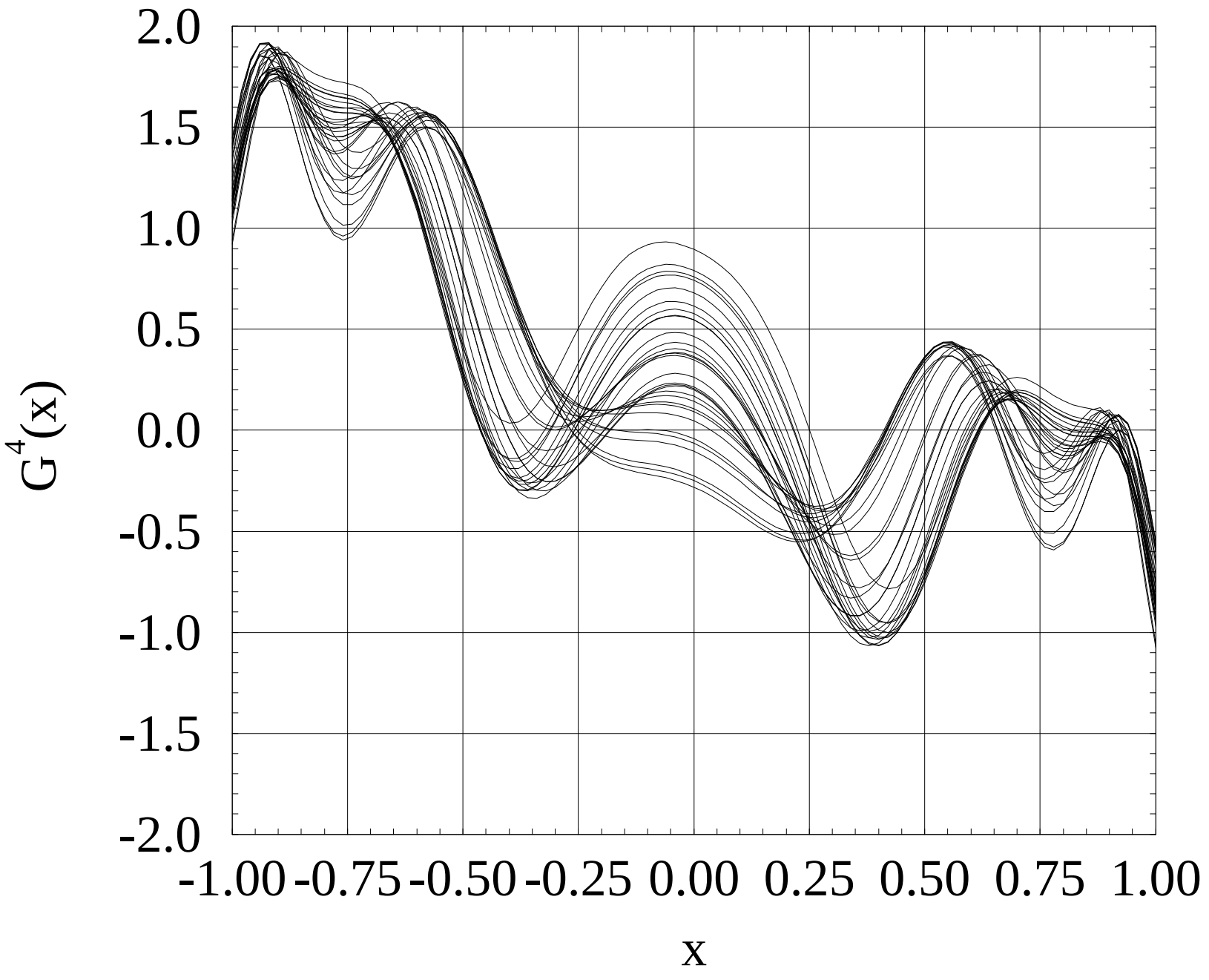

Frederick H. Willeboordse and Kunihiko Kaneko Fig. 19b).

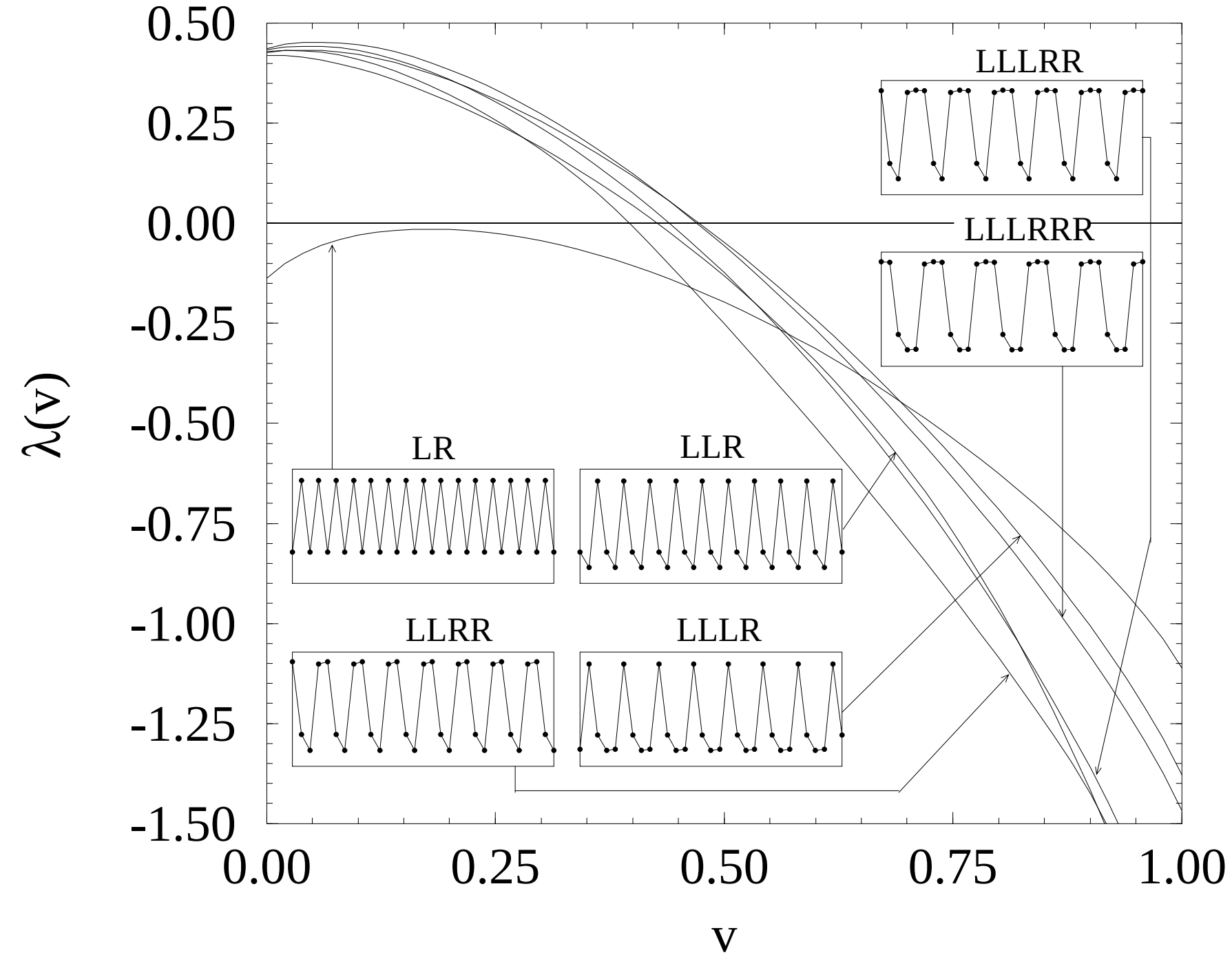

Frederick H. Willeboordse and Kunihiko Kaneko Fig. 20.

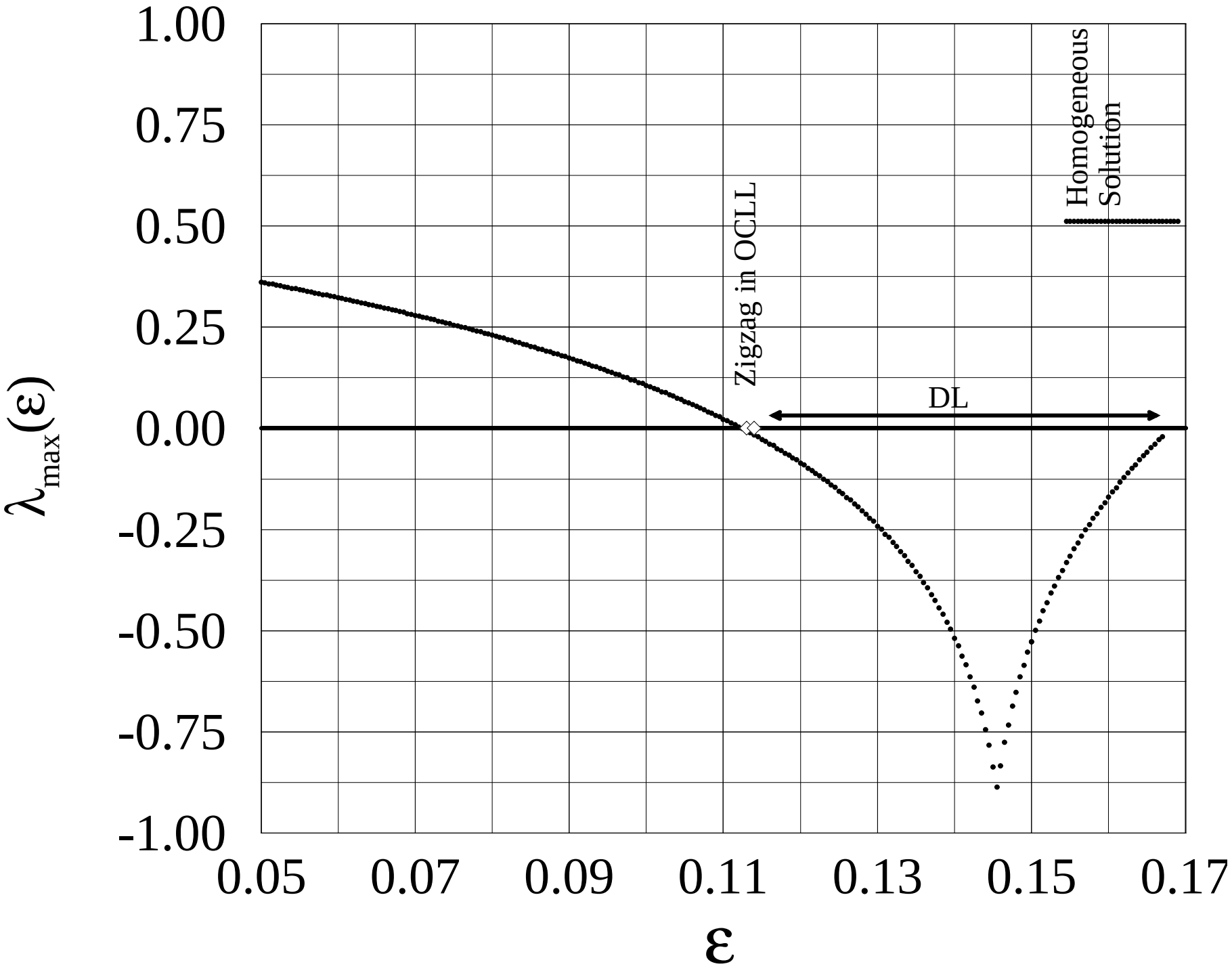

Frederick H. Willeboordse and Kunihiko Kaneko Fig. 21.

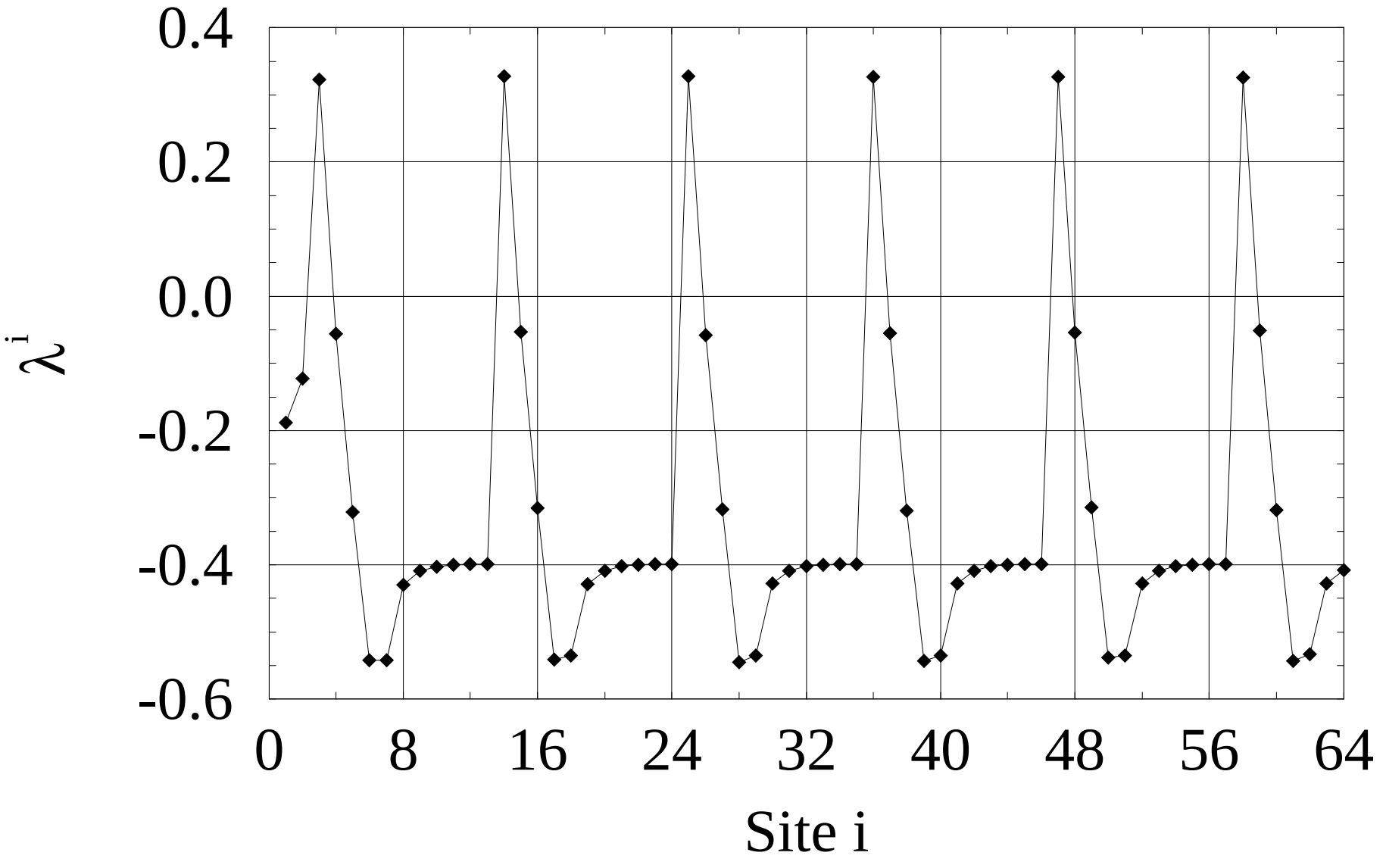

Frederick H. Willeboordse and Kunihiko Kaneko Fig. 22.

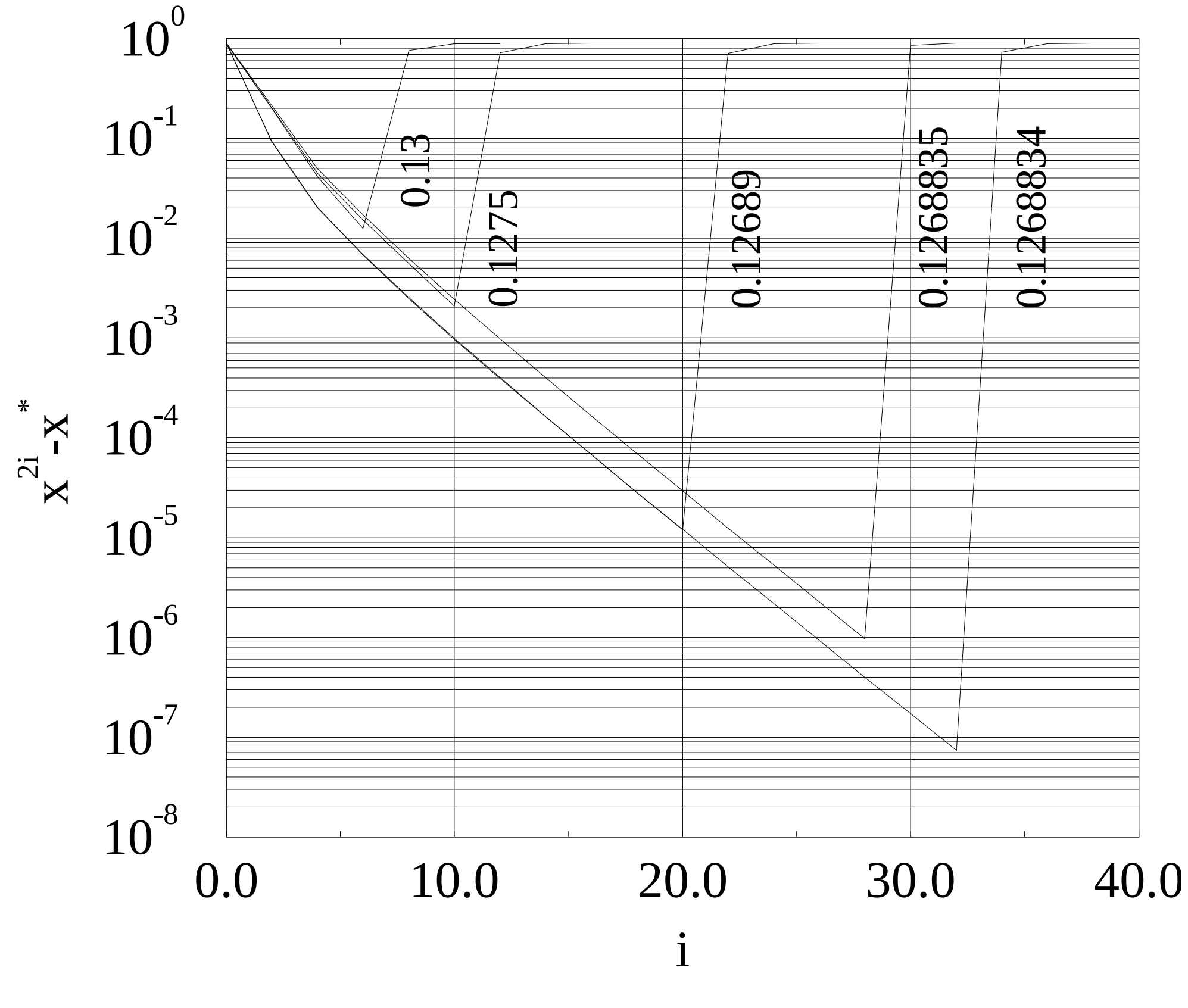

Frederick H. Willeboordse and Kunihiko Kaneko Fig. 23.

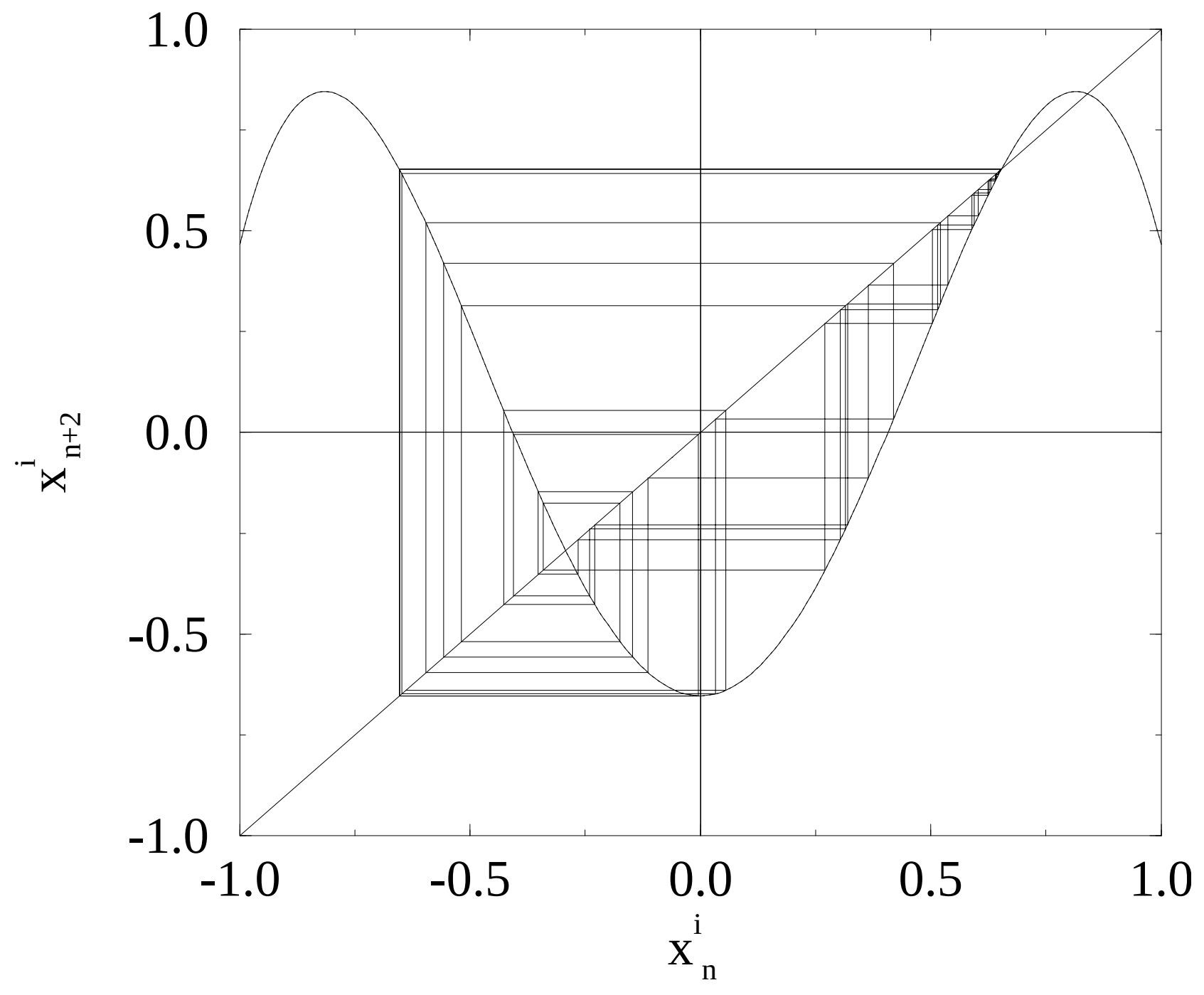

Frederick H. Willeboordse and Kunihiko Kaneko Fig. 24a).

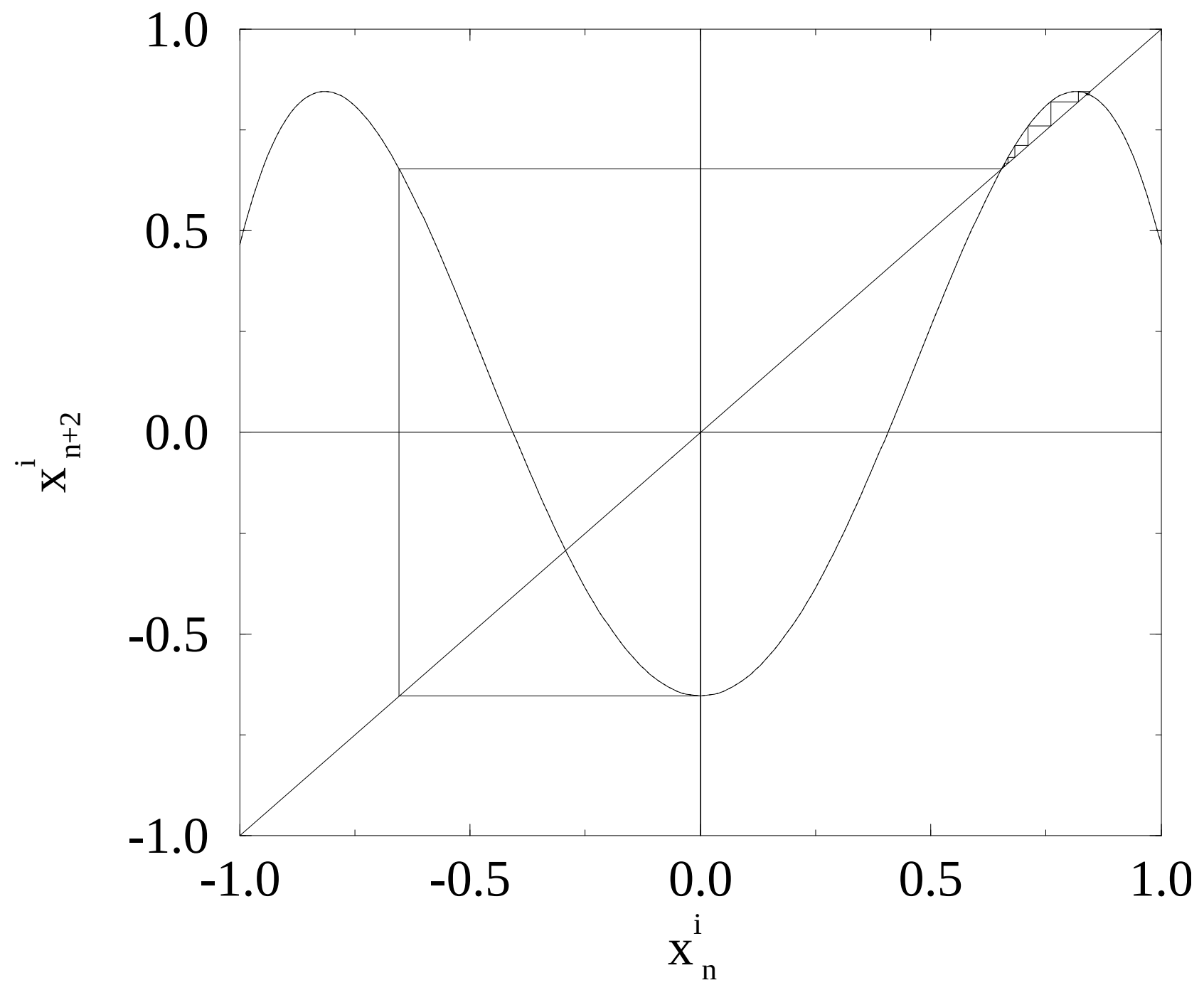

Frederick H. Willeboordse and Kunihiko Kaneko Fig. 24b).

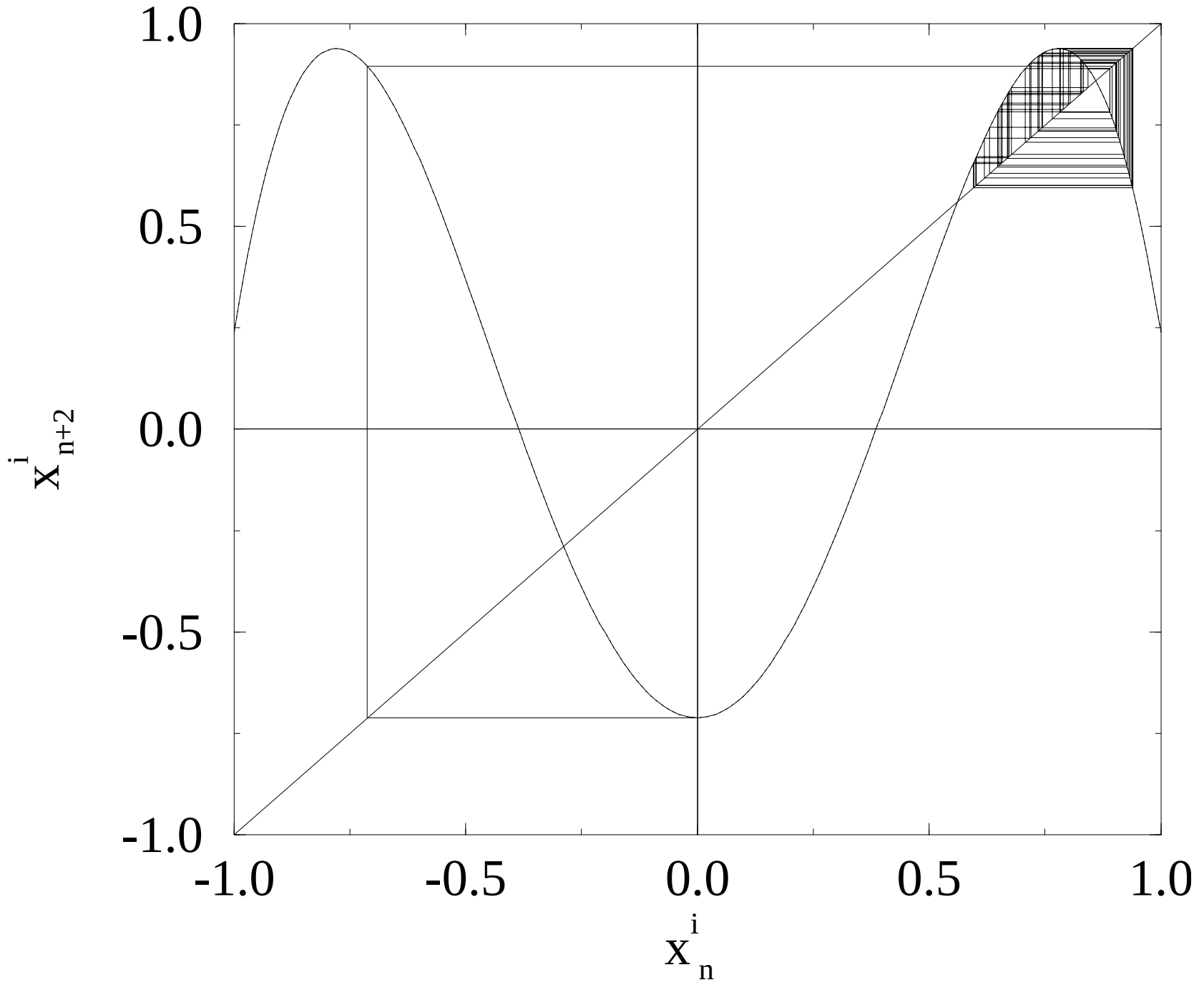

Frederick H. Willeboordse and Kunihiko Kaneko Fig. 25a).

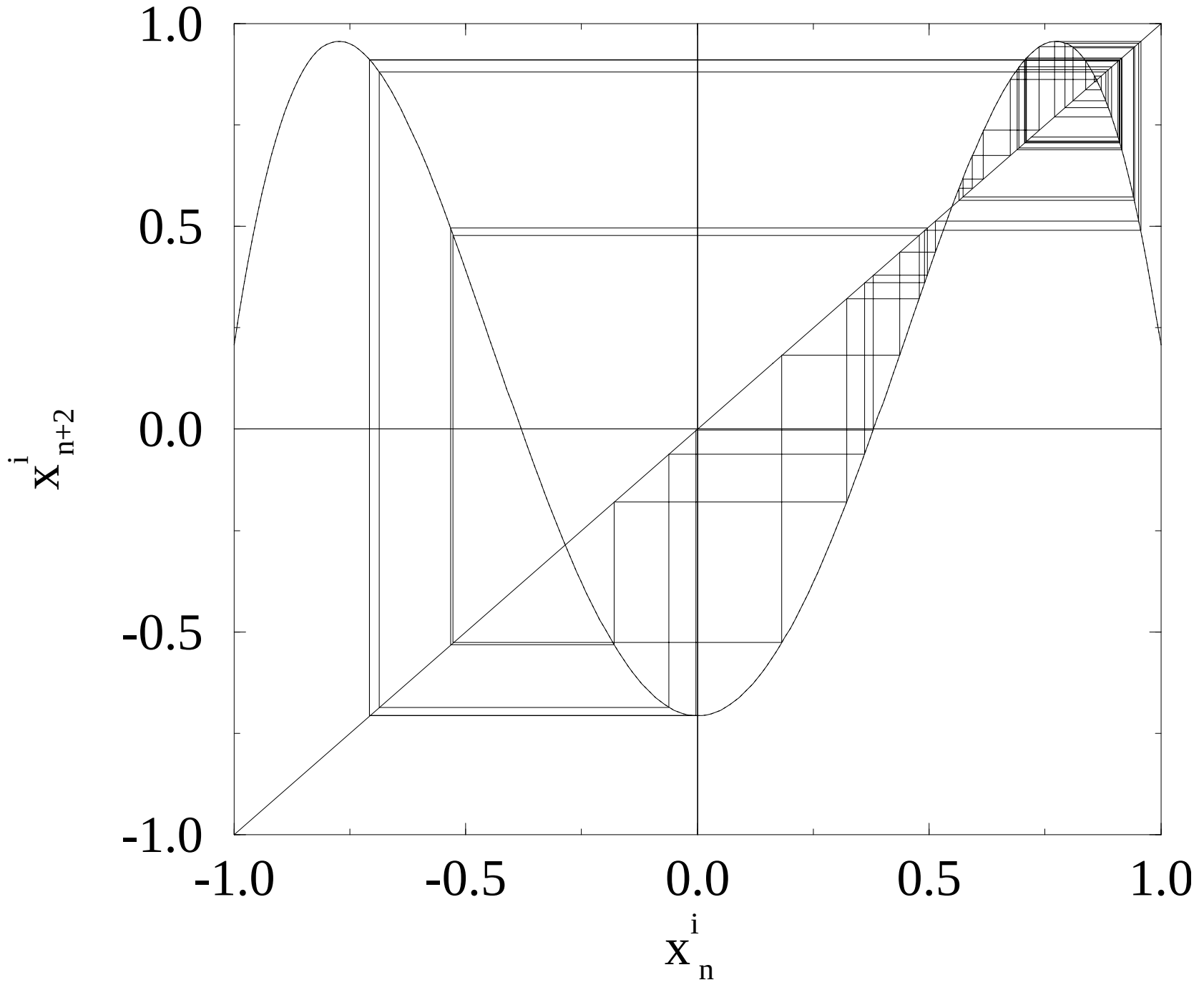

Frederick H. Willeboordse and Kunihiko Kaneko Fig. 25b).

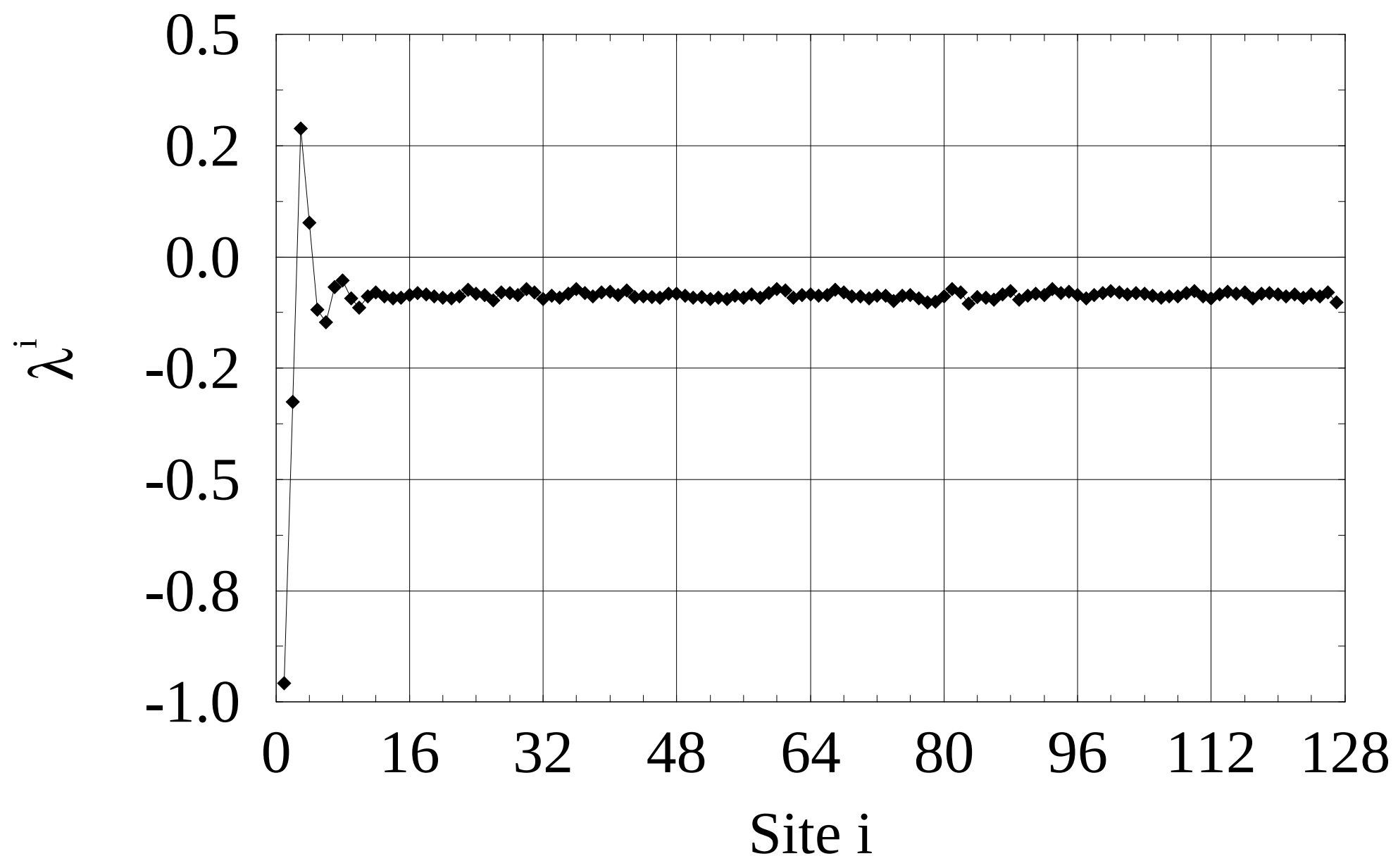

Frederick H. Willeboordse and Kunihiko Kaneko Fig. 26.

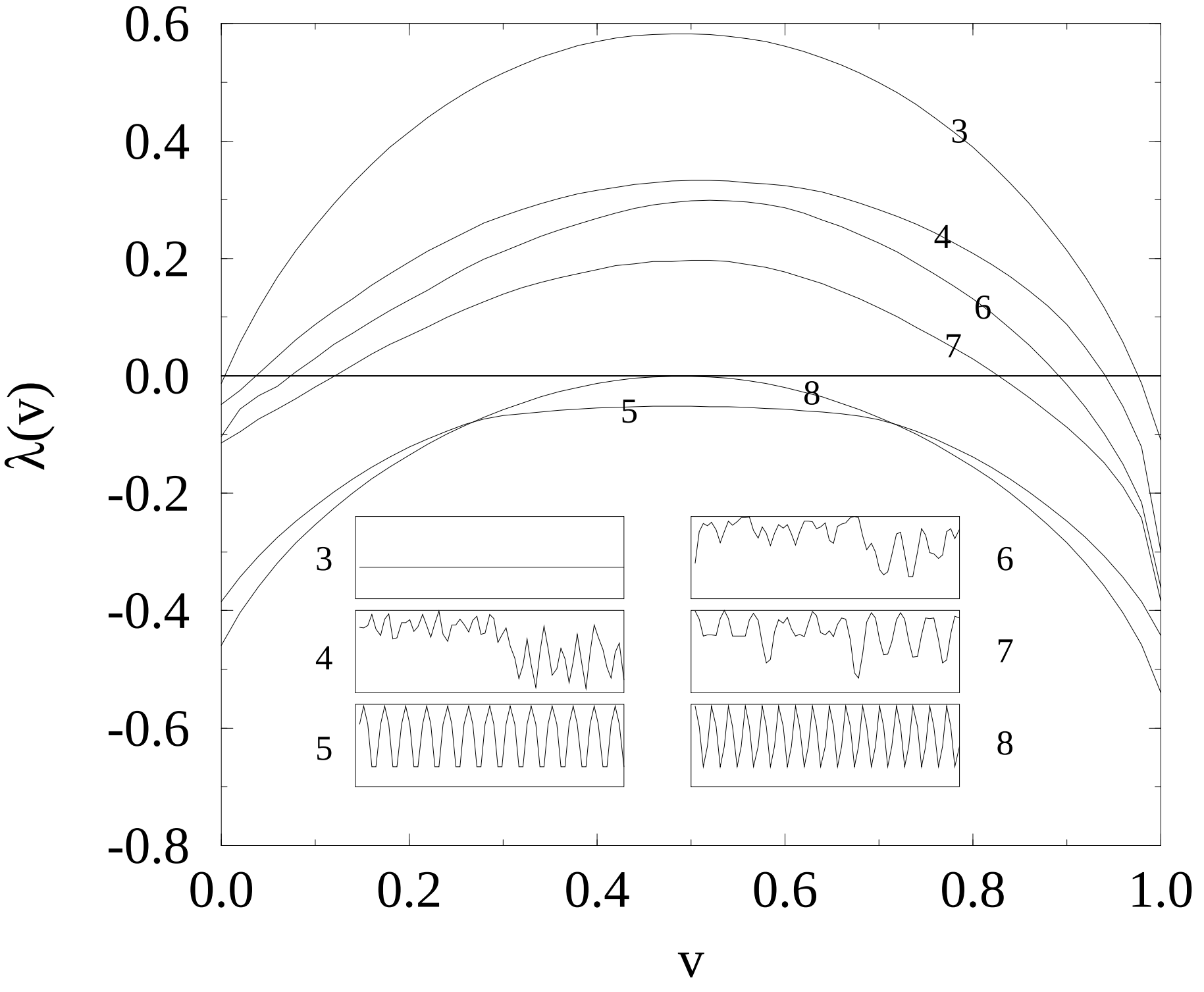

Frederick H. Willeboordse and Kunihiko Kaneko Fig. 27a).

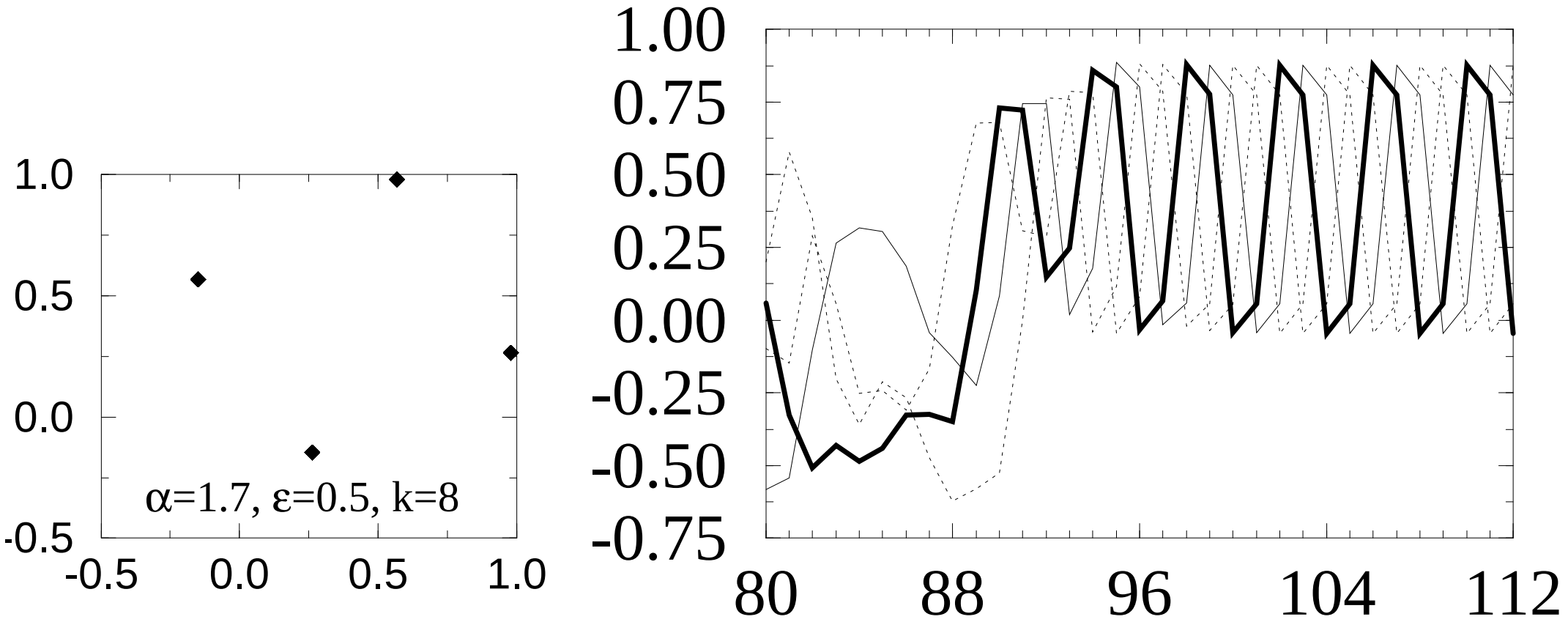

Frederick H. Willeboordse and Kunihiko Kaneko Fig. 27b).

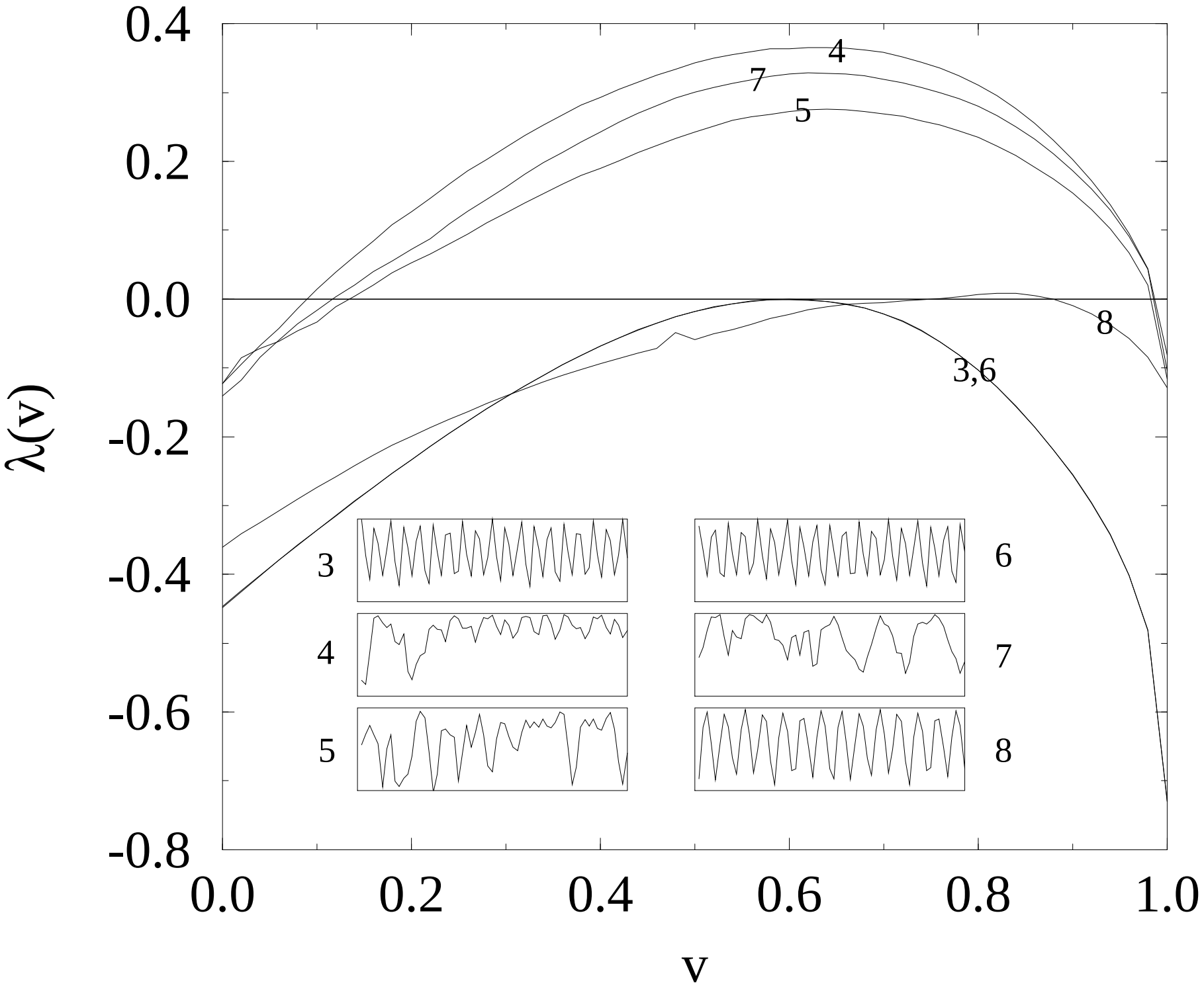

Frederick H. Willeboordse and Kunihiko Kaneko Fig. 27c).

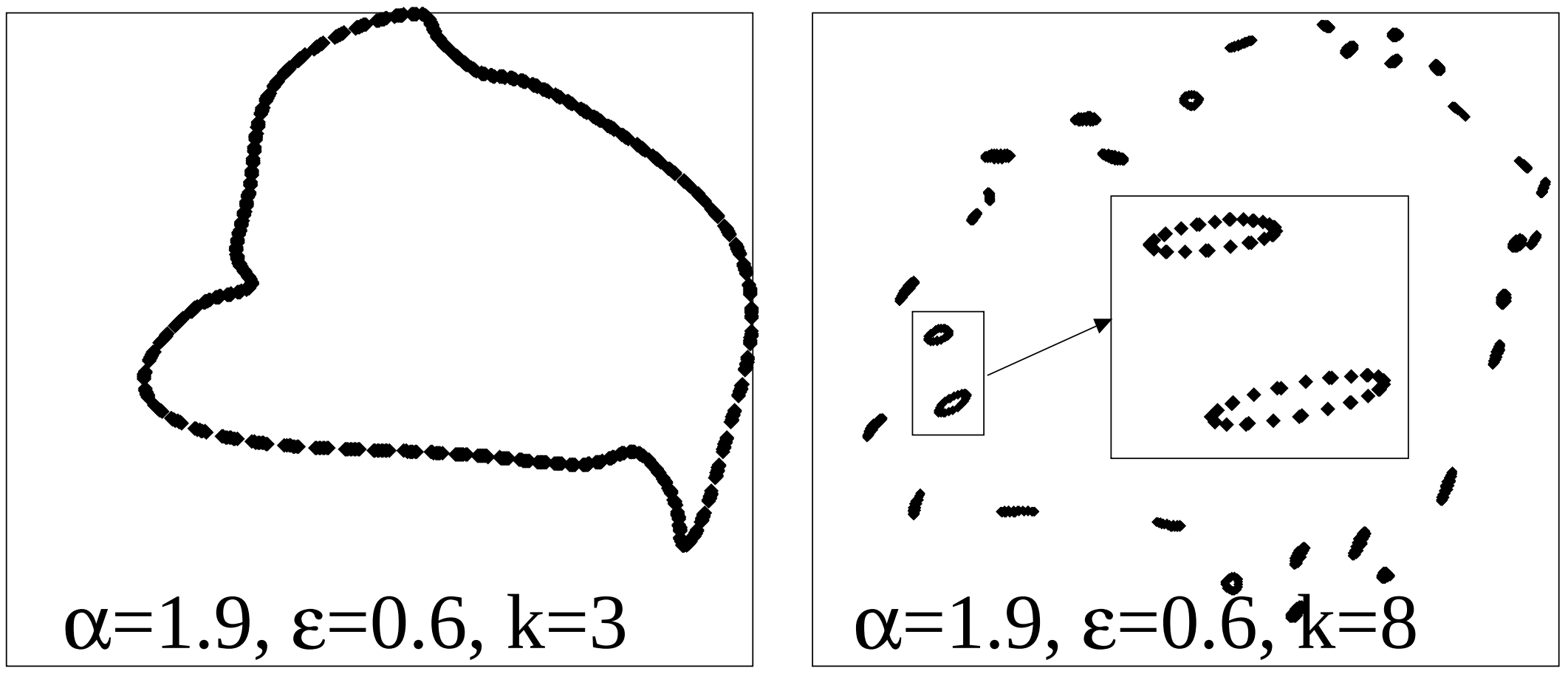

Frederick H. Willeboordse and Kunihiko Kaneko Fig. 27d).

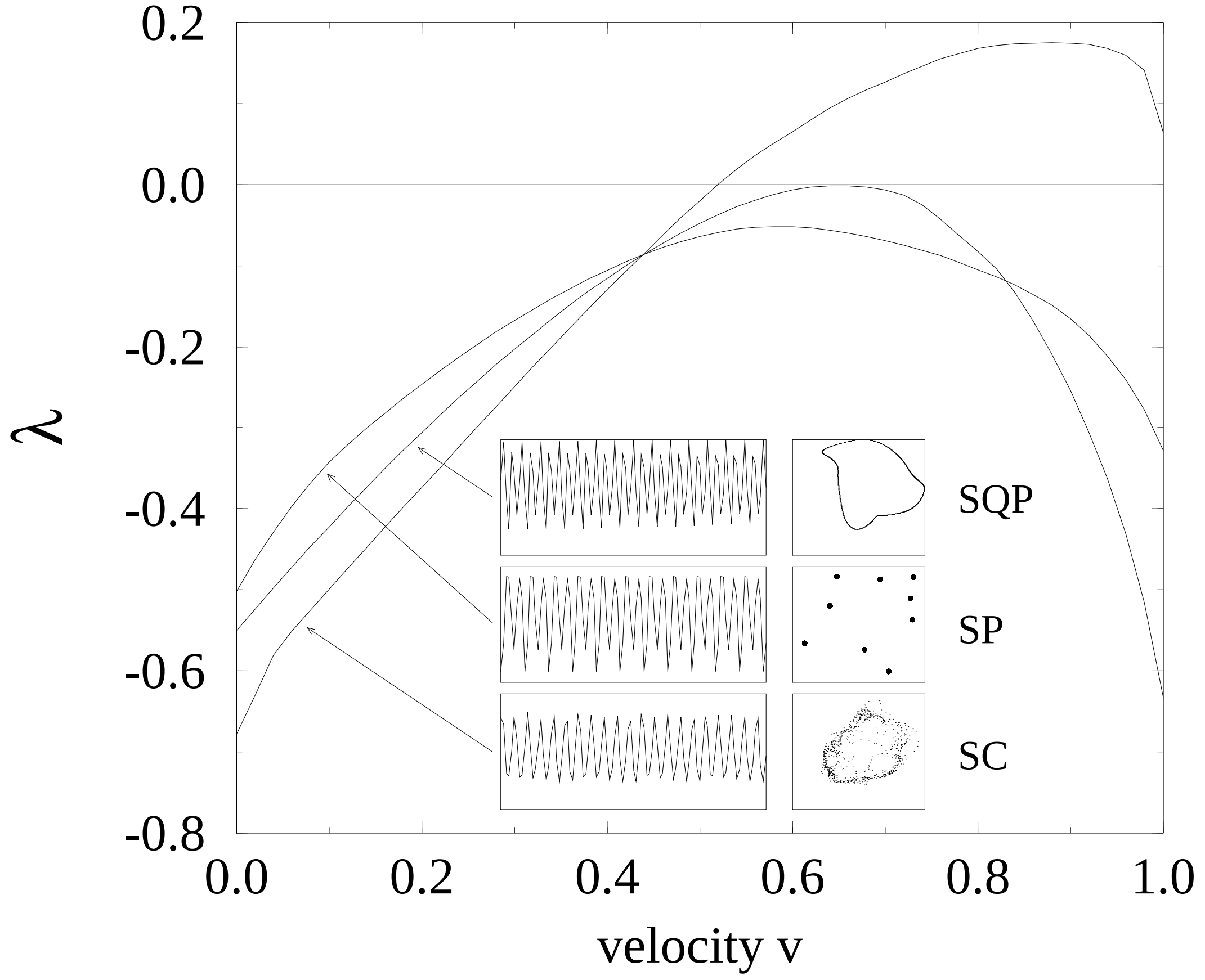

Frederick H. Willeboordse and Kunihiko Kaneko Fig. 28.

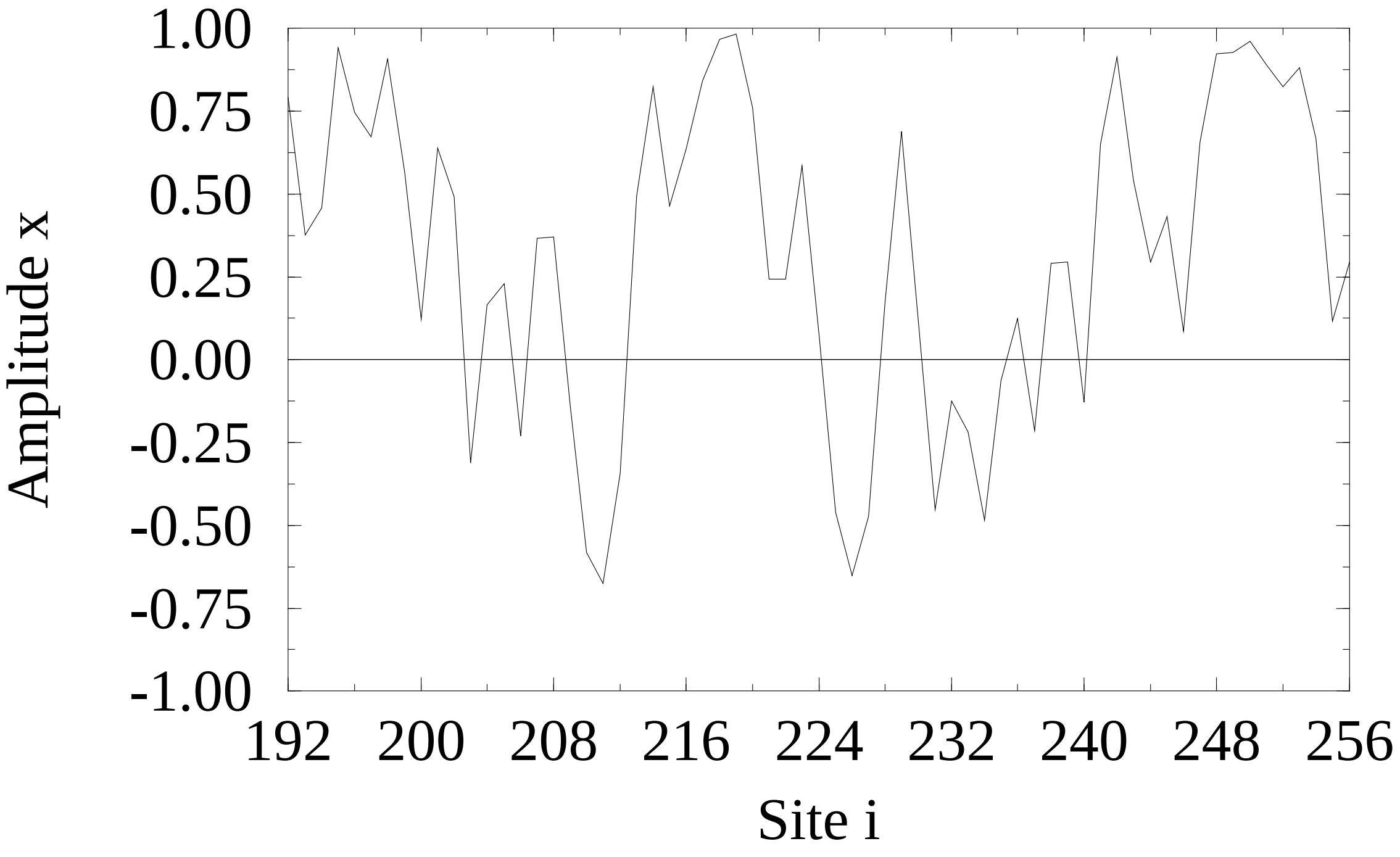

Frederick H. Willeboordse and Kunihiko Kaneko Fig. 29a).

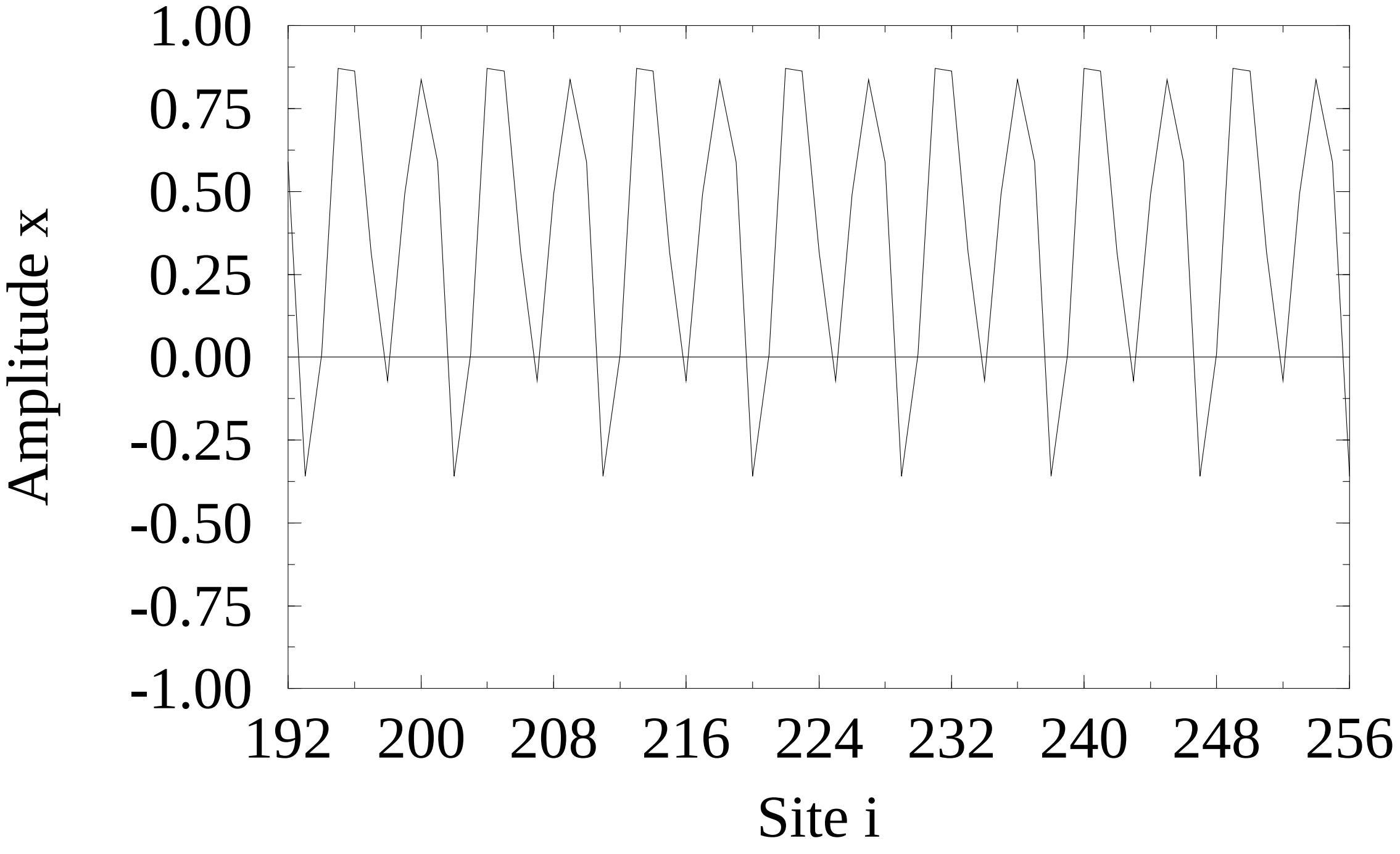

Frederick H. Willeboordse and Kunihiko Kaneko Fig. 29b).

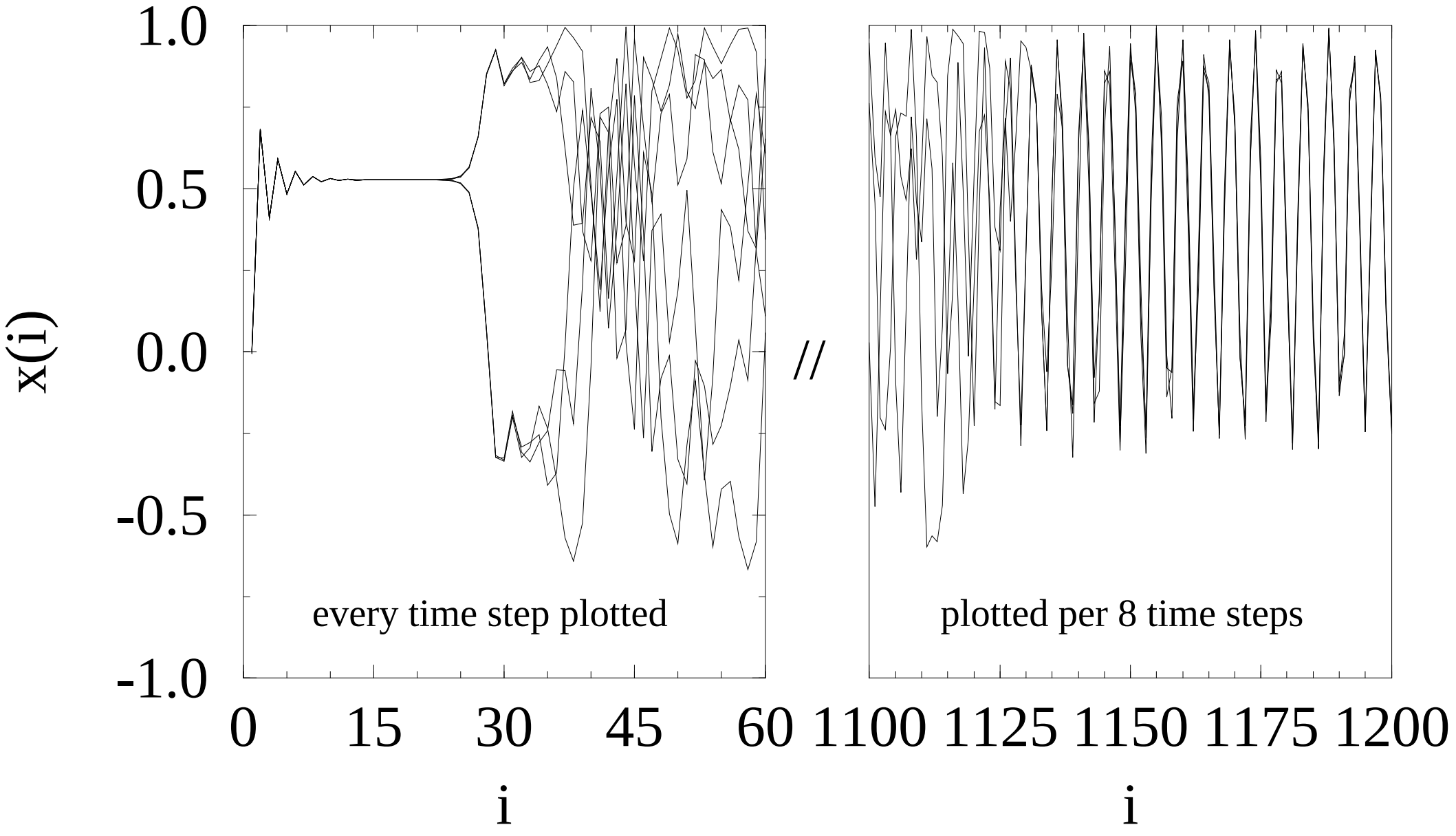

Frederick H. Willeboordse and Kunihiko Kaneko Fig. 30.